\documentstyle[twocolumn]{article}
\newcommand{\cqg}[1]{{\em Class.\ Quan.\ Grav.\ }{\bf #1}}
\newcommand{\grg}[1]{{\em Gen.\ Rel.\ Grav.\ }{\bf #1}}
\newcommand{\np}[1]{{\em Nucl.\ Phys.\ }{\bf #1}}
\newcommand{\pr}[1]{{\em Phys.\ Rev.\ }{\bf #1}}
\newcommand{\prl}[1]{{\em Phys.\ Rev.\ Lett.\ }{\bf #1}}
\newcommand{\pl}[1]{{\em Phys.\ Lett.\ }{\bf #1}}
\newcommand{\jmp}[1]{{\em J. Math.\ Phys.\ }{\bf #1}}
\newcommand{\jgp}[1]{{\em J. Geom.\ Phys.\ }{\bf #1}}
\newcommand{\cmp}[1]{{\em Commun.\ Math.\ Phys.\ }{\bf #1}}
\newcommand{\mpl}[1]{{\em Mod.\ Phys.\ Lett.\ }{\bf #1}}
\newcommand{\ijmp}[1]{{\em Int.\ J. Mod.\ Phys.\ }{\bf #1}}
\newcommand{\apny}[1]{{\em Ann.\ Phys.\ (N.Y.) }{\bf #1}}
\newcommand{\ncim}[1]{{\em Nuovo Cim.\ }{\bf #1}}

\newcommand{\pR}[1]{{\em Phys.\ Rept.\ }{\bf #1}}               

\newcommand{\atmp}[1]{{\em Adv.\ Theor.\ Math.\ Phys.\ }{\bf #1}}
\newcommand{\ap}[1]{{\em Appl.\ Phys.\ }{\bf #1}}
\newcommand{\lnp}[1]{{\em Lect.\ Notes\ Phys.\ }{\bf #1}}

\newcommand{\pw}[1]{{\em Phys.\ World\ }{\bf #1}}
\newcommand{\rpp}[1]{{\em Rept.\ Prog.\ Phys.\ }{\bf #1}}
\newcommand{\fp}[1]{{\em Found.\ Phys.\ }{\bf #1}}
\newcommand{\ahp}[1]{{\em Annales\ Henri\ Poincare\ }{\bf #1}}
\newcommand{\jp}[1]{{\em J.\ Phys.\ }{\bf #1}}
\newcommand{\ujp}[1]{{\em Ukr.\ J.\ Phys.\ }{\bf #1}}
\newcommand{\jhep}[1]{{\em JHEP\ }{\bf #1}}
\newcommand{\Ap}[1]{{\em Annalen\ Phys.\ }{\bf #1}}
\newcommand{\n}[1]{{\em Nature\  }{\bf #1}}
\newcommand{\sa}[1]{{\em Sci.\ Am.\  }{\bf #1}}
\newcommand{\tmp}[1]{{\em Theor.\ Math.\ Phys.\  }{\bf #1}}
\newcommand{\tmf}[1]{{\em Theor.\ Math.\ Fiz.\  }{\bf #1}}
\newcommand{\ijgmmp}[1]{{\em Int.\ J.\ Geom.\ Meth.\ Mod.\ Phys.\  }{\bf #1}}
\newcommand{\rmf}[1]{{\em Rev.\ Mex.\ Fis.\  }{\bf #1}}
\newcommand{\aipcp}[1]{{\em AIP\ Conf.\ Proc.\  }{\bf #1}}
\newcommand{\npps}[1]{{\em Nucl.\ Phys.\ Proc.\ Suppl.\  }{\bf #1}}
\newcommand{\hpa}[1]{{\em Helv.\ Phys.\ Acta\ }{\bf #1}}
\newcommand{\lrr}[1]{{\em Living Rev.\ Rel.\ }{\bf #1}}

\newcommand{\pra}[1]{{\em Pramana\ }{\bf #1}}
\newcommand{\nc}[1]{{\em Nuovo Cim.\ }{\bf #1}}
\newcommand{\ac}[1]{{\em Acta Cosmologica }{\bf #1}}

\date{August, 2005}

\begin{document}
\onecolumn
\title{Bibliography of Publications related to \\
Classical Self-dual variables \\and Loop Quantum Gravity
\footnote{Previous title: Bibliography of Publications related to
 Classical and Quantum Gravity in terms of Connection and Loop
Variables. Even older title: Bibliography of publications related
to classical and quantum gravity in terms of the Ashtekar
variables.}}
\author{
       Last updated by, \\
       Alejandro Corichi\footnote{corichi@nucleares.unam.mx}
        and Alberto Hauser. \\
       Instituto de Ciencias Nucleares, Universidad Nacional
       Aut\'onoma de M\'exico,\\ A. Postal 70-543, M\'exico D.F.,
       M\'exico.
}

\maketitle

\begin{abstract}
This bibliography attempts to give a comprehensive overview of all
the literature related to what is known as the Ashtekar-Sen
connection and the Rovelli-Smolin loop variables, from which the
program currently known as {\it Loop Quantum Gravity} emerged. The
original version was compiled by Peter H\"ubner in 1989, and it
has been subsequently updated by Gabriela Gonz\'alez, Bernd
Br\"ugmann, Monica Pierri, Troy Schilling, Christopher Beetle,
Alejandro Corichi and Alberto Hauser. The criteria for inclusion
in this list are the following: A paper in the classical theory is
included if it deals with connection variables for gravity. If the
paper is in the quantum domain, it is included when it is related
directly with gravity using connection/loop variables, with
mathematical aspects of connections, or when it introduces
techniques that might be useful for the construction of the (loop)
quantum theory of gravity. Information about additional
literature, new preprints, and especially corrections are always
welcome.
\end{abstract}

\newpage
\twocolumn

\section*{Pointers}


Here are some suggestions, intended to serve as entry points into
the literature.

First of all, for a complete and authoritative presentation of
canonical gravity in the Ashtekar variables there is of course
Ashtekar's latest book \cite{AA:2} which appeared in 1991.

Rather complete reviews of canonical gravity in the Ashtekar
variables can be found in Rovelli \cite{rovelli1991}, Kodama
\cite{kodama} and Smolin \cite{smolin1992}. For a critical
appraisal of canonical quantum gravity see Kucha{\v r}
\cite{kuchar}. An overview over different approaches to quantum
gravity is given by Isham in \cite{isham}.

Some now classic treatments of the two most prominent viewpoints
towards LQG, namely the `connection' and `loop-spin networks '
representations are given by Ashtekar et. al. \cite{alm2t} on one
side, and De Pietri and Rovelli \cite{depietri1} on the other. A
{\it dialogue concerning the two chief World systems} is given in
\cite{depietri2}. Note that nowadays the distinction between
connection and loop representations is no longer an issue.

Let us now mention some of the most recent introductory literature
to loop quantum gravity. Firstly, there are several primer
introductions to the subject, written for different purposes. For
instance, there was for many years the canonical primer by Pullin
\cite{pullin}. Unfortunately, it is now somewhat dated. Good
introductions to spin networks and recoupling theory needed in LQG
are given by the primers by Rovelli \cite{Rovelli:1998gg} and
Major \cite{seth}. There are recent up-to-date accounts written
for non-experts that give nice motivation, historical perspective
and an account of recent and in progress work from two different
perspectives by Ashtekar \cite{AA:NJP} and Smolin
\cite{Smolin:2004sx}. There are also technical reviews that give
many details and are certainly a good read such as the one by
Ashtekar and Lewandowski \cite{AL:review}, Perez
\cite{Perez:2004hj},  Thiemann \cite{Thiemann:2002nj}, and (from
an outside perspective) by Nicolai {\it et al}
\cite{Nicolai:2005mc}.

Several monographs have been written, including some recent and
updated. These monographs approach and present the subject from
different perspectives depending, of course, on the authors own
taste. From these, it is worth mentioning two. The first one by
Rovelli is physically motivated but is not so heavy in its
mathematical treatment, and can be found in the Book
\cite{Rovelli:2004tv}. A mathematically precise treatment, but not
for the faint of heart is given by the monograph by Thiemann
\cite{Thiemann:2001yy}. There have been also several nice reviews
that motivate and give a birdseye view of the subject by Rovelli
\cite{Rovelli:1999hz}, \cite{Rovelli:1997yv} and Pullin
\cite{pullin2}. Finally, there are several accounts on comparisons
between loop quantum gravity and other approaches, such as string
theory. On chronological order, we have a review by Rovelli
\cite{Rovelli:1997qj}, an entertaining dialog also by Rovelli
\cite{Rovelli:2003wd} and a critical assessment by Smolin
\cite{Smolin:2003rk}.

It is generally regarded that LQG has had three main achievements:
i) Quantization of geometric quantities, ii) Black hole entropy
and iii) Singularity avoidance in cosmology and collapse. All of
these achievements are discussed in the review articles, but
perhaps the best place to look at are the original references.

For the quantization of geometrical quantities the original
reference is by Rovelli and Smolin \cite{RS-geo} in the ``spin
network representation" and by Ashtekar-Lewandowski in the
connection representation" \cite{AL-Geo}, for the area operator
and in \cite{RS-geo} and \cite{Lew-vol} for the volume operator.
There are also operators associated to length \cite{length} and
angles \cite{angles}.

Black holes in loop quantum gravity were first considered by
Rovelli in \cite{BH-rovelli}. A systematic treatment of the
boundary conditions and the quantum geometry of the horizon was
given in \cite{abck,ack,abk}. Recently, a mistake in the original
entropy computation was corrected in \cite{lew-bh} and
\cite{meissner}. In particular, this corrects the value of the
Barbero-Immirzi parameter, a free parameter of the theory. The
existence of this parameter was first pointed out by Barbero
\cite{barbero} and its physical significance by Immirzi
\cite{immirzi}. The BH calculation was suggested out as a way of
computing its value in \cite{abck}. Other proposals include
asymptotic quasi-normal modes \cite{qnm}, and ``effective field
theory methods" \cite{eft1,eft2}.

Loop Quantum Cosmology (LQC) was born as a symmetry reduction of
the full theory in the work by Bojowald \cite{lqc1} (For some
comments regarding this reduction see \cite{tt-lqc} and
\cite{bojo:response}). The curvature at the big bang is bounded
and the absence of the initial singularity arises naturally when
the dynamical evolution continues across the `would be
singularity' \cite{bigbang}. A possible mechanism for inflation
was suggested in \cite{inflation}. A nice review of these main
results is given in \cite{lqc-rev}.
\newpage

\section*{Web Pages}

Nowadays there are several pages that contain information about
loop quantum gravity and spin foams, maintained by several people.
First, there is the Wikipedia entry:

\noindent
{\tt http://en.wikipedia.org/wiki/Loop\_quantum\_
gravity}

\noindent There is the page maintained by Dan Christensen,

\noindent {\tt http://jdc.math.uwo.ca/spin-foams/}

\noindent Seth Major,

\noindent {\tt http://academics.hamilton.edu/physics/smajor/\
resources.html}

\noindent and John Baez:

\noindent {\tt http://math.ucr.edu/home/baez/QG.html}

\noindent where information about LQG and new references can be
found. There is an interesting guide to bibliography on different
topics by Bombelli,

\noindent {\tt http://www.phy.olemiss.edu/\~{}luca/list.html}

\noindent Finally, the URL for {\em this} guide is,

\noindent {\tt
http://www.nucleares.unam.mx/\~{}corichi/lqgbib.pdf}

\newpage

\section*{Books and Dissertations}

\begin{enumerate}

\bibitem{AA:1} Abhay Ashtekar and {invited contributors}. {\em New
Perspectives in Canonical Gravity}. Lecture Notes.  Napoli, Italy:
Bibliopolis, February 1988. [Errata published as Syracuse
University preprint by Joseph D. Romano
  and Ranjeet S. Tate.]

\bibitem{AA:2} Abhay Ashtekar.
 {\em Lectures on non-perturbative canonical gravity.}
 (Notes prepared in collaboration with R. Tate).
Advanced Series in Astrophysics and Cosmology-Vol. 6. Singapore:
World Scientific, 1991.

\item J.C. Baez. {\em Knots and Quantum Gravity}. Oxford U. Press.
(1994). Also at:
{\tt http://math.ucr.edu/home/baez/kqg.html}

\item J.C. Baez and J. Muniain. {\em Gauge Fields, Knots, and
Gravity}. World Scientific Press (1994).

\item M. Bojowald. {\em Quantum geometry and symmetry}, Ph.D.
Thesis, RWTH Aachen, 2000.

\item R. Borissov. {\em Quantization of Gravity: In search of the
space of physical states}. Ph.D. Thesis, Temple U. (1997).

\item O. Bostr\"om. {\em Classical aspects on the road to quantum
gravity}. Ph.D. Thesis, Institute of Theoretical Physics,
G\"oteborg (1994).

\item B. Br\"ugmann. {\em On the constraints of quantum general
relativity in the loop representation.} Ph.D. Thesis, Syracuse
University (May 1993)

\item R. Capovilla. {\em The self-dual spin connection as the
fundamental gravitational variable.} Ph.D. Thesis, University of
Maryland (1991).

\item A.~Corichi, {\em Interplay Between Topology, Gauge Fields And
Gravity}, Ph.D. Thesis, Penn State U. (1997).\
{\tt http://cgpg.gravity.psu.edu/archives/\ thesis/1997/corichi.pdf}

\item O.~Dreyer,
 {\em Isolated horizons and black hole entropy},\\
Ph.D. Thesis, Penn State University. \\
{\tt http://cgpg.gravity.psu.edu/archives/\ thesis/2001/dreyer.pdf}

\bibitem{isham} J. Ehlers and H. Friedrich, eds. {\em Canonical Gravity:
From Classical to Quantum}. Lecture Notes in Physics 434,
(Springer-Verlag, Berlin, 1995).

\item K. Ezawa. Nonperturbative Solutions for Canonical Quantum
Gravity: an Overview. Ph.D. Thesis, Osaka U (January 1996).
gr-qc/9601050.

\item G. F\"ul\"op. {\em Supersymmetries and Ashtekar's
Variables}. Licentiate Thesis, I.T.P. G\"oteborg (1993).

\item R. Gambini and J. Pullin. {\em Loops, Knots, Gauge Theory
and Quantum Gravity}. Cambridge, Cambridge University Press
(1996).

\item N. Grot. {\em Topics in loop quantum gravity},
Ph.D. Thesis, University of Pittsburgh. (1998). {\tt
http://artemis.phyast.pitt.edu/thesis/norbert.ps}

\item V. Husain. {\it Investigations on the canonical quantization
of gravity.} Ph.D. Thesis, Yale University (1989).

\item J. Iwasaki. {\em On Loop-Theoretic Frameworks of Quantum
Gravity}. Ph.D. Thesis, University of Pittsburgh (April 1994).
{\tt http://artemis.phyast.pitt.edu/thesis/iwasaki.pdf}

\item S. Koshti. {\em Applications of the Ashtekar variables in
Classical Relativity}. Ph. D. Thesis, University of Poona (June
1991).

\item K.~Krasnov, {\em Spin foam models}, Ph.D. Thesis, Penn State U. (1999).
{\tt http://cgpg.gravity.psu.edu/archives/\ thesis/1999/kirill.pdf}

\item Yi Ling, {\em Extending Loop Quantum Gravity to Supergravity}, Ph.D.
Thesis, Penn State U. (2002).\
{\tt http://cgpg.gravity.psu.edu/archives/\ thesis/2002/ling.pdf}

\item
  E.~R.~Livine,
  {\em Loop gravity and spin foam: Covariant methods for the non-perturbative
  quantization of general relativity}. (In French),
  arXiv:gr-qc/0309028.

\item Seth Major
{\em q-Quantum Gravity}, Ph. D. Thesis, Penn State U. (1997).\
{\tt http://cgpg.gravity.psu.edu/archives/\ thesis/1997/seth.pdf}

\item H.J. Matschull. {\em Kanonishe Formulierung von Gravitations
und Supergravitations Theorien}. Ph.D. Thesis, Hamburg University
(July 1994), ISSN 0418-983.

\item H.A, Morales-T\'ecotl. {\em On Spacetime and Matter at
Planck Lenght}. Ph. D. Thesis SISSA/ISAS (January 1994).

\item
D.~Oriti, Spin foam models of quantum space-time, (Cambridge U.,
DAMTP),. Nov 2003. 335pp. Ph.D. Thesis. e-Print Archive:
gr-qc/0311066

\item P. Peld\'an. {\em From Metric to Connection: Actions for
gravity, with generalizations}. Ph.D. Thesis I.T.P. G\"oteborg
(1993) ISBN 91-7032-817-X.

\item A. Perez, {\em Spin foam models for quantum gravity}, Ph.D. Thesis,
University of Pittsburgh (2001). {\tt
http://artemis.phyast.pitt.edu/thesis/perez.ps}

\item Paul. A. Renteln. {\em Non-perturbative approaches to
Quantum Gravity. } Ph.D. Thesis, Harvard University (1988).

\bibitem{Rovelli:2004tv}
C. Rovelli. {\em Quantum Gravity}, Cambridge U. Press (2004).
For an online version see:\\
{\tt http://www.cpt.univ-mrs.fr/\~{}rovelli/book.pdf}

\item D. Rayner. {\em New variables in canonical quantisation and
quantum gravity.} Ph.D. Thesis, University of London (1991).

\item J. D.  Romano.
 {\em Geometrodynamics vs. Connection Dynamics (in the context
of (2+1)- and (3+1)-gravity)}.
 Ph.D. Thesis, Syracuse University (1991), see also gr-qc/9303032

\item H. Sahlmann. {\em Coupling Matter to Loop Quantum Gravity}.
Ph.D. Thesis, Universitaet Potsdam (2002).

\item V.O. Soloviev. {\em Boundary values as Hamiltonian
Variables. I. New Poisson brackets}. Ph.D.
IHEP93-48, \jmp{34}, 5747 (hep-th/9305133)).

\item C. Soo. {\em Classical and quantum gravity with Ashtekar
variables.} Ph.D. Thesis, Virginia Polytechnic Institute and State
University. VPI-IHEP-92-11 (July 1992)

\item R.S. Tate. {\em An algebraic approach to the quantization of
constrained systems: finite dimensional examples.} Ph.D. Thesis,
Syracuse University (Aug. 1992), gr-qc/9304043

\item T. Thiemann. {\em On the canonical quantization of gravity
in the Ashtekar framework}. Ph.D. Thesis, Aachen T. Hochschule,
1993.

\item T. Thiemann, {\em Modern non-perturbative quantum general
relativity}, Cambridge U. Press (2005).

\item J. Willis,
{\em On the Low-Energy Ramifications and a Mathematical Extension
of Loop Quantum Gravity}, Ph.D. Thesis, Penn State U. (2004).
{\tt http://cgpg.gravity.psu.edu/archives/\ thesis/2004/willis\_thesis.pdf}

\item O. Winkler

\item J. Wisniewski, {\em 2+1 General Relativity: Classical and
Quantum}, Ph.D. Thesis, Penn State U. (2002).
{\tt http://cgpg.gravity.psu.edu/archives/\ thesis/2002/wisniewski.pdf}

\item J. A. Zapata, {\em A Combinatorial Approach To Quantum Gauge
Theories And Quantum Gravity}, Ph.D. Thesis, Penn State U. (1998).\\
{\tt http://cgpg.gravity.psu.edu/archives/\ thesis/1998/zapata.pdf}

\item J.J. Zegwaard. {\em The Loop Representation for Canonical
Quantum Gravity and its Interpretation}. Ph.D. Thesis, Utrecht
University (January 1994). ISBN 90-393-0070-4.

\newpage

\newpage

\section*{Papers}

\section*{1977}

\item J.~F.~Pleba\`nski, On the separation of Einstein Substructure,
\jmp{12}, (1977), 2511.

\section*{1980}

\item Paul Sommers.
 Space spinors.
 {\em J. Math. Phys.} {\bf 21}(10):2567--2571, October 1980.

\section*{1981}

\item Amitabha Sen.
 On the existence of neutrino ``zero-modes'' in vacuum spacetimes.
 {\em J. Math. Phys.} {\bf 22}(8):1781--1786, August 1981.

\item R.~Giles. The reconstruction of gauge potentials from Wilson
loops. \pr{D24}: 2160-2168 (1981).

\section*{1982}

\item Abhay Ashtekar and G.T. Horowitz.
 On the canonical approach to quantum gravity.
 {\em Phys. Rev.} {\bf D26}: 3342--3353, 1982.

\item Amitabha Sen.
 Gravity as a spin system.
 {\em Phys. Lett. } {\bf B119}:89--91, December 1982.

\section*{1984}

\item A. Ashtekar.
 On the {H}amiltonian of general relativity.
 {\em Physica} {\bf A124}:51--60, 1984.

\item A. Ashtekar and G.T. Horowitz. Phase space of general
relativity revisited: A canonical choice of time and
simplification of the Hamiltonian. \jmp{25}:1473-1480, (1984).

\item E.~T. Newman.
 Report of the workshop on classical and quantum alterate theories of
  gravity.
 In B.~Bertotti, F.~{de Felice}, and A.~Pascolini, editors, {\em The
  Proceedings of the 10th International Conference on General Relativity and
  Gravitation}, Amsterdam, 1984.

\section*{1986}

\item A. Ashtekar.
 New variables for classical and quantum gravity.
 {\em Phys. Rev. Lett.} {\bf 57}(18):2244--2247, November 1986.

\item A. Ashtekar.
 Self-duality and spinorial techniques in the canonical approach to
  quantum gravity.
 In C.~J. Isham and R.~Penrose, editors, {\em Quantum Concepts in
  Space and Time}, pages 303--317. Oxford University Press, 1986.

\item Robert~M. Wald.
 Non-existence of dynamical perturbations of {S}chwarzschild with
  vanishing self-dual part.
 {\em Class. Quan. Grav.} {\bf 3}(1):55--63, January 1986.


\section*{1987}

\item A. Ashtekar.
 New {H}amiltonian formulation of general relativity.
 {\em Phys. Rev. } {\bf D36}(6):1587--1602, September 1987.

\item A. Ashtekar.
 {E}instein constraints in the {Y}ang-{M}ills form.
 In G.~Longhi and L~Lusanna, editors, {\em Constraint's Theory and
  Relativistic Dynamics}, Singapore, 1987. World Scientific.

\item Abhay Ashtekar, Pawel Mazur, and Charles~G. Torre.
 {BRST} structure of general relativity in terms of new variables.
 {\em Phys. Rev. } {\bf D36}(10):2955--2962, November 1987.

\item John~L. Friedman and Ian Jack.
 Formal commutators of the gravitational constraints are not
  well-defined: A translation of {A}shtekar's ordering to the {S}chr{\"o}dinger
  representation.
 {\em Phys. Rev. } {\bf D37}(12):3495--3504, June 1987.

\item Kazuo Ghoroku. New variable formalism of higher derivative
gravity. \pl{B194}:535-538, 1987

\item Ted Jacobson and Lee Smolin.
 The left-handed spin connection as a variable for canonical gravity.
 {\em Phys. Lett. } {\bf B196}(1):39--42, September 1987.

\item Joseph Samuel.
 A {L}agrangian basis for {A}shtekar's reformulation of canonical
  gravity.
 {\em Pram{\=a}na-J Phys.} {\bf 28}(4):L429-L432, April 1987.

\item N.~C. Tsamis and R.~P. Woodard.
 The factor ordering problem must be regulated.
 {\em Phys. Rev.}  {\bf D36}(12):3641--3650, December 1987.

\newpage

\section*{1988}

\item Abhay Ashtekar.
 A $3+1$ formulation of {E}instein self-duality.
 In J.~Isenberg, editor, {\em Mathematics and General Relativity},
  Providence, 1988. American Mathematical Society.

\item Abhay Ashtekar.
 Microstructure of space-time in quantum gravity.
 In K.~C. Wali, editor, {\em Proceedings of the Eight Workshop in
  Grand Unification}, Singapore, 1988. World Scientific.

\item Abhay Ashtekar.
 New perspectives in canonical quantum gravity.
 In B.~R. Iyer, A.~Kembhavi, J.~V. Narlikar, and C.~V. Vishveshwara,
  editors, {\em Highlights in Gravitation and Cosmology}. Cambridge University
  Press, 1988.

\item Abhay Ashtekar, Ted Jacobson, and Lee Smolin.
 A new characterization of half-flat solutions to {E}instein's
  equation.
 {\em Commun. Math. Phys.} {\bf 115}:631--648, 1988.

\item I. Bengtsson. Ashtekar's variables. Goteborg-88-46 preprint
(November 1988). in Proc. XXIInd Int. Symp. Ahrenshoop on Theory of Elementary Particles, Ahrenshoop, 1988, Oct. 17-21, Ed. by E. Wieczorek, Inst. f. Hochenergiephysik Berlin-Zeuthen, PHE 88-13

\item Ingemar Bengtsson.
 Note on {A}shtekar's variables in the spherically symmetric case.
 {\em Class. Quan. Grav.} {\bf 5}(10):L139--L142, October 1988.

\item R. Gianvittorio, R. Gambini and A. Trias. \pr{D38} (1988)
702

\item J.~N. Goldberg.
 A {H}amiltonian approach to the strong gravity limit.
 {\em Gen. Rel. Grav.} {\bf 20}(9):881--891, September 1988.

\item J.~N. Goldberg.
 Triad approach to the {H}amiltonian of general relativity.
 {\em Phys. Rev. } {\bf D37}(8):2116--2120, April 1988.

\item Viqar Husain.
 The {$G_{\mbox{Newton}}\rightarrow\infty$} limit of quantum gravity.
 {\em Class. Quan. Grav.} {\bf 5}(4):575--582, April 1988.

\item Ted Jacobson.
 Fermions in canonical gravity.
 {\em Class. Quan. Grav.} {\bf 5}(10):L143--L148, October 1988.

\item Ted Jacobson.
 New variables for canonical supergravity.
 {\em Class. Quan. Grav.} {\bf 5}:923--935, 1988.

\item Ted Jacobson.
 Superspace in the self-dual representation of quantum gravity.
 In J.~Isenberg, editor, {\em Mathematics and General Relativity},
  Providence, 1988. American Mathematical Society.

\item Ted Jacobson and Lee Smolin.
 Covariant action for {A}shtekar's form of canonical gravity.
 {\em Class. Quan. Grav.} {\bf 5}(4):583--594, April 1988.

\item Ted Jacobson and Lee Smolin.
 Nonperturbative quantum geometries.
 {\em Nucl. Phys.} {\bf B299}(2):295--345, April 1988.

\item Hideo Kodama.
 Specialization of {A}shtekar's formalism to {B}ianchi cosmology.
 {\em Prog. Theor. Phys.} {\bf 80}(6):1024--1040, December 1988.

\item Carlo Rovelli, Loop Space Representation, in {\em
New Perspectives in canonical gravity}, Ref.\cite{AA:1}.

\item Carlo Rovelli and Lee Smolin.
 Knot theory and quantum gravity.
 {\em Phys. Rev. Lett.} {\bf 61}:1155--1158, 1988.

\item Joseph Samuel.
 Gravitational instantons from the {A}shtekar variables.
 {\em Class. Quan. Grav.} {\bf 5}:L123--L125, 1988.

\item Lee Smolin.
 Quantum gravity in the self-dual representation.
 In J.~Isenberg, editor, {\em Mathematics and General Relativity},
  Providence, 1988. American Mathematical Society.

\item C.~G. Torre.
 The propagation amplitude in spinorial gravity.
 {\em Class. Quan. Grav.} {\bf 5}:L63--L68, 1988.

\item Edward Witten.
 (2+1) dimensional gravity as an exactly soluble system.
 {\em Nucl. Phys.} {\bf B311}(1):46--78, December 1988.

\newpage

\section*{1989}

\item Abhay Ashtekar.
 Non-pertubative quantum gravity: A status report.
 In M.~Cerdonio, R.~Cianci, M.~Francaviglia, and M.~Toller, editors,
  {\em General Relativity and Gravitation}. Singapore: World Scientific, 1989.

\item Abhay Ashtekar.
 Recent developments in {H}amiltonian gravity.
 In B.~Simon, I.~M. Davies, and A.~Truman, editors, {\em The
  Proceedings of the {IX}th International Congress on Mathematical Physics},
Swansea UK, July 1988.(Bristol, UK: Adam Hilger, 1989).

\item Abhay Ashtekar.
 Recent developments in quantum gravity.
 In E.~J. Fenyves, editor, {\em Proceedings of the Texas Symposium on
  Relativistic Astrophysics}. New York Academy of Science, 1989.

\item Abhay Ashtekar.
 Recent Developments in Quantum Gravity. {\it Annals of the New York
Academy of Sciences} {\bf 571}, 16-26. December 1989.

\item Abhay Ashtekar, A.~P. Balachandran, and S.~G. Jo.
 The {CP}-problem in quantum gravity.
 {\em Int. Journ. Theor. Phys.} {\bf A4}:1493--1514, 1989.

\item Abhay Ashtekar, Viqar Husain, Carlo Rovelli, Joseph Samuel,
and Lee Smolin.
 $2+1$ quantum gravity as a toy model for the $3+1$ theory.
 {\em Class. Quan. Grav.} {\bf 6}:L185--L193, 1989.

\item Abhay Ashtekar and Joseph~D. Romano.
 {C}hern-{S}imons and {P}alatini actions and ($2+1$)-gravity.
 {\em Phys. Lett. } {\bf B229}(1,2):56--60, October 1989.

\item Abhay Ashtekar, Joseph~D. Romano, and Ranjeet~S. Tate.
 New variables for gravity: Inclusion of matter.
 {\em Phys. Rev. } {\bf D40}(8):2572--2587, October 1989.

\item Abhay Ashtekar and Joseph~D. Romano.
 Key ($3+1$)-equations in terms of new variables (for numerical
  relativity).
 Syracuse University Report (1989).

\item Ingemar Bengtsson.
 {Y}ang-{M}ills theory and general relativity in three and four
  dimensions.
 {\em Phys. Lett. } {\bf B220}:51--53, 1989.

\item Ingemar Bengtsson.
 Some remarks on space-time decomposition, and degenerate metrics, in
  general relativity.
 {\em Int. J.  Mod. Phys. } {\bf A4}(20):5527--5538,
  1989.

\item Riccardo Capovilla, John Dell, and Ted Jacobson.
 General relativity without the metric.
 {\em Phys. Rev. Lett.} {\bf 63}(21):2325--2328, November 1989.

\item Steven Carlip.
 Exact quantum scattering in 2+1 dimensional gravity.
 {\em Nucl. Phys.} {\bf B324}(1):106--122, 1989.

\item B. P. Dolan.
 On the generating function for Ashtekar's canonical transformation.
 {\em Phys. Lett. } {\bf B233}(1,2):89-92 , December 1989.

\item Tevian Dray, Ravi Kulkarni, and Joseph Samuel.
 Duality and conformal structure.
 {\em J. Math. Phys.} {\bf 30}(6):1306--1309, June 1989.

\item N.~N. Gorobey and A.~S. Lukyanenko.
 The closure of the constraint algebra of complex self-dual gravity.
 {\em Class. Quan. Grav.} {\bf 6}(11):L233--L235, November 1989.

\item M.~Henneaux, J.~E. Nelson, and C.~Schomblond.
 Derivation of {A}shtekar variables from tetrad gravity.
 {\em Phys. Rev. } {\bf D39}(2):434--437, January 1989.

\item A. Herdegen. Canonical gravity from a variation principle in
a copy of a tangent bundle. {\it Class.  Quan. Grav.} {\bf
6}(8):1111-24, (1989).

\item G. T. Horowitz. Exactly soluble diffeomorphism invariant
theories. {\it Commun. Math. Phys.} {\bf 125}(3): 417-37, 1989.

\item Viqar Husain.
 Intersecting loop solutions of the {H}amiltonian constraint of
  quantum general relativity.
 {\em Nucl. Phys.} {\bf B313}:711--724, 1989.

\item Viqar Husain and Lee Smolin.
 Exactly solvable quantum cosmologies from two {K}illing field
  reductions of general relativity.
 {\em Nucl. Phys.} {\bf B327}:205--238, 1989.

\item V.~Khatsymovsky.
 Tetrad and self-dual formulation of {R}egge calculus.
 {\em Class. Quan. Grav.} {\bf 6}(12):L249--L255, December 1989.

\item Sucheta Koshti and Naresh Dadhich.
 Degenerate spherical symmetric cosmological solutions using
  {A}shtekar's variables.
 {\em Class. Quan. Grav.} {\bf 6}:L223--L226, 1989.

\item Stephen~P. Martin.
 Observables in 2+1 dimensional gravity.
 {\em Nucl. Phys.} {\bf 327}(1):78--204, November 1989.

\item L.~J. Mason and E.~T. Newman.
 A connection between {E}instein and {Y}ang-{M}ills equations.
 {\em Commun. Math. Phys.} {\bf 121}(4):659--668, 1989.

\item J.~E. Nelson and T.~Regge.
 Group manifold derivation of canonical theories.
 {\em Int. J. Mod. Phys.} {\bf A4},2021 (1989).

\item Paul Renteln and Lee Smolin.
 A lattice approach to spinorial quantum gravity.
 {\em Class. Quan. Grav.} {\bf 6}:275--294, 1989.

\item Amitabha Sen and Sharon Butler.
 The quantum loop.
 {\em The Sciences}:32--36, November/December 1989.

\item L. Smolin.
 Invariants of links and critical points of the {C}hern-{S}imon path
  integrals.
 {\em Mod. Phys. Lett.} {\bf A4}:1091--1112, 1989.

\item L. Smolin. Loop representation for quantum gravity in 2+1
dimensions. In the {\em Proceedings of the John's Hopkins
Conference on Knots, Topology and Quantum Field Theory}, ed. L.
Lusanna (World Scientific, Singapore 1989)

\item Sanjay~M. Wagh and Ravi~V. Saraykar.
 Conformally flat initial data for general relativity in {A}shtekar's
  variables.
 {\em Phys. Rev. } {\bf D39}(2):670--672, January 1989.

\item Edward Witten.
 Gauge theories and integrable lattice models.
 {\em Nucl. Phys.} {\bf B322}(3):629--697, August 1989.

\item Edward Witten.
 Topology-changing amplitudes in (2+1) dimensional gravity.
 {\em Nucl. Phys.} {\bf B323}(1):113--122, August 1989.

\newpage

\section*{1990}

\item C. Aragone and A. Khouder .
 Vielbein gravity in the light-front gauge.
 {\em Class. Quan. Grav.} {\bf 7}:1291--1298, 1990.

\item Abhay Ashtekar. Old problems in the light of new variables.
In {\em Proceedings of the Osgood Hill Conference on Conceptual
Problems in Quantum Gravity}, eds. A. Ashtekar and J. Stachel
(Birkh\"auser, Boston 1991)

\item Abhay Ashtekar.
 Self duality, quantum gravity, {W}ilson loops and all that.
 In N.~Ashby, D.~F. Bartlett, and W.~Wyss, editors, {\em Proceedings
  of the 12th International Conference on General Relativity and Gravitation}.
  Cambridge University Press, 1990.

\item Abhay Ashtekar and Jorge Pullin.
 {B}ianchi cosmologies: A new description.
 {\em Proc. Phys. Soc. Israel} {\bf 9}:65-76 (1990).

\item Abhay Ashtekar.
 Lessons from 2+1 dimensional quantum gravity.
In {\em "Strings 90"} edited by R. Arnowitt et al (Singapore:
World Scientific, 1990).

\item J.~Ambjorn and Y.~M.~Makeenko, Properties Of Loop Equations
For The Hermitean Matrix Model And For Two-Dimensional Quantum
Gravity, \mpl{A5}, 1753 (1990).

\item Ingemar Bengtsson.
 A new phase for general relativity?
 {\em Class. Quan. Grav.} {\bf 7}(1):27--39, January 1990.

\item Ingemar Bengtsson.
 P, T, and the cosmological constant.
 {\em Int. J.  Mod. Phys. } {\bf A5}(17):3449-3459 (1990).

\item Ingemar Bengtsson.
 Self-Dual Yang-Mills fields and Ashtekar variables.
 {\em Class. Quan. Grav.} {\bf 7}:L223-L228 (1990)

\item Ingemar Bengtsson and P. Peld{\' a}n. Ashtekar variables,
the theta-term, and the cosmological constant. {\em Phys. Lett.}
{\bf B244}(2): 261-64, 1990.

\item M.~P. Blencowe.
 The {H}amiltonian constraint in quantum gravity.
 {\em Nuc. Phys.} {\bf B341}(1):213, 1990.

\item L.~Bombelli and R.~J. Torrence.
 Perfect fluids and {A}shtekar variables, with applications to
  {K}antowski-{S}achs models.
 {\em Class.Quan. Grav.} {\bf 7}:1747 (1990).

\item Riccardo Capovilla, John Dell, and Ted Jacobson.
 Gravitational instantons as {SU(2)} gauge fields.
 {\em Class.Quan. Grav.} {\bf 7}(1):L1--L3, January 1990.

\item Steven Carlip.
 Observables, gauge invariance and time in 2+1 dimensional gravity.
 {\em Phys. Rev.} {\bf D42}, 2647-2654 (October 1990).

\item S. Carlip and S. P. de Alwis.
 Wormholes in (2+1)-gravity.
 {\em Nuc. Phys.} {\bf B337}:681-694, June 1990.

\item G. Chapline. Superstrings and Quantum Gravity. {\em Mod.
Phys. Lett.}{\bf A5}:2165-72 (1990).

\item F.~David, Loop Equations And Nonperturbative Effects In
Two-Dimensional Quantum Gravity, \mpl{A5}, 1019 (1990).

\item R. Floreanini and R. Percacci.
 Canonical algebra of GL(4)-invariant gravity.
 {\em Class.Quan. Grav.} {\bf 7}:975--984, 1990.

\item R. Floreanini and R. Percacci.
 Palatini formalism and new canonical variables for GL(4)-invariant
gravity. {\em Class. Quan. Grav.} {\bf 7}:1805-18, 1990.

\item R. Floreanini and R. Percacci.
 Topological pregeometry.
 {\em Mod. Phys. Lett.} {\bf A5}:2247-51, 1990.

\item Takeshi Fukuyama and Kiyoshi Kaminura.
 Complex action and quantum gravity.
{\em Phys. Rev.} {\bf D41}:1105-11, February 1990.

\item G.~Gonzalez and J.~Pullin.
 {BRST} quantization of 2+1 gravity.
 {\em Phys. Rev. } {\bf D42}(10): 3395-3400 (1990).
[Erratum: {\em Phys. Rev.} {\bf 43}:2749, April 1991].

\item N.~N. Gorobey and A.~S. Lukyanenko.
 The {A}shtekar complex canonical transformation for supergravity.
 {\em Class. Quan. Grav.} {\bf 7}(1):67--71, January 1990.

\item C. Holm.
 Connections in Bergmann manifolds.
 {\em Int. Journ. Theor. Phys.} {\bf A29}(1):23-36, January 1990.

\item V. Husain and K. Kucha{\v r}.
 General covariance, the New variables, and dynamics without dynamics.
{\em Phys. Rev.} {\bf D42}(12)4070-4077 (December 1990).

\item Viqar Husain and Jorge Pullin.
 Quantum theory of space-times with one Killing field.
 {\em Modern Phys. Lett. } {\bf A5}(10):733-741, April 1990.

\item K. Kamimura and T. Fukuyama. Ashtekar's formalism in 1st
order tetrad form. {\em Phys. Rev.}  {\bf D41}(6): 1885-88, 1990.

\item H. Kodama.
 Holomorphic wavefunction of the universe.
 {\em Phys. Rev.} {\bf D42}:2548-2565 (October 1990).

\item Sucheta Koshti and Naresh Dadhich.
 On the self-duality of the {W}eyl tensor using {A}shtekar's
  variables.
 {\em Class. Quan. Grav.} {\bf 7}(1):L5--L7, January 1990.

\item Noah Linden.
 New designs on space-time foams.
 {\em Physics World} {\bf 3}(3):30-31, March 1990.

\item N.Manojlovic.
 Alternative loop variables for canonical gravity.
 {\em Class. Quan. Grav.} {\bf 7}:1633-1645. (1990).

\item E.~W. Mielke.
 Generating functional for new variables in general relativity and
  {P}oincare gauge theory.
 {\em Phys. Lett.} {\bf A149}:345-350 (1990).

\item E.~W. Mielke.
 Positive gravitational energy proof from complex variables?
 {\em Phys. Rev.} {\bf D42}(10): 3338-3394 (1990).

\item Peter Peld\'{a}n.
 Gravity coupled to matter without the metric.
{\em Phys. Lett.} {\bf B248}(1,2): 62-66 (1990).

\item D.~Rayner.
 A formalism for quantising general relativity using non-local
  variables.
 {\em Class. Quan. Grav.} {\bf 7}(1):111--134, January 1990.

\item D.~Rayner. Hermitian operators on quantum general relativity
loop space.
 {\em Class. Quan. Grav.} {\bf 7}(4):651--661, April 1990.

\item Paul Renteln.
 Some results of {SU}(2) spinorial lattice gravity.
 {\em Class. Quan. Grav.} {\bf 7}(3):493--502, March 1990.

\item D.C. Robinson and C. Soteriou.
 Ashtekar's new variables and the vacuum constraint equations.
 {\em Class. Quan. Grav.} {\bf 7}(11): L247-L250 (1990).

\item Carlo Rovelli and Lee Smolin.
 Loop representation of quantum general relativity.
 {\em Nuc. Phys.} {\bf B331}(1): 80-152, February 1990.

\item M.~Seriu and H.~Kodama.
 New canonical formulation of the {E}instein theory.
 {\em Prog. Theor. Phys.} {\bf 83}(1):7-12, January 1990.

\item Lee Smolin.
 Loop representation for quantum gravity in $2+1$ dimensions.
 In {\em Proceedings of the 12th John Hopkins Workshop: Topology and
  Quantum Field Theory} (Florence, Italy), 1990.

\item C. G. Torre.
 Perturbations of gravitational instantons.
 {\em Phys. Rev.} {\bf D41}(12) : 3620-3621, June 1990.

\item C.~G. Torre.
 A topological field theory of gravitational instantons.
 {\em Phys. Lett } {\bf B252}(2):242-246 (1990).

\item C. G. Torre.
 On the linearization stability of the conformally
(anti)self dual  {E}instein equations,
 {\em J. Math. Phys.} {\bf 31}(12): 2983-2986 (1990).

\item H. Waelbroeck.
 2+1 lattice gravity.
 {\em Class. Quan. Grav.} {\bf 7}(1): 751--769, January 1990.

\item M. Waldrop. Viewing the Universe as a Coat of Chain Mail.
{\em Science} {\bf 250}:1510-1511 (1990).

\item R. P. Wallner
 New variables in gravity theories.
 {\em Phys. Rev. } {\bf D42}(2):441-448 ,July  1990.

\item R.S. Ward.
 The SU($\infty$) chiral model and self-dual vacuum spaces.
 {\em Class. Quan. Grav.} {\bf 7}:L217-L222 (1990).

\newpage

\section*{1991}

\item V. Aldaya and J. Navarro-Salas. New solutions of the
hamiltonian and diffeomorphism constraints of quantum gravity from
a highest weight loop representation. {\em Phys. Lett.} {\bf
B259}:249-55, April 1991.

\item Abhay Ashtekar. Old problems in the light of new variables.
In {\em Proceedings of the Osgood Hill Conference on Conceptual
Problems in Quantum Gravity}, eds. A. Ashtekar and J. Stachel
(Birkh\"auser, Boston 1991)

\item Abhay Ashtekar. The winding road to quantum gravity. In {\em
Proceedings of the Osgood Hill Conference on Conceptual Problems
in Quantum Gravity}, eds. A. Ashtekar and J. Stachel
(Birkh\"auser, Boston 1991)

\item Abhay Ashtekar. Canonical Quantum Gravity. In {\em The
Proceedings of the 1990 Banff Workshop on Gravitational Physics},
edited by R. Mann (Singapore: World Scientific, 1991), and in the
{\em Proceedings of SILARG VIII Conference}, edited by M.
Rosenbaum and M. Ryan (Singapore: World Scientific 1991).

\item A. Ashtekar, C. Rovelli and L. Smolin. Gravitons and loops.
{\em Phys. Rev.}{\bf D44}(6):1740-55, 15 September 1991.

\item A. Ashtekar and J. Samuel. Bianchi cosmologies: the role of
spatial topology. \cqg{8} (1991) 2191--215

\item J.~W.~Barrett. Holonomy and path structures in general relativity
and Yang-Mills theory, {\em Int.\ J.\ Theor.\ Phys.}\  {\bf 30},
1171 (1991).

\item I. Bengtsson.
 The cosmological constants.
{\em Phys. Lett.} {\bf B254}:55-60, 1991.

\item I. Bengtsson. Self-duality and the metric in a family of
neighbours of Einstein's equations. {\em J. Math. Phys.}{\bf 32}
(Nov. 1991) 3158--61

\item I. Bengtsson. Degenerate metrics and an empty black hole.
\cqg{8}, 1847 (1991), Goteborg-90-45 (December 1990).

\item Peter G. Bergmann and Garrit Smith. Complex phase spaces and
complex gauge groups in general relativity.
 {\em Phys. Rev.} {\bf D43}:1157-61,  February 1991.

\item L. Bombelli. Unimodular relativity, general covariance,
time, and the Ashtekar variables. In {\em Gravitation. A Banff
Summer Institute}, eds.~R. Mann and P. Wesson (World Scientific
1991) 221--32

\item L. Bombelli, W.E. Couch and R.J.Torrence. Time as spacetime
four--volume and the Ashtekar variables. \pr{D44}:2589--92 (15.
Oct. 1991)

\item B. Br\"{u}gmann.
 The method of loops applied to lattice gauge theory.
{\em Phys. Rev. } {\bf D43}:566-79, January 1991.

\item B. Br\"{u}gmann and J. Pullin.
 Intersecting N loop solutions of the Hamiltonian constraint
of Quantum Gravity.
 {\em Nuc. Phys.} {\bf B363}:221-44, September 1991.

\item R. Capovilla, J. Dell, T. Jacobson and L. Mason.
 Self dual forms and gravity.
{\em Class. Quan. Grav.} {\bf 8}:41-57, January 1991.

\item R. Capovilla, J. Dell and T. Jacobson.
 A pure spin-connection formulation of gravity.
{\em Class. Quan. Grav.} {\bf 8}:59-74, January 1991. [Erratum,
{\bf 9}, 1839 (1992)].

\item Steven Carlip.
 Measuring the metric in 2+1 dimensional quantum gravity.
{\em Class. Quan. Grav.} {\bf 8}:5-17, January 1991.

\item S. Carlip and J. Gegenberg. Gravitating topological matter
in 2+1 dimensions. {\em Phys. Rev.}{\bf D44}(2):424-28, 15 July
1991.

\item L. Crane. 2-d physics and 3-d topology. {\em Commun. Math.
Phys.} {\bf 135}:615-640, January 1991.

\item L.~Crane,
  Conformal field theory, spin geometry, and quantum gravity,
  \pl{B259}, 243 (1991).

\item N. Dadhich, S. Koshti and A. Kshirsagar.
 On constraints of pure connection formulation of General Relativity
for non-zero cosmological constant. {\em Class. Quan. Grav.} {\bf
8}:L61-L64, March 1991.

\item B.~P. Dolan.
 The extension of chiral gravity to {SL}(2,{C}).
In {\em Proceedings of the 1990 Banff Summer School on
gravitation}, ed. by R. Mann (World Scientific, Singapore 1991)

\item R. Floreanini and R. Percacci.
 GL(3) invariant gravity without metric.
 {\em Class. Quan. Grav.}{\bf 8}(2):273-78, February 1991.

\item G. Fodor and Z. Perjes. Ashtekar variables without
hypersurfaces. {\em Proc. of Fifth Sem. Quantum Gravity, Moscow}
(Singapore: World Scientific 1991) 183--7

\item H. Fort and R. Gambini. Lattice QED with light fermions in
the P representation. IFFI preprint, 90-08. \pr{D44}:1257-1262,
1991.

\item T. Fukuyama and K. Kamimura. Schwarzschild solution in
Ashtekar formalism. \mpl{A6} (1991) 1437--42

\item R. Gambini. Loop space representation of quantum general
relativity and the group of loops. {\em Phys. Lett.} {\bf
B255}:180-88, February 1991.

\item J.N. Goldberg. Self-dual Maxwell field on a null cone.
\grg{23} (December 1991) 1403--1413

\item J.N. Goldberg, E.T. Newman, and C. Rovelli. On Hamiltonian
systems with first class constraints. \jmp{32}(10) (1991) 2739--43

\item J. Goldberg, D.C. Robinson and C. Soteriou. Null surface
canonical formalism. In {\em  Gravitation and Modern Cosmology},
ed. Zichichi (Plenum Press, New York, 1991)

\item J. Goldberg, D.C.Robinson and C. Soteriou. A canonical
formalism with a self-dual Maxwell field on a null surface. In
{\em 9th Italian Conference on General Relativity and
Gravitational Physics (P.G.  Bergmann Festschrift)}, ed. R. Cianci
et al (World Scientific, Singapore 1991)

\item G. Harnett. Metrics and dual operators.  \jmp{32} (1991) 84-91.
Florida Atlantic University preprint, 1991.

\item G.T. Horowitz. Topology change in classical and quantum
gravity. {\em Class. Quan. Grav.} {\bf 8}:587-601, April 1991.

\item V. Husain. Topological quantum mechanics. {\em Phys. Rev.}
{\bf D43}:1803-07, March 1991.

\item H. Ikemori. Introduction to two form gravity and Ashtekar
formalism. YITP-K-922 preprint (March 1991). {\em Tokyo Quantum
Gravity}:7-88, 1991.

\item C. J. Isham. Loop Algebras and Canonical Quantum Gravity. To
appear in Contemporary Mathematics,edited by M. Gotay, V. Moncrief
and J. Marsden (American Mathematical Society, Providence, 1991).

\item K. Kamimura, S. Makita and T. Fukuyama . Spherically
symmetric vacuum solution in Ashtekar's formulation of gravity.
\mpl{A6} (30. Oct. 1991) 3047--53

\item C. Kozameh, W. Lamberti, and E.T. Newman Holonomy and the
Einstein equations, \ap{206}, 193-220 (1991).

\item C. Kozameh and E.T. Newman. The O(3,1) Yang-mills equations
and the Einstein equations. {\em Gen. Rel. Grav.} {\bf 23}:87-98,
January 1991.

\item H. C. Lee and Z. Y. Zhu. Quantum holonomy and link
invariants. {\em Phys. Rev.}{\bf D44}(4):R942-45, 15 August 1991.

\item R. Loll.
 A new quantum representation for canonical gravity and SU(2)
Yang-Mills theory. \np{B350} (1991) 831--60

\item Y.~Makeenko, Loop equations in matrix models and in 2--D
quantum gravity, \mpl{A6}:1901 (1991).

\item E. Mielke, F. Hehl. Comment on ``General relativity without
the metric''. \prl{67} (Sept.\ 1991) 1370

\item V. Moncrief and M. P. Ryan. Amplitude-real-phase exact
solutions for quantum mixmaster universes. {\em Phys. Rev.}{\bf
D44}, (1991), 2375.

\item C.~Nayak.
 The loop space representation of 2+1 quantum gravity: physical
  observables,variational principles,  and the issue of time.
{\em Gen. Rel. Grav.} {\bf 23}:661-70, June 1991.

\item C.~Nayak. Einstein-Maxwell theory in 2+1 dimensions. {\em
Gen. Rel. Grav.}{\bf 23}:981-90, September 1991.

\item H. Nicolai. The canonical structure of maximally extended
supergravity in three dimensions. \np{B353} (April 1991) 493

\item P. Peld{\' a}n. Legendre transforms in Ashtekar's theory of
gravity. \cqg{8} (Oct. 1991) 1765--83

\item P. Peld{\' a}n. Non-uniqueness of the ADM Hamiltonian for
gravity. \cqg{8} (Nov. 1991) L223--7

\item C.~Rovelli,  ``Knots and Physics", {\em Encyclopaedia Britannica},
        1991 Yearbook of Science and Technology, feature article.

\bibitem{rovelli1991} C. Rovelli. Ashtekar's formulation of general relativity
and loop-space non-perturbative quantum gravity : a report. {\em
Class. Quan. Grav.}{\bf 8}(9): 1613-1675, September 1991.

\item Carlo Rovelli.
 Holonomies and loop representation in quantum gravity.
In {\em The Newman Festschrift}, ed. by A. Janis and J. Porter.
(Birkh{\" a}user, Boston 1991)

\item Joseph Samuel.
 Self-duality in Classical Gravity.
In {\em The Newman Festschrift}, ed. by A. Janis and J. Porter.
(Birkh{\" a}user, Boston 1991)

\item Lee Smolin.
 Nonperturbative quantum gravity via the loop representation.
In {\em Conceptual Problems of Quantum Gravity}, eds. A. Ashtekar
and J. Stachel (Birkh\"auser, Boston, 1991)

\item G. t'Hooft. A chiral alternative to the vierbein field in
general relativity. \np{B357}:211-221, 1991.

\item C. G. Torre. A deformation theory of self-dual Einstein
spaces.
in {\em Mathematical aspects of classical field theory}.
Proceedings of the AMS-IMS-SIAM Joint Summer Research
Conference held at the University ofWashington, Seattle,
Washington, July 20–26, 1991.
Edited by Mark J. Gotay, Jerrod E. Marsden [Jerrold Eldon Marsden]
and Vincent Moncrief.
Contemporary Mathematics, 132, p 611.
SU-GP-91/8-7, Syracuse University preprint, 1991.

\item S. Uehara. A note on gravitational and SU(2) instantons with
Ashtekar variables. \cqg{8}:L229--34 (Nov. 1991)

\item M. Varadarajan. Non-singular degenerate negative energy
solution to the Ashtekar equations. \cqg{8} (Nov. 1991) L235--40

\item K. Yamagishi and G.F. Chapline. Induced 4-d self-dual
quantum gravity: $\hat{W}_{\infty}$ algebraic approach. {\em
Class. Quan. Grav.}{\bf 8}(3):427-46, March 1991.

\item J.~Zegwaard, Gravitons in loop quantum gravity,
\np{B378}:288 (1992).

\item J. Zegwaard. Representations of quantum general relativity
using Ashtekar's variables. {\em Class. Quan. Grav.}{\bf 8} (July
1991) 1327--37

\newpage

\section*{1992}

\item A. Ashtekar. Loops, gauge fields and gravity. In {\em
Proceedings of the VIth Marcel Grossmann meeting on general
relativity}, eds.\ H. Sato and T. Nakamura (World Scientific,
1992), and in {\em Proceedings of the VIIIth Canadian conference
on general relativity and gravitation}, edited by G. Kunstater et
al (World Scientific, Singapore 1992)

\item A. Ashtekar and C. Isham. Representations of the holonomy
algebras of gravity and non-abelian gauge theories. \cqg{9} (June
1992) 1433--85

\item A. Ashtekar and C. Isham. Inequivalent observable algebras:
a new ambiguity in field quantisation. \pl{B274} (1992) 393--398

\item A. Ashtekar and J.D. Romano. Spatial infinity as a boundary
of space-time. \cqg{9} (April 1992) 1069--100

\item A. Ashtekar and C. Rovelli. Connections, loops and quantum
general relativity. \cqg{9} suppl. (1992) S3--12

\item A. Ashtekar and C. Rovelli. A loop representation for the
quantum Maxwell field. \cqg{9} (May 1992) 1121--50

\item A. Ashtekar, C. Rovelli and L. Smolin. Self duality and
quantization. {\em J. Geom. Phys.} {\bf 8} (1992) 7--27

\item A. Ashtekar, C. Rovelli and L. Smolin. Weaving a classical
geometry with quantum threads. \prl{69} (1992) 237--40

\item J.C. Baez. Link invariants of finite type and perturbation
theory. Lett. Math. Phys {\bf 26} (1992) 43--51.

\item I. Bengtsson and O. Bostr\"om. Infinitely many cosmological
constants. \cqg{9} (April 1992) L47--51

\item I. Bengtsson and P. Peldan. Another `cosmological' constant.
\ijmp{A7} (10 March 1992) 1287--308

\item O. Bostr\"om. Degeneracy in loop variables; some further
results. \cqg{9} (Aug. 1992) L83--86

\item B. Br\"{u}gmann, R. Gambini and J. Pullin. Knot invariants
as nondegenerate quantum geometries. \prl{68} (27 Jan. 1992)
431--4

\item B. Br\"{u}gmann, R. Gambini and J. Pullin. Knot invariants
as nondegenerate states of four-dimensional quantum gravity. In
{\em Proceedings of the Twentieth International conference on
Differential Geometric Methods in Theoretical Physics, Baruch
College, City University of New York,1-7 June 1991}, S. Catto, A.
Rocha eds. (World Scientific, Singapore 1992)

\item B. Br\"{u}gmann, R. Gambini and J. Pullin. Jones polynomials
for intersecting knots as physical states of quantum gravity.
\np{B385} (Oct.\ 1992) 587--603

\item R. Capovilla. Nonminimally coupled scalar field and Ashtekar
variables. \pr{D46}:1450 (Aug. 1992)

\item R. Capovilla.
 Generally covariant gauge theories.
 UMDGR 90-253 Preprint, May 1990, \np{B373}:233-246, (1992).

\item R. Capovilla and T. Jacobson. Remarks on pure spin
connection formulation of gravity. Maryland preprint UMDGR-91-134,
{\em Mod. Phys. Lett} {\bf A7}:1871-1877, (1992).

\item L. N. Chang and C. P. Soo. Ashtekar's Variables and the
Topological Phase of Quantum Gravity. In {\em Proceedings of the
Twentieth International conference on Differential Geometric
Methods in Theoretical Physics, Baruch College, City University of
New York,1-7 June 1991}, S. Catto, A. Rocha eds. (World
Scientific, Singapore 1992)

\item L. N. Chang and C. P. Soo. BRST cohomology and invariants of
four-dimensional gravity in Ashtekar's variables.
\pr{D46}:4257--62 (Nov. 1992)

\item S. Carlip. (2+1)-dimensional Chern-Simons gravity as a Dirac
square root. \pr{D45} (1992) 3584--90

\item Y.M. Cho, K.S. Soh, J.H. Yoon and Q.H. Park. Gravitation as
gauge theory of diffeomorphism group. \pl{B286}:251-255, 1992.

\item R.~Dijkgraaf, H.~Verlinde and E.~Verlinde, Loop Equations
And Virasoro Constraints In Nonperturbative 2-D Quantum Gravity,
\np{B348}, 435 (1991).

\item G. Esposito. Mathematical structures of space-time.
Cambridge preprint DAMTP-R-gols,
{\em Fortsch. Phys.} {\bf 40} (1992) 1-30.

\item F.~Fucito and M.~Martellini, Loop Equations And Kdv
Hierarchy In 2-D Quantum Gravity, Int.\ J.\ Mod.\ Phys.\
\ijmp{A7}, 2285 (1992).

\item G. F\"ul\"op. Transformations and BRST-charges in 2+1
dimensional gravitation. gr-qc/9209003, \mpl{A7}:3495--3502 (1992)

\item T. Fukuyama. Exact Solutions in Ashtekar Formalism. In {\em
Proceedings of the VIth Marcel Grossmann meeting on general
relativity}, eds.\ H. Sato and T. Nakamura (World Scientific,
1992)

\item T. Fukuyama, K. Kamimura and S. Makita. Metric from
non-metric action of gravity. \ijmp{D1}:363--70 (1992)

\item A. Giannopoulos and V. Daftardar. The direct evaluation of
the Ashtekar variables for any given metric using the algebraic
computing system STENSOR. \cqg{9}:1813--22 (July 1992)

\item J. Goldberg. Quantized self-dual Maxwell field on a null
surface. \jgp{8}:163--172 (1992)

\item J. Goldberg. Ashtekar variables on null surfaces. In {\em
Proceedings of the VIth Marcel Grossmann meeting on general
relativity}, eds.\ H. Sato and T. Nakamura (World Scientific,
1992)

\item J.N. Goldberg, J. Lewandowski, and C. Stornaiolo. Degeneracy
in loop variables. \cmp{148}:377--402 (1992)

\item J.N. Goldberg, D.C. Robinson and C. Soteriou. Null
hypersurfaces and new variables. \cqg{9} (May 1992) 1309--28

\item J. Horgan. Gravity quantized? {\em Scientific American}
(Sept.\ 1992) 18--20

\item G. Horowitz. Ashtekar's approach to quantum gravity.
University of California preprint, 1991. in Strings and Symmetries 1991, proceedings of a conference held in Stony Brook, May 21-25, 1991 (World Scientific 1992) (602 pages) (editors N. Berkovits, H. Itoyama, K. Schoutens, A. Sevrin, W. Siegel, P. van Nieuwenhuizen and J. Yamron).

\item V. Husain. 2+1 gravity without dynamics. \cqg{9} (March
1992) L33--36

\item T. Jacobson and J.D. Romano. Degenerate Extensions of
general relativity. \cqg{9} (Sept. 1992) L119--24

\item A. Kheyfets and W.A. Miller. E. Cartan moment of rotation in
Ashtekar's self-dual representation of gravitation. \jmp{33} (June
1992) 2242--2248.

\item C. Kim, T. Shimizu and K. Yushida. 2+1 gravity with spinor
field. \cqg{9} (1992) 1211-16

\bibitem{kodama} H. Kodama. Quantum gravity by the complex canonical
formulation. gr-qc/9211022, \ijmp{D1} (1992) 439

\item S. Koshti. Massless Einstein Klein-Gordon equations in the
spin connection formulation. \cqg{9} (1992) 1937--42

\item J. Lewandowski. Reduced holonomy group and Einstein's
equations with a cosmological constant. \cqg{9} (Oct. 1992)
L147--51

\item R. Loll. Independent SU(2)-loop variables and the reduced
configuration space of SU(2)-lattice gauge theory. \np{B368}
(1992) 121--42

\item R. Loll. Loop approaches to gauge field theory. Syracuse
SU-GP-92/6-2, in {\em Memorial Volume for M.K. Polivanov, Teor.
Mat. Fiz.} {\bf 91} (1992)

\item A. Magnon. Ashtekar variables and unification of
gravitational and electromagnetic interactions. \cqg{9} suppl.
(1992) S169--81

\item J. Maluf. Self-dual connections, torsion and Ashtekar's
variables. \jmp{33} (Aug.\ 1992) 2849--54

\item J.W. Maluf. Symmetry properties of Ashtekar's formulation of
canonical gravity. Nuovo Cimento {\bf 107} (July 1992) 755--

\item N. Manojlovi{\' c} and A. Mikovi{\' c}. Gauge fixing and
independent canonical variables in the Ashtekar formalism of
general relativity. \np{B382}:148--70 (June 1992)

\item N. Manojlovi{\' c} and A. Mikovi{\' c}. Ashtekar Formulation
of (2+1)-gravity on a torus. \np{B385} (July 1992) 571--586

\item E.W. Mielke. Ashtekar's complex variables in general
relativity and its teleparallelism equivalent. \apny{219} (1992)
78--108

\item E.T. Newman and C. Rovelli. Generalized lines of force as
the gauge-invariant degrees of freedom for general relativity and
Yang-Mills theory. \prl{69} (1992) 1300--3

\item P. Peld\'an. Connection formulation of (2+1)-dimensional
Einstein gravity and topologically massive gravity. \cqg{9} (Sept.
1992) 2079--92

\item L. Smolin. The ${\rm G_{Newton}\rightarrow 0}$ limit of
Euclidean quantum gravity. \cqg{9} (April 1992) 883--93

\bibitem{smolin1992} L. Smolin. Recent developments in
nonperturbative quantum gravity. In {\em Proceedings of the XXII
Gift International Seminar on Theoretical Physics, Quantum Gravity
and Cosmology, June 1991, Catalonia, Spain} (World Scientific,
Singapore 1992)

\item V. Soloviev. Surface terms in Poincare algebra in Ashtekar's
formalism. In {\em Proceedings of the VIth Marcel Grossmann
meeting on general relativity}, eds.\ H. Sato and T. Nakamura
(World Scientific, 1992)

\item Vladimir Soloviev. How canonical are Ashtekar's variables?
\pl{B292}:30, 1992.

\item R.S. Tate. Polynomial constraints for general relativity
using real geometrodynamical variables. \cqg{9} (Jan. 1992)
101--19

\item C.G. Torre. Covariant phase space formulation of
parametrized field theory. \jmp{33} (Nov. 1992) 3802--12

\item R.P. Wallner. Ashtekar's variables reexamined. \pr{D46}
(Nov. 1992) 4263--4285

\item J. Zegwaard. Gravitons in loop quantum gravity. \np{B378}
(July 1992) 288--308

\newpage

\section*{1993}

\item A. Ashtekar. Recent developments in classical and quantum
theories of connections including general relativity. In {\em
Advances in Gravitation and Cosmology}, eds.\ B. Iyer, A.
Prasanna, R. Varma and C. Vishveshwara (Wiley Eastern, New Delhi
1993)

\item A. Ashtekar, lecture notes by R.S. Tate. Physics in loop
space. In {\em Quantum gravity, gravitational radiation and large
scale structure in the universe}, eds. B.R.\ Iyer, S.V. Dhurandhar
and K. Babu Joseph (1993)

\item A. Ashtekar and J. Lewandowski. Completeness of Wilson loop
functionals on the moduli space of $SL(2,C)$ and
$SU(1,1)$-connections. gr-qc/9304044, \cqg{10} (June 1993) L69--74

\item A. Ashtekar, R.S. Tate and C. Uggla. Minisuperspaces:
observables, quantization and singularities. Int. J. Mod. Phys.
{\bf D2}, 15--50 (1993).

\item A. Ashtekar, R.S. Tate and C. Uggla. Minisuperspaces:
symmetries and quantization. In {\em Misner Festschrift}, edited
by B.L. Hu, M. Ryan and C.V. Vishveshwara (Cambridge University
Press, 1993)

\item J.C. Baez. Quantum gravity and the algebra of tangles.
\cqg{10} (April 1993) 673--94

\item I. Bengtsson. Some observations on degenerate metrics.
\grg{25} (Jan. 1993) 101--12

\item I. Bengtsson. Strange Reality:  Ashtekar's variables with
Variations. Theor. Math. Phys. {\bf 95} (May 1993) 511

\item I. Bengtsson. Neighbors of Einstein's equations --- some new
results. G\"oteborg preprint ITP92-35 \cqg{10}:1791-1802, 1993.

\item J. Birman. New points of view in knot theory. {\em Bull.
AMS}{\bf 28} (April 1993) 253--287

\item B.~Br\"ugmann, Bibliography of publications related to
classical and quantum gravity in terms of the Ashtekar variables,
arXiv:gr-qc/9303015.

\item B. Br\"ugmann, R. Gambini and J. Pullin. How the Jones
polynomial gives rise to physical states of quantum general
relativity. \grg{25} (Jan.\ 1993) 1--6

\item B. Br\"{u}gmann and J. Pullin. On the constraints of quantum
gravity in the loop representation. \np{B390} (Feb.\ 1993)
399--438

\item R. Capovilla, J. Dell and T. Jacobson. The initial value
problem in light of Ashtekar's variables. In {\em Directions in
General Relativity, vol. 2}, eds. B.~L.~Hu and T.~A.~Jacobson
(Cambridge University Press, 1993), pp. 66-77. gr-qc/9302020

\item R. Capovilla and Jerzy Pleba\'nski. Some exact solutions of
the Einstein field equations in terms of the self-dual spin
connection. \jmp{34} (Jan.\ 1993) 130--138

\item S. Carlip. Six ways to quantize (2+1)-dimensional gravity.
gr-qc/9305020, {\em Canadian Gen. Rel.} (1993), 215

\item C. Di Bartolo, R. Gambini and J. Griego. The Extended Loop
Group:  An Infinite Dimensional Manifold Associated With the Loop
Space.IFFI/93.01, gr-qc/9303010 \cmp{158} (Nov. 1993) 217--40

\item R.~Gambini, The Loop representation in gauge theories and
quantum gravity, In {\em Proceedings at the 1993 Workshop on Particles and
Fields}, R. Huerta, M.A. Perez and L.F. Urrutia, eds. (World Scientific,
Singapore, 1994) p.162. arXiv:hep-th/9403006.

\item R. Gambini and J. Pullin. Quantum Einstein-Maxwell fields: a
unified viewpoint from the loop representation. hep-th/9210110,
\pr{D47} (June 1993) R5214--8

\item D.~Giulini, Ashtekar variables in classical general
relativity, in "Canonical gravity: from classical to quantum" Eds J Ehlers and H Friedrich, Springer (1994). arXiv:gr-qc/9312032.

\item D.E. Grant. On self-dual gravity. gr-qc/9301014, \pr{D48}
(Sept. 1993) 2606--12

\item V. Husain. Ashtekar variables, self-dual metrics and
$W_\infty$. \cqg{10}:543--50 (March 1993)

\item V. Husain. General covariance, loops, and matter.
gr-qc/9304010, \pr{D47} (June 1993) 5394--9

\item V. Husain. Faraday Lines and Observables for the
Einstein-Maxwell Theory. \cqg{10} (1993) L233--L237

\item G. Immirzi. The reality conditions for the new canonical
variables of general relativity. \cqg{10} (Nov. 1993) 2347--52

\item T. Jacobson and J.D. Romano. The Spin Holonomy Group in
General Relativity. \cmp{155} (July 1993) 261--76

\item C. Kiefer. Topology, decoherence, and semiclassical gravity.
gr-qc/9306016, \pr{D47} (June 1993) 5413--21

\bibitem{kuchar} K. Kuchar. Canonical quantum gravity. gr-qc/9304012. In
{\em General Relativity and Cosmology 1992}, R.J Gleiser, C
Kozameh, O.M. Moreshi eds. (IOP Publishing, 1993).

\item H. Kunitomo and T. Sano. The Ashtekar formulation for
canonical $N=2$ supergravity. Prog. Theor. Phys. suppl. (1993) 31

\item K. Kamimura and T. Fukuyama. Massive analogue of
Ashtekar-CDJ action. Vistas in astronomy {\bf 37} (1993) 625--

\item S. Lau. Canonical variables and quasilocal energy in general
relativity. gr-qc/9307026, \cqg{10} (Nov. 1993) 2379--99

\item H.Y. Lee, A. Nakamichi and T. Ueno. Topological two-form
gravity in four dimensions. \pr{D47} (Feb.\ 1993) 1563--68

\item J. Lewandowski. Group of loops, holonomy maps, path bundle
and path connection. \cqg{10} (1993) 879--904

\item R. Loll. Lattice gauge theory in terms of independent Wilson
loops. In {\em Lattice 92}, eds J. Smit and P. van Baal, \np{B}
(Proc.\ Suppl.) {\bf 30} (March 1993)

\item R. Loll. Loop variable inequalities in gravity and gauge
theory. \cqg{10} (Aug. 1993) 1471--76

\item R. Loll. Yang-Mills theory without Mandelstam constraints.
\np{B400} (1993) 126--44

\item J. Louko. Holomorphic quantum mechanics with a quadratic
Hamiltonian constraint. gr-qc/9305003, \pr{D48} (Sept. 1993)
2708--27

\item J. Maluf. Degenerate triads and reality conditions in
canonical gravity. \cqg{10} (April 1993) 805--9

\item N. Manojlovi{\' c} and G.A. Mena Marug{\' a}n.
Nonperturbative canonical quantization of nimisuperspace models:
Bianchi types I and II. gr-qc/9304041, \pr{D48} (Oct. 1993)
3704--19

\item N. Manojlovi{\' c} and A. Mikovi{\' c}. Canonical analysis
of Bianchi models in the Ashtekar formulation. \cqg{10} (March
1993) 559--74

\item D.M. Marolf. Loop representations for $2+1$ gravity on a
torus. \cqg{10} (Dec. 1993) 2625--47

\item D. Marolf. An illustration of 2+1 gravity loop transform
troubles. gr-qc/9305015, {\em Canadian Gen. Rel.} (1993) 256.

\item H.-J. Matschull. Solutions to the Wheeler-DeWitt constraint
of canonical gravity coupled to scalar matter fields.
gr-qc/9305025, \cqg{10}:L149-L154, (1993).

\item H. Nicolai and H.J. Matschull. Aspects of Canonical Gravity
and Supergravity. \jgp{11}:15-62, (1993).

\item O. Obreg{\'o}n, J. Pullin, M.P. Ryan. Bianchi cosmologies:
New variables and a hidden supersymmetry. gr-qc/9308001, \pr{D48}
(Dec. 1993) 5642--47

\item Peter Peld\'an. Unification of gravity and Yang-Mills theory
in 2+1 dimensions. \np{B395} (1993) 239--62

\item A. Rendall. Comment on a paper of Ashtekar and Isham.
\cqg{10} (March 1993) 605--8

\item A. Rendall. Unique determination of an inner-product by
adjointness relations in the algebra of quantum observables.
\cqg{10} (Nov. 1993) 2261--69

\item J. Romano. Geometrodynamics vs. connection dynamics.
gr-qc/9303032, \grg{25} (Aug.\ 1993) 759--854

\item J. Romano. Constraint algebra of degenerate relativity.
gr-qc/9306034, \pr{D48} (Dec. 1993) 5676--83

\item C. Rovelli. Area is length of Ashtekar's triad field.
\pr{D47} (Feb.\ 1993) 1703--5

\item C. Rovelli. Basis of the Ponzano-Regge-Turaev-Viro-Ooguri
quantum-gravity model is the loop representation basis. \pr{D48}
(Sept. 1993) 2702--07

\item C. Rovelli. A generally covariant quantum field theory and a
prediction on quantum measurements of geometry. \np{B405} (Sept.
1993) 797--815

\item T. Sano and J. Shiraishi. The non-perturbative canonical
quantization of the $N=1$ supergravity. \np{B410} (Dec. 1993)
423--47

\item L. Smolin. What can we learn from the study of
non-perturbative quantum general relativity? gr-qc/9211019, in
{\em General Relativity and Cosmology 1992}, R.J Gleiser, C
Kozameh, O.M. Moreshi eds. (IOP Publishing, 1993).

\item L. Smolin. Time, measurement and information loss in quantum
cosmology. in {\em Directions in General Relativity, Proceedings,
Simposium, College Park, USA, May 1993} B.L. Hu and T. Jacobson
(eds),Cambridge U. Press, 1993. gr-qc/9301016,

\item R.S. Tate. Constrained systems and quantization. In {\em
Quantum gravity, gravitational radiation and large scale structure
in the universe}, eds. B.R.\ Iyer, S.V. Dhurandhar and K. Babu
Joseph (1993)

\item T. Thiemann. On the solution of the initial value
constraints for general relativity coupled to matter in terms of
Ashtekar's variables. \cqg{10} (Sept. 1993) 1907--21

\item T. Thiemann and H.A. Kastrup. Canonical quantization of
spherically symmetric gravity in Ashtekar's self-dual
representation. \np{B399} (June 1993) 211--58

\item D.A. Ugon, R. Gambini and P. Mora. Link invariants for
intersecting loops. \pl{B305} (May 1993) 214--22

\item J. Zegwaard. Physical interpretation of the loop
representation for non-perturbative quantum gravity. \cqg{10}
suppl. (Dec. 1993) S273--6

\item J. Zegwaard. The weaving of curved geometries. \pl{B300}
(Feb. 1993) 217--222

\newpage

\section*{1994}

\item D. Armand-Ugon, R. Gambini, J. Griego, and L. Setaro.
Classical loop actions of gauge theories, hep-th/9307179, \pr{D50}
(1994) 5352

\item A. Ashtekar. Overview and outlook. CGPG-94/1-1,
gr-qc/9403038, in  J Ehlers and H. Friedrich, eds, {\em Canonical
Gravity: From Classical to Quantum}. Springer-Verlag Berlin
(1994).

\item A. Ashtekar and J. Lee. Weak field limit of General
Relativity in terms of new variables: A Hamiltonian framework.
CGPG-94/8-3, \ijmp{D3}:675-694, (1994).

\item A. Ashtekar and J. Lewandowski. Representation theory of
analytic holonomy $C^{*}$-algebras. In {\em Knots and Quantum
Gravity}, ed. J. Baez, Oxford U. Press, 1994. gr-qc/9311010.

\item A. Ashtekar and R. Loll. New loop representations for $2+1$
gravity. CGPG-94/5-1, gr-qc/9405031, \cqg{11}:2417-2434, 1994

\item A. Ashtekar, D. Marolf and J. Mour\~ao. Integration on the
space of connections modulo gauge transformations. {\em Proc.
Lanczos International Centenary Conference}. J. Brown et al (eds),
SIAM, Philadelphia, 1994. CGPG-94/3-4, gr-qc/9403042.

\item A. Ashtekar and R.S. Tate. An algebraic extension of Dirac
quantization:  Examples. CGPG-94/6-1, gr-qc/9405073. \jmp{35}
(1994), 6434

\item A. Ashtekar and M. Varadarajan. A striking property of the
gravitational Hamiltonian. CGPG-94/8-3, gr-qc/9406040, \pr{D52}
(1994), 4944

\item J. Baez. Generalized Measures in Gauge Theory. {\em Lett.
Math. Phys.} {\bf 31} (1994) 213--223

\item J.C. Baez. Strings, loops, knots, and gauge fields. in {\em
Knots and Quantum Gravity, Proceedings, Workshop, Riverside,
1993}. J.C. Baez (ed), Clarendon, Oxford U.K. (1994).
hep-th/9309067.

\item J.C. Baez. Diffeomorphism-invariant generalized measures on
the space of connections modulo gauge transformations. in {\em The
Proceedings of the Conference on Quantum Topology}, L. Crane and
D. Yetter (eds) World Scientific, Singapore, 1994. hep-th/9305045.

\item J.F. Barbero G. Real-polynomial formulation of general
relativity in terms of connections. \pr{D49} (June 1994) 6935--38

\item J.F. Barbero. General Relativity as a Theory of 2
Connections. CGPG-93/9-5, gr-qc/9310009, \ijmp{D3} (1994) 379--392

\item J.F. Barbero G. and M. Varadarajan. The phase space of
$(2+1)$-dimensional gravity in the Ashtekar formulation. \np{B415}
(Mar. 1994) 515--530, gr-qc/9307006.

\item C. Di Bartolo, R. Gambini, J. Griego and J. Pullin. Extended
loops:  A new arena for nonperturbative quantum gravity. \prl{72}
(June 1994) 3638--41

\item Y. Bi and J. Gegenberg Loop variables in topological
gravity. gr-qc/9307031, \cqg{11} (Apr. 1994) 883--96

\item R. Borissov. Weave states for plane gravitational waves.
\pr{D49} (Jan. 1994) 923--29

\item O. B\"ostrom. Loop variables and degeneracy. in {\em
Proceedings of the VII J.A. Swieca Summer School on Particles and
Fields, Sao Paolo, Brazil, 1993}, World Scientific (1994).

\item B. Br\"ugmann. Loop Representations. MPI-Ph/93-94,
gr-qc/9312001.In  J Ehlers and H. Friedrich, eds, {\em Canonical
Gravity: From Classical to Quantum}. Springer-Verlag Berlin
(1994).

\item R. Capovilla and J. Guven. Super-Minisuperspace and New
Variables. CIEA-GR-9401, gr-qc/9402025,   \cqg{11} (1994) 1961--70

\item R. Capovilla and O. Obregon No quantum Superminisuperspace
with $\Lambda\neq 0$. \pr{D49} (1994), 6562

\item S. Carlip. Geometrical structures and loop variables in
$2+1)$-dimensional gravity. In {\em Knots and Quantum Gravity},
ed. J. Baez, Oxford U. Press, 1994. UCD-93-30, gr-qc/9309020.

\item S.M. Carroll and G.B. Field. Consequences of propogating
torsion in connection-dynamic theories of gravity.
\pr{D50}:3867-3873, 1994. gr-qc/9403058.

\item S. Chakraborty and P. Peld\'an. Towards a unification of
gravity and Yang-Mills theory. CGPG-94/1-3, gr-qc/9401028,
\prl{73} (1994) 1195.

\item S. Chakraborty and P. Peld\'an. Gravity and Yang-Mills
theory: two faces of the same theory? CGPG-94/2-2, gr-qc/9403002,
{\em Int. J. of Mod. Phys.}{\bf D}3 (1994) 695.

\item L.N. Chang and C. Soo. Superspace Dynamics and perturbations
around `emptiness'. \ijmp{D3}:529 (1994).

\item L. Crane. Topological field theory as the key to quantum
gravity. In {\em Knots and Quantum Gravity}, ed. J. Baez, Oxford
U. Press, 1994.

\item G. Esposito and H.A. Morales-T\'ecotl. Self Dual Action for
Fermionic Fields and Gravitation. gr-qc/9506073, {\em Nuov.
Cimiento} {\bf 109B}:973-982 (1994).

\item G. Esposito, H.A. Morales-T\'ecotl and G. Pollifrone.
Boundary Terms for Massless Fermionic Fields. gr-qc/9506075, {\em
Found. Phys. Lett.} {\bf 7}:303-308 (1994).

\item S. Frittelli, S. Koshti, E.T. Newman and C. Rovelli.
Classical and quantum dynamics of the Faraday lines of force.
\pr{D49} (June 1994) 6883--91

\item G. F{\" u}l{\" o}p. About a super-Ashtekar-Renteln ansatz.
gr-qc/9305001, \cqg{11} (Jan. 1994) 1--10

\item R. Gambini and J. Pullin. The Gauss linking number in
quantum gravity. In {\em Knots and Quantum Gravity}, ed. J. Baez,
Oxford U. Press, 1994.

\item D. Giulini. Ashtekar variables in classical general
relativity. THEP-93/31, gr-qc/9312032. In  J Ehlers and H.
Friedrich, eds, {\em Canonical Gravity: From Classical to
Quantum}. Springer-Verlag Berlin (1994).

\item J.N. Goldberg and D.C. Robinson. Linearized constraints in
the connection representation: Hamilton-Jacobi solution.
\pr{D50}:6338-6343, 1994. gr-qc/9405030.

\item S. Hacyan. Hamiltonian Formulation of General Relativity in
terms of Dirac Spinors. \grg{26} (Jan. 1994) 85--96

\item J. Hallin. Representations of the SU(N) T-Algebra and the
Loop Representation in 1+1 Dimensions.  \cqg{11} 1615--1629

\item H.-L. Hu. $W_{1+\infty}$, KP and loop representation of four
dimensional gravity. \pl{B324} (Apr. 1994) 293--98

\item V. Husain. Self-dual gravity and the chiral model. \prl{72}
(Feb. 1994) 800--03

\item V. Husain. Self-dual gravity as a two-dimensional theory and
conservation laws. \cqg{11} (Apr. 1994) 927--38

\item V. Husain. Observables for spacetimes with two Killing field
symmetries. Alberta-Thy-55.93, gr-qc/9402019, \pr{D50}, (1994),
6207

\item G. Immirzi. Regge Calculus and Ashtekar Variables. \cqg{11}
(1994) 1971--79 gr-qc/9402004

\item J. Iwasaki and C. Rovelli. Gravitons from Loops --
Nonperturbative Loop-Space Quantum-Gravity Contains the
Graviton-Physics Approximation.  \cqg{11} (1994) 1653--76

\item H.A. Kastrup and T. Thiemann. Spherically symmetric gravity
as a completely integrable system. PITHA 93-35, gr-qc/9401032,
\np{B425}:665-685, (1994).

\item L.H. Kauffmann. Vassiliev invariants and the loop states in
quantum gravity. In {\em Knots and Quantum Gravity}, ed. J. Baez,
Oxford U. Press, 1994. arXiv:gr-qc/9310035.

\item J. Lewandowski. Topological Measure and Graph-Differential
Geometry on the Quotient Space of Connections. gr-qc/9406025,
\ijmp{D3} (1994) 207--210

\item J. Lewandowski. Differential geometry for the space of
connections modulo gauge transformations. In {\em the proceedings
of  Cornelius Lanczos International Centenary Conference}. J.
Brown, et al (eds), SIAM Phyladelphya, 1994. gr-qc/9406026

\item R. Loll. The loop formulation of gauge theory and gravity.
In {\em Knots and Quantum Gravity}, ed. J. Baez, Oxford U. Press,
1994.

\item R. Loll. Gauge theory and gravity in the loop formulation.
CGPG-94/1-1. In  J Ehlers and H. Friedrich, eds, {\em Canonical
Gravity: From Classical to Quantum}. Springer-Verlag Berlin
(1994).

\item J. Louko and D. Marolf. Solution space of 2+1 gravity on
${\bf R} \times T^2$ in Witten's connection formulation.
gr-qc/9308018, \cqg{11} (1994), 315

\item J. M{\" a}kel{\" a}. Phase space coordinates and the
Hamiltonian constraint of Regge calculus. \pr{D49} (Mar. 1994)
2882--96

\item H.-J. Matschull. About loop states in supergravity.
DESY-94-037, gr-qc/9403034, \cqg{11}:2395-2410, (1994).

\item H.-J. Matschull and H. Nicolai. Canonical quantum
supergravity in three dimensions. gr-qc/9306018, \np{B411} (Jan.
1994) 609--46

\item G.A. Mena Marug\'an, Reality conditions for Lorentzian and
Euclidean gravity in the Ashtekar formulation, \ijmp{D3}, 513-528
(1994) [gr-qc/9311020].

\item G.A. Mena Marug{\' a}n. Reality conditions in
non-perturbative quantum cosmology. \cqg{11} (Mar. 1994) 589--608

\item G. Modanese. Wilson Loops in 4-Dimensional Quantum-Gravity.
\pr{D49} (1994) 6534--42

\item H.A. Morales-T{\' e}cotl and C. Rovelli. Fermions in quantum
gravity. \prl{72} (June 1994) 3642--45

\item O.~Moritsch, M.~Schweda, T.~Sommer, L.~Tataru and
H.~Zerrouki, BRST cohomology of Yang-Mills gauge fields in the
presence of gravity in Ashtekar variables, arXiv:hep-th/9409081.

\item J.M. Nester, R.-S. Tung and Y.Z. Zhang. Ashtekar's new
variables and positive energy. \cqg{11} (Mar. 1994) 757--66

\item J.A. Nieto, O. Obreg\'on and J. Socorro. The gauge theory of
the de-Sitter group and Ashtekar formulation. IFUG-94-001,
gr-qc/9402029, \pr{D50} (1994), 3583

\item P. Peld{\' a}n. Actions for gravity, with generalizations:
a review. gr-qc/9305011, \cqg{11} (May. 1994) 1087--1132

\item Peter Peld\'an. Ashtekar's variables for arbitrary gauge
group. G\"oteborg ITP 92-17, \pr{D46}, R2279.

\item P. Peld\'an. Real formulations of complex gravity and a
complex formulation of real gravity. GCPG-94/4-6, gr-qc/9405002,
\np{B430}, (1994), 460

\bibitem{pullin}
 J. Pullin. Knot theory and quantum gravity: a primer. in {\em
Fifth Mexican School of Particles and Fields}. J.L. Lucio and M.
Vargas (eds), AIP Conference Proceedings 317, AIP Press (1994).
hep-th/9301028.

\item A. Rendall. Adjointness relations as a criterion for
choosing an inner product. MPA-AR-94-1, gr-qc/9403001.  In  J
Ehlers and H. Friedrich, eds, {\em Canonical Gravity: From
Classical to Quantum}. Springer-Verlag Berlin (1994).

\item C. Rovelli and L. Smolin. The physical Hamiltonian in
nonperturbative quantum gravity. \prl{72} (Jan. 1994) 446--49

\item D.C. Salisbury and L.C. Shepley. A connection approach to
numerical relativity. \cqg{11}:2789-2806, 1994. gr-qc/9403040.

\item P. Schaller, J. Strobl. Canonical quantization of
two-dimensional gravity with torsion and the problem of time.
\cqg{11} (Feb. 1994) 331--46

\item L. Smolin Finite Diffeomorphism-Invariant Observables in
Quantum Gravity. SU-GP-93/1-1, gr-qc/9302011, \pr{D49} (1994)
4028--40

\item T.~A.~Schilling, Bibliography of publications related to
classical and quantum gravity in terms of the Ashtekar variables,
arXiv:gr-qc/9409031.

\item J. Tavares. Chen integrals, generalized loops and loop
calculus. \ijmp{A9}:4511-4548, 1994.

\item T. Thiemann. Reduced Phase-Space Quantization of Spherically
Symmetrical Einstein-Maxwell Theory Including a Cosmological
Constant. \ijmp{D3} (1994) 293-298

\newpage

\section*{1995}

\item D. Armand-Ugon, R. Gambini and P. Mora. Intersecting braids
and intersecting knot theory. {\em Journal of Knot theory and its
ramifications} {\bf 4}:1, 1995. hep-th/9309136.

\item A. Ashtekar. Mathematical problems of non-perturbative
quantum general relativity. Published in: {\em Gravitation and
Quantization (Les Houches, Session LVIII, 1992}. Ed. B. Julia
(Elsevier, Amsterdam, 1995)

\item A. Ashtekar. Recent Mathematical Developments in Quantum
General Relativity. In {\em The Proceedings of PASCOS-94}. K.C.
Wali (ed), World Scientific, 1995. gr-qc/9411055.

\item A. Ashtekar and J. Lewandowski. Differential geometry on the
space of connections via graphs and projective limits.
CGPG-94/12-4, hep-th/9412073, {\em J. Geom. Phys.} {\bf 17}
(1995), 191

\item A. Ashtekar and J. Lewandowski. Projective techniques and
functional integration. CGPG/94/10-6, gr-qc/9411046, \jmp{36}
(1995), 2170

\bibitem{alm2t} A. Ashtekar, J. Lewandowski, D. Marolf, J. Mour\~ao and T.
Thiemann. Quantization of diffeomorphism invariant theories of
connections with local degrees of freedom. gr-qc/9504018, \jmp{36}
(1995), 519

\item G. Au. The quest for quantum gravity. {\em Current Science},
{\bf 69}:499-518, 1995. gr-qc/9506001.

\item J.C. Baez. Link invariants, holonomy algebras and functional
integration. {\em Jour. Funct. Analysis} {\bf 127}:108, 1995.

\item J. C. Baez, J. P. Muniain, and Dardo D. Piriz. Quantum
gravity Hamiltonian for manifolds with boundary. gr-qc/9501016,
\pr{D52} (1995), 6840

\item J. F. Barbero G. Solving the constraints of general
relativity.\cqg{12}:L5-L10, 1995.
 CGPG-94/11-3, gr-qc/9411016.

\item J. F. Barbero G. Reality Conditions and Ashtekar Variables:
a Different Perspective, \pr{D51}:5498-5506, 1995. gr-qc/9410013.

\bibitem{barbero} J. F. Barbero G. Real Ashtekar Variables for Lorentzian
Signature Space-times. \pr{D51}:5507 -5510 (1995). gr-qc/9410014.

\item J. F. Barbero and M. Varadarajan. Homogeneous
(2+1)-Dimensional Gravity in the Ashtekar Formulation,
\np{B456}:355-376, 1995. gr-qc/9507044.

\item C. Di Bartolo, R. Gambini, and J. Griego. Extended loop
representation of quantum gravity, gr-qc/9406039, \pr{D51} (Jan.
1995) 502

\item C. Di Bartolo, R. Gambini, J. Griego and J. Pullin. The
space of states of quantum gravity in terms of loops and extended
loops:  some remarks. \jmp{36}:6511, 1995.
 gr-qc/9503059.

\item P.A. Blaga, O. Moritsch and H. Zerrouki. Algebraic structure
of gravity in Ashtekar's variables. TUW 94-15, hep-th/9409046,
\pr{D51} (Mar. 1995) 2792

\item B. Br\"ugmann. On a geometric derivation of Witten's
identity for Chern-Simons theory via loop deformations. {\em Int.
J. Theor. Phys.} {\bf 34}:145, 1995. hep-th/9401055.

\item G. Esposito and C. Stornaiolo. Space-time covariant form of
Ashtekar's constraints. \ncim{110B}:1137-1152, 1995.
gr-qc/9506008.

\item K. Ezawa. Combinatorial solutions to the Hamiltonian
constraint in (2+1)-dimensional Ashtekar gravity. gr-qc/9506043,
\np{B459}:355-392 (1995).

\item T.J. Foxon. Spin networks, Turaev - Viro theory and the loop
representation. \cqg{12} (April 1995) 951

\item R. Gambini, A. Garat and J. Pullin. The constraint algebra
of quantum gravity in the loop representation. CGPG-94/4-3,
gr-qc/9404059, {\em Int. J. Mod. Phys.} {\bf D4} (1995) 589

\item H. Garcia-Compean and T. Matos. Solutions in Self-dual
Gravity constructed Via Chiral Equations. \pr{D 52}:4425-4429,
1995.
 hep-th/9409135.

\item J.N. Goldberg and C. Soteriou. Canonical General Relativity
on a null surface with coordinate and gauge fixing.
\cqg{12}:2779-2798, 1995. gr-qc/9504043.

\item G. Gonzalez and R. Tate. Classical analysis of Bianchi types
I and II in Ashtekar variables. \cqg{12}:1287-1304, 1995.
 gr-qc/9412015.

\item I. Grigentch and D.V. Vassilevich. Reduced phase space
quantization of Ashtekar's gravity on de Sitter background.
\ijmp{D4}:581-588, 1995. gr-qc/9405023.

\item V. Husain. The affine symmetry of self-dual gravity.
\jmp{36}:6897, 1995.
 hep-th/9410072.

\item R.A. d'Inverno and J.A. Vickers. 2+2 decomposition of
Ashtekar variables.  \cqg{12} (Mar. 1995) 753

\item J. Iwasaki and C. Rovelli. Gravitons as embroidery on the
weave. {\em Int. J. Mod. Phys}{\bf D1} (1995) 533

\item R. Loll. Independent loop invariants for $2+1$ gravity.
CGPG-94/7-1, gr-qc/9408007, \cqg{12}:1655-1662, (1995).

\item R.~Loll, Making quantum gravity calculable, \ac{21}:131
(1995) [arXiv:gr-qc/9511080].

\item R. Loll. Quantum Aspects of $2+1$ Gravity. gr-qc/9503051,
\jmp{36}:6494-6509, (1995).

\item R. Loll. Non-perturbative solutions for lattice quantum
gravity. gr-qc/9502006, \np{B444}:614-640, (1995).

\item R. Loll. The volume operator in discretized quantum gravity.
gr-qc/9506014, \prl{75}:3048-3051, (1995).

\item R. Loll, J.M. Mour\~ao and J.N. Tavares. Generalized
coordinates on the phase space of Yang-Mills theory. CGPG-94/4-2,
gr-qc/9404060, \cqg{12}:1191-1198, (1995).

\item J. Louko. Chern-Simons functional and the no-boundary
proposal in Bianchi IX quantum cosmology. \pr{D51}:586-590, 1995.
gr-qc/9408033.

\item S. Major and L. Smolin. Cosmological histories for the new
variables. \pr{D51}:5475 (1995) gr-qc/9402018

\item D. Marolf and J.M. Mour\~ao. On the support of the
Ashtekar-Lewandowski measure. CGPG-94/3-1, hep-th/ 9403112, {\em
Comm. Math. Phys.}{\bf 170} (1995), 583

\item H.J. Matschull. Tree-dimansional Canonical Quantum Gravity.
gr-qc/9506069, \cqg{12}:2621-2704, (1995).

\item H.J. Matschull. New Representation and a Vacuun State for
Canonical Quantum Gravity. gr-qc/9412020, \cqg{12}:651-676,
(1995).

\item G.A. Mena Marug\'an. Is the exponential of the Chern-Simons
action a normalizable physical state? CGPG-94/2-2, gr-qc/9402034,
\cqg{12} (Feb. 1995) 435

\item S. Mizoguchi. The Geroch group in the Ashtekar formulation.
\pr{D51}:6788-6802, 1995. gr-qc/9411018.

\item H.A. Morales-T\'ecotl and C. Rovelli. Loop Space
Representation of Quantum Fermions and Gravity. \np{B451} (1995)
325.

\item R. de Pietri and C. Rovelli. Eigenvalues of the Weyl
Operator as Observables of General Relativity. \cqg{12}:1279-1286
(1995).

\item J. Pullin. Recent Developments in Canonical Quantum Gravity.
CAM-94 Physics Meeting, in AIP Conf. Proc {\bf 342}, 459, ed.
Zepeda A
 (AIP Press, Woodbury, New York), 1995.

\item M. Reisenberger. New Constraints for Canonical General
Relativity. \np{B457}:643-687, 1995. gr-qc/9505044.

\item C. Rovelli. Outline of a generally covariant quantum field
theory and a quantum theory of gravity. gr-qc/9503067,
\jmp{36}:6529-6547 (1995).

\item C. Rovelli and L. Smolin. Spin Networks and Quantum Gravity.
gr-qc/9505006, \pr{D52}:5743-5759.

\bibitem{RS-geo}C. Rovelli and L. Smolin. Discreteness of area and
volume in quantum gravity. \np{B442}:593-622 (1995). Erratum:
\np{B456}:734 (1995). gr-qc/9411005,

\item L. Smolin. Linking TQFT and Nonperturbative Quantum Gravity.
gr-qc/9505028, \jmp{36}:6417-6455 (1995).

\item L. Smolin and C. Soo. The Chern-Simons invariant as the
natural time variable for classical and quantum cosmology.
CGPG-94/4-1, gr-qc/9405015, \np{B449}:289-316 (1995).

\item C. Soo. Self-dual variables, positive semi-definite action,
and discrete transformations in 4-d quantum gravity.
gr-qc/9504042, \pr{D52}:3484-3493 (1995).

\item I. A. B. Strachan. The symmetry structure of the
anti-self-dual Einstein hierarchy. \jmp{36}:3566-3573, 1995.
 hep-th/9410047.

\item T. Thiemann. Generalized boundary conditions for general
relativity for the asymptotically flat case in terms of Ashtekar's
variables. \cqg{12} (Jan. 1995) 181

\item T. Thiemann. Complete quantization of a diffeomorphism
invariant field theory. \cqg{12} (Jan. 1995) 59

\item T. Thiemann The reduced phase space of spherically symmetric
Einstein-Maxwell theory including a cosmological constant.
\np{B436} (Feb. 1995) 681

\item T. Thiemann. An account of transforms on $\overline{{\cal
A}/{\cal G}}$. {\em Acta Cosmologica} {\bf 21}:145-167, 1995.
gr-qc/9511050.

\item M.~Tsuda, T.~Shirafuji and H.~J.~Xie, Ashtekar variables and
matter coupling, arXiv:gr-qc/9501021.

\item
  M.~Tsuda, T.~Shirafuji and H.~J.~Xie,
  General considerations of matter coupling with the selfdual connection,
  \cqg{12}, 3067 (1995), arXiv:gr-qc/9505019.

\item R.S. Tung and T. Jacobson. Spinor one forms as gravitational
potentials \cqg{12}:L51-L55, 1995. gr-qc/9502037.

\item R. P. Wallner. A new form of Einstein's equations,
\jmp{36} (1995) 6937-6969.

\newpage
\section*{1996}

\item D. Armand-Ugon, R. Gambini, O. Obregon and J. Pullin.
Towards a loop representation for quantum canonical supergravity.
hep-th/9508036, \np{B460} (1996), 615

\item J. M. Aroca, H. Fort and R. Gambini The Path Integral for
The Loop Representation of Lattice Gauge Theories.
\pr{D54}:7751-7756, 1996. hep-th/9605068

\item A. Ashtekar. A Generalized Wick Transform for Gravity.
gr-qc/9511083, \pr{D53}:2865-2869, (1996),

\item A. Ashtekar and J. Lewandowski. Quantum Field Theory of
Geometry,  in {\em Proceedings, Conference on Historical
Examination and Philosophical Reflections on Foundations of
Quantum Field Theory, Boston, MA, 1996}. hep-th/9603083

\item A. Ashtekar, J. Lewandowski, D. Marolf, J. Mour\~ao and T.
Thiemann. A manifestly gauge-invariant approach to quantum
theories of gauge fields. in {\em Geometry of Constrained
Dynamical Systems. Proceedings, Conference, Cambridge, UK, 1994}.
J.M. Charap (ed), Cambridge University Press, 1996. hep-th/9408108

\item A. Ashtekar, J. Lewandowski, D. Marolf, J. Mour\~ao and T.
Thiemann. Coherent State Transforms for Spaces of Connections.
{\em J. Functional Analysis} {\bf 135}:519-551, 1996.
 gr-qc/9412014.

\item J.~F.~Barbero and M.~P.~.~Ryan, Minisuperspace Examples of
Quantization Using Canonical Variables of the Ashtekar Type:
Structure and Solutions, \pr{D53}:5670 (1996)
[arXiv:gr-qc/9510030].

\item J. C. Baez. Spin networks in gauge theory. {\em Advances in
Mathematics} {\bf 117}:253, 1996. gr-qc/9411007.

\item J. C. Baez. Spin networks in nonperturbative quantum
gravity. In {\em Interface of Knot Theory and Physics}, L.
Kauffman (ed), American Mathematical Society, Providence, Rhode
Island, 1996. gr-qc/9504036.

\item J.C. Baez. Knots and Quantum Gravity: Progress and Prospects.
in {\sl Proceedings of the Seventh Marcel Grossman Meeting on General
Relativity}, ed.\ Robert T.\ Jantzen and G.\ Mac Keiser,
World Scientific Press, Singapore, 1996, pp.\ 779--797. gr-qc/9410018.

\item J. F. Barbero G. From Euclidean to Lorenzian General
Relativity: The Real Way, \pr{D54}:1492-1499, 1996. gr-qc/9605066

\item R. De Pietri. Spin Networks and Recoupling in Loop Quantum
Gravity, in {\em Proceedings of the 2nd Conference on
Constrained Dynamics and Quantum Gravity, Santa Margherita, Italy,
1996}. gr-qc/9701041.

\item S.~Frittelli, L.~Lehner and C.~Rovelli, The complete
spectrum of the area from recoupling theory in loop  quantum
gravity, \cqg{13}:2921 (1996) [arXiv:gr-qc/9608043].

\item J. F. Barbero and M. P. Ryan. Minisuperspace Examples of
Quantization Using Canonical Variables of the Ashtekar Type:
Structure and Solutions. \pr{D53}:5670-5681, 1996. gr-qc/9510030.

\item M. Barreira, M. Carfora and C. Rovelli. Physics with
nonperturbative quantum gravity: radiation from a quantum black
hole. \grg{28}:1293-1299, 1996. gr-qc/9603064.

\item R. Borissov, S. Major and L. Smolin, The Geometry of Quantum
Spin Networks. \cqg{13}:3183-3196, 1996. gr-qc/9512043.

\item B.~Brugmann, On the constraint algebra of quantum gravity in
the loop representation, \np{B474}, 249 (1996)
[arXiv:gr-qc/9512036].

\item L. N. Chang and Chopin Soo. Chiral fermions, gravity and
GUTs. CGPG-94/9-3, hep-th/9411064, \pr{D53}:5682-5691 (1996).

\item L. Chang and C. Soo. The standard model with gravity
couplings. \pr{D53}, 1996, pp.5682-5691. CGPG-94/6-2.

\bibitem{depietri1} R. De Pietri and C. Rovelli. Geometry Eigenvalues and
Scalar Product from Recoupling Theory in Loop Quantum Gravity.
gr-qc/9602023, \pr{D54}:2664-2690 (1996).

\item Ch. Devchand and V. Ogievetsky. Self-dual Gravity Revisited.
\cqg{13}, 2515-2536 (1996).
JINR-E2-94-342, hep-th/9409160.

\item C. Di Bartolo. The Gauss constraint in the extended loop
representation. \pl{B389} (1996) 661-664. gr-qc/9607014.

\item K. Ezawa. A Semiclassical Interpretation of the Topological
Solutions for Canonical Quantum Gravity. gr-qc/9512017,
\pr{D53}:5651-5663 (1996).

\item K. Ezawa. Multi-plaquette solutions for discretized Ashtekar
gravity. gr-qc/9510019, {\em Mod. Phys. Lett.} {\bf A}:349-356
(1996).

\item K. Ezawa. Ashtekar's formulation for $N=1,2$ supergravities
as ``constrained" BF theories. hep-th/9511047, {\em Prog. Theor.
Phys.} {\bf 95}:863 (1996).

\item S. Fritelli, L. Lehner and C. Rovelli. The complete spectrum
of the area from recoupling theory in loop quantum gravity.
\cqg{13}:2921-2932,1996.  gr-qc/9608043.

\item R. Gambini and J. Pullin. A rigorous solution of the quantum
Einstein equations. \pr{D54}:5935-5938, 1996. gr-qc/9511042.

\item R. Gambini and J. Pullin. Knot theory and the dynamics of
quantum general relativity. \cqg{13}:L125, 1996. gr-qc/9603019.

\item H. Garcia-Compean, L. E. Morales and J. F. Plebanski. A Hopf
algebra structure in self-dual gravity. {\em Rev. Mex. Fis.} {\bf
42}:695, 1996. hep-th/9410154.

\item H. Garcia-Compean, J. Plebanski and M. Przanowski. From
Principal chiral model to selfdual gravity. \mpl{A11}:663-674,
1996.

\item H. Garcia-Compean, J. Plebanski and M. Przanowski Further
remarks on the chiral model approach to self dual
gravity.\pl{219A}:249-256, 1996. hep-th/9512013

\item J.N. Goldberg. Generalized Hamilton-Jacobi transformations:
gauge and diffeomorphism constraint. \pr{D54}:4997-5001, 1996.

\item J. Griego. Extended knots and the space of states of quantum
gravity. \np{B473}:291-307, 1996. gr-qc/9601007.

\item J. Griego. The Kauffman Bracket and the Jones Polynomial in
Quantum Gravity. \np{B467}:332-354, 1996. gr-qc/9510050.

\item J. Griego. Is the Third Coefficient of the Jones Knot
Polynomial a Quantum State of Gravity? \pr{D53}:6966-6978, 1996.
gr-qc/9510051.

\item N. Grot and C. Rovelli. Moduli-space of knots with
intersections. gr-qc/960410, \jmp{37}:3014-3021 (1996).

\item H. Ishihara, H. Kubotani and T. Fukuyama. Gravitational
Instantons in Ashtekar's Formalism. \ijmp{A11}:2707-2720, 1996.
gr-qc/9509009.

\item T. Jacobson. 1+1 sector of 3+1 gravity. \cqg{13}:L111-L116,
1996. erratum-ibid {\bf 13}:3269, 1996. gr-qc/9604003.

\item S. Holst. Barbero's Hamiltonian derived from a generalized
Hilbert-Palatini action. \pr{D53}:5966-5969, 1996. gr-qc/9511026.

\item V. Husain. Einstein's equations and the chiral model.
gr-qc/9602050, \pr{D53} (1996), 4327

\item V. Husain. General Covariance, and Supersymmetry Without
Supersymmetry. \pr{D54}:7849-7856, 1996. hep-th/9609009.

\item G. Immirzi Quantizing Regge Calculus. \cqg{13}:2385-2394,
1996. gr-qc/9512040.

\item K. Krasnov. Quantum loop representation for fermions coupled
to Einstein-Maxwell field. \pr{D53}:1874-1888, 1996.

\item S. R. Lau. New Variables, the gravitational action, and
boosted quasilocal stress-energy-momentum. \cqg{13}:1509-1540,
1996. gr-qc/9504026.

\item L. Leal. Electric-Magnetic duality and the `Loop
Representation' in Abelian gauge theories. \mpl{A11}:1107-1114,
1996.

\item R. Loll. Spectrum of the volume operator in quantum gravity.
gr-qc/9511030 \np{B460}:143-154, (1996).

\item R. Loll. A real alternative to quantum gravity in loop
space. \pr{D54}:5381, 1996. gr-qc/9602041.

\item R. Loll, J. Mour\~ao, J. Tavares. Complexification of gauge
theories. {\em Journal of Geometry and Physics} {\bf 17} (1996) 1-24. hep-th/9307142

\item S. Major and L. Smolin. Quantum Deformation of Quantum
Gravity. \np{B473}:267-290, 1996. gr-qc/9512020.

\item H.~A.~Morales-Tecotl, L.~F.~Urrutia and J.~D.~Vergara,
Reality conditions for Ashtekar variables as Dirac constraints,
\cqg{13}:2933 (1996) [arXiv:gr-qc/9607044].

\item H.J. Matschull. Causal Structure and Diffeomorphisms in
Ashteker's Gravity. gr-qc/9511034, \cqg{13}:765-782, (1996).

\item G. Mena Marug\'an. Involutions on the algebra of physical
observables from reality conditions. \jmp{37}:196-205, 1996.
gr-qc/9506038.

\item H.A. Morales-T\'ecotl, L.F. Urrutia and J.D. Vergara.
Reality Conditions for Ashtekar Variables as Dirac Constraints.
\cqg{13}:2933-2940, 1996. gr-qc/9607044.

\item P. Peldan. Large Diffeomorphisms in (2+1)-Quantum Gravity on
the Torus. CGPG-95/1-1, gr-qc/9501020 \pr{D53} (1996), 3147

\item J. Plebanski and M. Przanowski. The Lagrangian of a
self-dual gravitational field as a limit of the SDYM Lagrangian.
\pl{212A}:22. 1996. hep-th/9605233.

\item C. Rovelli. Loop Quantum Gravity and Black hole Physics.
{\em Helv. Phys. Acta} {\bf 69}:582-611, 1996. gr-qc/9608032.

\bibitem{BH-rovelli} C. Rovelli. Black hole entropy from loop quantum gravity.
\prl{77}:3288-3291, 1996. gr-qc/9603063.

\item T.A. Schilling. Non-covariance of the generalized
holonomies:  Examples. \jmp{37}, 4041 (1996). CGPG-95/3-1, gr-qc/9503064.

\item T. Thiemann. Reality conditions inducing transforms for
quantum gauge field theory and quantum gravity.
\cqg{13}:1383-1404, 1996. gr-qc/9511057.

\item T. Thiemann. Anomaly-Free Formulation of Nonperturbative
Four-dimensional Lorentzian Quantum Gravity. \pl{B380}:257-264,
1996. gr-qc/9606088.

\item H. Waelbroeck and J.A. Zapata. 2+1 covariant lattice theory
and `t Hooft's formulation. \cqg{13}:1761-1768, 1996.
gr-qc/9601011.

\item G. Yoneda and H. Shinkai. Constraints and reality conditions
in the Ashtekar formulation of general relativity.
\cqg{13}:783-790, 1996. gr-qc/9602026.

\item J. A. Zapata. Topological lattice gravity using self dual
variables. \cqg{13}:2617-2634, 1996. gr-qc/9603030.

\newpage
\section*{1997}

\item D. Altschuler \& L. Freidel. Vasilev knot invariants and
Chern--Simons perturbation theory to all orders.
\cmp{187}:261--287, 1997. [q--alg/9603010]

\item A. Ashtekar. Geometric Issues in Quantum Gravity,
in {\em Geometric Issues in the Foundation of Science}, L. Mason
et al (eds.) (Oxford University Press, 1997). CGPG-96-61-4.

\item A. Ashtekar. Polymer Geometry at Planck Scale and Quantum
Einstein Equations. in {\em Proceedings of the 14th International
Conference on General Relativity and Gravitation}, M. Francaviglia
(ed) World Scientific, Singapore, 1997. hep-th/9601054.

\item A. Ashtekar. Quantum Mechanics of Riemannian Geometry. in
{\em the Proceedings of the Barcelona Workshop Fisica I Geometria and Pacific Conference on Gravitation and Cosmology}, edited by D. Jou (Institut D'Estudis Catan, Barcelona, 1997).

\item A.~Ashtekar and A.~Corichi, Gauss linking number and
electro-magnetic uncertainty principle, \pr{D56}: 2073 (1997)
[arXiv:hep-th/9701136].

\item A.~Ashtekar and A.~Corichi. Photon Inner Product and the
Gauss Linking Number. \cqg{14}:A43-A53,  1997. gr-qc/9608017.

\bibitem{AL-Geo} A. Ashtekar and J. Lewandowski. Quantum Theory of Geometry
I:  Area Operators. \cqg{14}:A55-A81,  1997. gr-qc/9602046.

\item A.~Ashtekar, J.~Lewandowski, D.~Marolf, J.~Mourao and
T.~Thiemann, SU($N$) quantum Yang-Mills theory in two dimensions:
A complete solution, \jmp{38}: 5453 (1997) [arXiv:hep-th/9605128].

\item J.C. Baez and S. Sawin. Functional Integration on Spaces of
Connections. {\em Jour.
Funct. Analysis} {\bf 150}, 1-27 (1997). q-alg/9507023.

\item I. Bengtsson and A. Kleppe. On chiral P forms. \ijmp{A12}, No. 19 (1997) 3397-3411.
hep-th/9609102.

\item G.~Barnich and V.~Husain, Geometrical representation of the
constraints of Euclidean general relativity, \cqg{14}: 1043 (1997)
[arXiv:gr-qc/9611030].

\item R.~Borissov, Operator calculations in loop quantum gravity,
\npps{57}:237 (1997).

\item R. Borissov. Regularization of the Hamiltonian constraint
and the closure of the constraint algebra. \pr{D55}:2059-2068,
1997. gr-qc/9411038.

\item R. Borissov Graphical evolution of spin network states.
\pr{D55}:6099--6111, 1997. [gr--qc/9606013]

\item R. Borissov,  R. De Pietri \& C. Rovelli.  Matrix elements
of Thiemann's Hamiltonian constraint in loop quantum gravity.
\cqg{14}:2793--2823, 1997. [gr--qc/9703090]

\item A.~Corichi and J.~A.~Zapata, On diffeomorphism invariance
for lattice theories, \np{B493}: 475 (1997) [arXiv:gr-qc/9611034].

\bibitem{depietri} R.~De Pietri, The equivalence between the connection and the
loop representation of quantum gravity, in {\em Jerusalem 1997, Recent developments in theoretical and experimental general relativity, gravitation, and relativistic field theories, Pt.B} 892-894. arXiv:gr-qc/9711021.

\bibitem{depietri2} R. De Pietri. On the relation between the connection and
the loop representation of quantum gravity. \cqg{14}: 53-69, 1997.
gr-qc/9605064.

\item R.~De Pietri, Spin networks and recoupling in loop quantum
gravity, \npps{57}: 251 (1997) [arXiv:gr-qc/9701041].

\item H.~Fort, R.~Gambini and J.~Pullin, Lattice knot theory and
quantum gravity in the loop representation, \pr{D56}: 2127 (1997)
[arXiv:gr-qc/9608033].

\item B.P. Dolan and K.P. Haugh. A Covariant Approach to
Ashtekar's Canonical Gravity,  \cqg{14}:477-488,  1997.

\item K. Ezawa  Nonperturbative solutions for canonical quantum
gravity:  An Overview. \pR{286}:271--348, 1997. [gr--qc/9601050]

\item H. Fort,  R. Gambini \& J. Pullin. Lattice knot theory and
quantum gravity in the loop representation. \pr{D56}:2127--2143,
1997. [gr--qc/9608033]

\item R. Gambini \& J. Pullin.  Variational derivation of exact
skein relations from Chern--Simons theories. \cmp{185}:621--640,
1997. [hep--th/9602165]

\item R. Gambini,  J. Griego \& J. Pullin. Chern--Simons states in
spin network quantum gravity. \mpl{B413}:260--266, 1997.
[gr--qc/9703042]

\item N. Grot \& C. Rovelli. Weave states in loop quantum gravity.
\grg{29}: 1039--1048, 1997.

\item Y.~Hashimoto, Y.~Yasui, S.~Miyagi and T.~Otsuka,
  Applications of the Ashtekar gravity to four dimensional hyperKaehler
  geometry and Yang-Mills instantons,
  \jmp{38}, 5833 (1997)
  [arXiv:hep-th/9610069].

\item V. Husain \& S. Major. Gravity and BF theory defined in
bounded regions. \np{B500}: 381--401, 1997. [gr--qc/9703043]

\item G. Immirzi. Quantum gravity and Regge calculus. in {\em 2nd
Meeting on Constrained Dynamics and Quantum Gravity}, Santa
Margherita, Italy, 1996. \npps{57} (1997) 65-72.
  gr-qc/9701052.

\bibitem{immirzi} G.~Immirzi, Real and complex connections for canonical
gravity, \cqg{14}: L177 (1997) [arXiv:gr-qc/9612030].

\item V.~M.~Khatsymovsky,
  Ashtekar Constraint Surface as Projection of Hilbert-Palatini One,
  \pl{B394}, 57 (1997)
  [arXiv:gr-qc/9604053].

\item K. Krasnov. Geometrical entropy from loop quantum gravity.
\pr{D55}:3505,  1997. gr-qc/9603025.

\bibitem{Lew-vol} J. Lewandowski. Volume and Quantizations.
\cqg{14}:71-76, 1997. gr-qc/9602035.

\item J.~Lewandowski and J.~Wisniewski, 2+1 sector of 3+1 gravity,
\cqg{14}: 775 (1997) [arXiv:gr-qc/9609019].

\item R.~Loll, Further results on geometric operators in quantum
gravity, \cqg{14}: 1725 (1997) [arXiv:gr-qc/9612068].

\item R.~Loll, Imposing det $E > 0$ in discrete quantum gravity,
 \pl{B399} (1997) 227-232. gr-qc/9703033.

\item R.~Loll. Still on the way to quantizing gravity in Dremin, I.M., and Semikhatov, A.M., eds., Proceedings of the 2nd International Sakharov Conference on Physics, 280283, (World Scientic, Singapore, 1997).
gr-qc/9701032.

\item R. Loll Wilson loop coordinates for $2+1$ gravity.
in Bassan, M. et al., ed., Proceedings of the 12th Italian Conference on General Relativity and Gravitational Physics, 280-283, (World Scientific, Singapore, 1997). Preprint
CGPG-94/8-1.

\item F. Markopoulou \& L. Smolin. Causal evolution of spin
networks. \np{B508}:409--430, 1997. [gr--qc/9702025]

\item D. Marolf,  J. Mourao \& T. Thiemann.  The Status of
diffeomorphism superselection in Euclidean (2+1) gravity.
\jmp{38}:4730--4740, 1997. [gr--qc/9701068]

\item G.A. Mena Marugan, Canonical quantization of the Gowdy model,
\pr{D56}, 908-919 (1997) [gr-qc/9704041].

\item M. Montesinos--Velasquez,  H.A. Morales-- Tecotl \& Tonatiuh
Matos. Fermion mass gap in the loop representation of quantum
gravity. \cqg{14}:L135--L142, 1997. [gr--qc/9704066]

\item D. E. Neville. Open-flux solutions to the constraints for
plane gravity waves. \pr{D55}: 766-780,  1997. gr-qc/9511061.

\item D. E. Neville. Closed flux solutions to the quantum
constraints for plane gravity waves. \pr{D55}: 2069-2075,  1997.
gr-qc/9607053.

\item J.~A.~Nieto, J.~Socorro and O.~Obregon,
  Gauge theory of supergravity based only on a selfdual spin connection,
  \prl{76}, 3482 (1996).

\item J. Pullin. Canonical quantum gravity with new variables and
loops: a report in {\em Gravitation and Cosmology}, S. Dhurandhar
and T. Padmanabhan (Eds), Kluwer Academic Publishing, Amsterdam 1997,
pp 199-210. gr-qc/9606061.

\item M. Reisenberger. A Left-Handed Simplicial Action for
Euclidean General Relativity. \cqg{14}, 1753-1770 (1997).
gr-qc/9609002.

\item M.~P.~Reisenberger and C.~Rovelli, ``Sum over surfaces"' form
of loop quantum gravity, \pr{D56}:3490 (1997)
[arXiv:gr-qc/9612035].

\item C.~Rovelli, ``Quantum gravity as a `sum over
        surfaces",  {\em Nuclear Physics B (Proc Suppl)}
        {\bf 57} (1997) 28-43.

\item L. Smolin. Experimental Signatures of Quantum Gravity.
 in {\em The Proceedings of the 1994 Drexel Symposium on
Quantum Theory and Measurement}, World Scientific, 1997
gr-qc/9503027.

\item  V.~O.~Solovev,
  The algebra independence of boundary conditions in the Ashtekar
  formalism,''
  {\em Theor.\ Math.\ Phys.}\  {\bf 112}, 906 (1997)
  [{\em Teor.\ Mat.\ Fiz.}\  {\bf 112N1}, 142 (1997)]
  [arXiv:gr-qc/9809058].

\item T. Thiemann. An axiomatic approach to quantum gauge field
theory. {\em Banach Center Publications} Vol. 39, 1997
Symplectic Singularities and Geometry of Gauge Fields
Editors of the volume: Robert Budzynski, Stanislaw Janeczko, Witold Kondracki, Alfred F. Kunzle p 389-403. hep-th/9511122.

\item T.~C.~Toh, The loop algebra of quantum gravity, \hpa{70}:417
(1997).

\item J.~A.~Zapata, Combinatorial space from loop quantum gravity,
\grg{30}: 1229 (1998) [arXiv:gr-qc/9703038].

\item J.~A.~Zapata, A combinatorial approach to diffeomorphism
invariant quantum gauge theories, \jmp{38}: 5663 (1997)
[arXiv:gr-qc/9703037].

\item G.~Yoneda, H.~a.~Shinkai and A.~Nakamichi,
  A trick for passing degenerate points in Ashtekar formulation,
  \pr{D56}, 2086 (1997)
  [arXiv:gr-qc/9704081].

\newpage
\section*{1998}

\item  S.~Y.~Alexandrov and D.~V.~Vassilevich,
  Path integral for the Hilbert-Palatini and Ashtekar gravity,
  \pr{D58}, 124029 (1998)
  [arXiv:gr-qc/9806001].

\item G.~Amelino-Camelia, On the area operators of the
Husain-Kuchar-Rovelli model and  canonical/loop quantum gravity,
\mpl{A13}:1155 (1998) [arXiv:gr-qc/9804063].

\item J.M. Aroca,   H. Fort \& R. Gambini. World sheet formulation
for lattice staggered fermions. \pr{D57}:3701--3710, 1998

\item J.M. Aroca, H. Fort and  R. Gambini. On the path integral
loop representation of (2+1) lattice non-Abelian theory. \pr{D58}, 045007, 1998.
arXiv: hep-lat/9703007.

\bibitem{abck} A. Ashtekar,   J. Baez,   A. Corichi \& K. Krasnov.
Quantum geometry and black hole entropy. \prl{80}: 904--907, 1998.
[gr-qc/9710007]

\item A. Ashtekar, A. Corichi \& J. A. Zapata. Quantum theory of
geometry III: Noncommutativity of riemannian structures. \cqg{15}:
2955--2972,  1998  e--Print Archive: gr--qc/9806041

\item J.~C.~Baez, Degenerate solutions of general relativity from
topological field  theory, \cmp{193}: 219 (1998)
[arXiv:gr-qc/9702051].

\item J.C. Baez \& K.V. Krasnov.  Quantization of diffeomorphism
invariant theories with fermions. \jmp{39}:1251--1271,  1998.
[hep--th/9703112]

\item J.W. Barrett. The Classical evaluation of relativistic spin
networks.  \atmp{2}:593--600,  1998. [math.qa/9803063]

\item J.W. Barrett \& L. Crane.  Relativistic spin networks and
quantum gravity. \jmp{39}:3296--3302,  1998. [gr--qc/9709028]

\item S.~Chakraborty and N.~Chakravarty, A study of
Kantowski-Sachs model in Ashtekar variables, \ijmp{A13}:4931
(1998).

\item A. Corichi \& K.V. Krasnov.  Ambiguities in loop
quantization:  Area versus electric charge. \mpl{A13}:1339--1346,
1998.

\item R.~De Pietri, Canonical `loop' quantum gravity and spin foam
models, in Proceeding of the XXIII SIGRAV conference, Monopoli (Italy), September 1998.  arXiv:gr-qc/9903076.

\item R. Gambini,   J. Griego \& J. Pullin.  A Spin network
generalization of the Jones polynomial and Vassilev invariants.
\mpl{B425}:41--47,  1998. [q--alg/9711014]

\item R. Gambini,   J. Griego \&,   J. Pullin.  Vassiliev
invariants:  A New framework for quantum gravity.
\np{B534}:675--696,  1998. [gr--qc/9803018]

\item R. Gambini,   J. Lewandowski,   D. Marolf \& J. Pullin. On
the consistency of the constraint algebra in spin network quantum
gravity. \ijmp{D7}:97--109,  1998. [gr--qc/9710018]

\item R. Gambini,   H.A. Morales--Tecotl,   L.F. Urrutia \& J.D.
Vergara.  Loop variables for compact two--dimensional quantum
electrodynamics. \pr{D57}:3711--3724,  1998. [hep--th/9712096]

\item R.~Gambini and J.~Pullin, Does loop quantum gravity imply
Lambda = 0?, \pl{B437}:279 (1998) [arXiv:gr-qc/9803097].

\item
  L.~J.~Garay and G.~A.~Mena Marugan, Thiemann transform for gravity
  with matter fields,'' \cqg{15}, 3763 (1998)
  [arXiv:gr-qc/9805010].

\item H. Garcia--Compean, J. Plebanski and M. Przanowski. Geometry
associated with self-dual Yang-Mills and the chiral model approach
to selfdual gravity. {\em Acta Physica Polonica} {\bf B29}, 549-571, 1998.
hep-th/9702046.

\item C.M. Granzow \& G. Mahler.  Quantum trajectories of
interacting pseudo spin networks. \ap{B67}:733--741,  1998.
[quant--ph/9901004]

\item  M.~S.~Iriondo, E.~O.~Leguizamon and O.~A.~Reula,
  On the dynamics of Einstein's equations in the Ashtekar formulation,
  {\em Adv.\ Theor.\ Math.\ Phys.}\  {\bf 2}, 1075 (1998)
  [arXiv:gr-qc/9804019].

\item K.V. Krasnov.  Note on the area spectrum in quantum gravity.
\cqg{15}:L47--L53,  1998. [gr--qc/9803074]

\item K.V. Krasnov.  On Quantum statistical mechanics of
Schwarzschild black hole. \grg{30}:53--68,  1998. [gr--qc/9605047]

\item K.V. Krasnov. On the constant that fixes the area spectrum
in canonical quantum gravity. \cqg{15}:L1--L4,  1998.
[gr--qc/9709058]

\item J. Lewandowski \& D. Marolf. Loop constraints:  A Habitat
and their algebra. \ijmp{D7}:299--330,  1998. [gr--qc/9710016]

\item R.~Loll,
  Discrete approaches to quantum gravity in four dimensions,
  {\em Living Rev.\ Rel.}\  {\bf 1}, 13 (1998)
  [arXiv:gr-qc/9805049].

\item
  R.~Loll,
  On the diffeomorphism-commutators of lattice quantum gravity,
  \cqg{15}, 799 (1998)
  [arXiv:gr-qc/9708025].

\item F. Markopoulou \& L. Smolin. Nonperturbative dynamics for
abstract (p,  q) string networks. \pr{D58}:084033,  1998.
[hep--th/9712148]

\item F. Markopoulou \& L. Smolin. Quantum geometry with intrinsic
local causality. \pr{D58}:084032,  1998. [gr--qc/9712067]

\item
  G.~A.~Mena Marugan,
  Geometric interpretation of Thiemann's generalized Wick transform,
  {\em Grav.\ Cosmol.}\  {\bf 4}, 257 (1998)
  [arXiv:gr-qc/9705031].

\item
  E.~W.~Mielke,
  Chern-Simons solution of the Ashtekar constraints for the  teleparallelism
  equivalent of gravity,''
  {\em Acta Phys.\ Polon.}\ B {\bf 29}, 871 (1998).

\item A. Mikovic \& N. Manojlovic. Remarks on the reduced phase
space of (2+1)--dimensional gravity on a torus in the Ashtekar
formulation. \cqg{15}:3031--3039,  1998. [gr--qc/9712011]

\item M. Montesinos \& C. Rovelli. The Fermionic contribution to
the spectrum of the area operator in nonperturbative quantum
gravity. \cqg{15}:3795--3801,  1998. [gr--qc/9806120]

\bibitem{Rovelli:1997yv} C.~Rovelli, Loop quantum gravity, \lrr{1}:1 (1998)
[arXiv:gr-qc/9710008].

\bibitem{Rovelli:1997qj}
C. Rovelli, Strings, loops and others: A critical survey
 of the present approaches to quantum gravity, in
 {\em Gravitation and Relativity: At the turn of the Millennium, 15th International Conference on General Relativity and Gravitation}. Edited by Naresh Dadhich and Jayant Narlikar. Published by Inter-University Centre for Astronomy and Astrophysics, 1998., p.281.
  {\it Preprint} gr-qc/9803024

\item C. Rovelli \& T. Thiemann. The Immirzi parameter in quantum
general relativity. \pr{D57}: 1009--1014,  1998. [gr--qc/9705059]

\item L. Smolin. The future of spin networks, in {\em
Geometric Issues in the Foundation of Science}, L. Mason et al
(eds.) (Oxford University Press, 1998). qr-qc/9702030.

\bibitem{length} T.~Thiemann, A length operator for canonical quantum
gravity, \jmp{39}, 3372 (1998) [arXiv:gr-qc/9606092].

\item T.~Thiemann, Closed formula for the matrix elements of the
volume operator in  canonical quantum gravity, \jmp{39}: 3347
(1998) [arXiv:gr-qc/9606091].

\item T. Thiemann. Quantum spin dynamics (QSD). \cqg{15}:839--873,
1998. [gr--qc/9606089]

\item T. Thiemann. Quantum spin dynamics (QSD) 2.
\cqg{15}:875--905,  1998. [gr--qc/9606090]

\item T. Thiemann. QSD 3:  Quantum constraint algebra and physical
scalar product in quantum general relativity. \cqg{15}:1207--1247,
1998. [gr--qc/9705017]

\item T. Thiemann. QSD 4:  (2+1) Euclidean quantum gravity as a
model to test (3+1) Lorentzian quantum gravity.
\cqg{15}:1249--1280,  1998. [gr--qc/9705018]

\item T. Thiemann. QSD 5:  Quantum gravity as the natural
regulator of matter quantum field theories. \cqg{15}:1281--1314,
1998. [gr--qc/9705019]

\item T. Thiemann. QSD 6:  Quantum Poincare algebra and a quantum
positivity of energy theorem for canonical quantum gravity.
\cqg{15}:1463--1485,  1998. [gr--qc/9705020]

\item T. Thiemann. Kinematical Hilbert spaces for Fermionic and
Higgs quantum field theories. \cqg{15}:1487--1512,  1998.
[gr--qc/9705021]

\item T. Thiemann. The inverse loop transform, \jmp{39} (1998) 1236-1248.
hep-th/9601105.

\item M.~Zagermann, Metric Versus Ashtekar Variables in Two
Killing Field Reduced Gravity, \cqg{15}:1367 (1998)
[arXiv:gr-qc/9710133].

\newpage
\section*{1999}

\item J.M. Aroca  \& Y. Kubyshin. Calculation of wilson loops in
two--dimensional yang--mills theories. Jan 1999. 45pp. Published
in \apny{ 283}:11--56,  2000 e--Print Archive: hep--th/9901155.

\item M.~Arnsdorf and R.~S.~Garcia, Existence of spinorial states
in pure loop quantum gravity, \cqg{16}, 3405 (1999)
[arXiv:gr-qc/9812006].

\item A. Ashtekar. Classical \& quantum physics of isolated horizons:
a brief overview. \lnp{541}:50--70, 2000 Also in *Polanica 1999,
Towards quantum gravity* 50--70

\item A. Ashtekar. Interface of general relativity,   quantum
physics and statistical mechanics:  some recent developments.
\Ap{9}:178--198,  2000  e--Print Archive: gr--qc/9910101

\item J.C. Baez \& J.W. Barrett. The Quantum tetrahedron in
three--dimensions and four--dimensions. \atmp{3}:815--850,  1999.
[gr--qc/9903060]

\item
  J.~F.~Barbero and M.~Varadarajan,
  A Comment on the Degrees of Freedom in the Ashtekar Formulation for 2+1
  Gravity,
  Class.\ Quant.\ Grav.\  {\bf 16}, 3765 (1999)
  [arXiv:gr-qc/9909036].

\item J.W. Barrett,   M.Rocek \& R.M. Williams. A Note on area
variables in Regge calculus. \cqg{16}:1373--1376,  1999.
[gr--qc/9710056]

\item J.W. Barrett \& R.M. Williams. The Asymptotics of an
amplitude for the four simplex. \atmp{3}:209--215,  1999.
[gr--qc/9809032]

\item R. Borissov \& S. Gupta. Propagating spin modes in canonical
quantum gravity. \pr{D60}:024002,  1999. [gr--qc/9810024]

\item R. De Pietri \& L. Freidel.  SO(4) Plebanski action and
relativistic spin foam model. \cqg{16}:2187--2196,  1999.
[gr--qc/9804071]

\item  C.~Fleischhack,
  A new type of loop independence and SU(N) quantum Yang-Mills theory in  two
  dimensions,
  \jmp{40}, 2584 (1999),
  [J.\ Math.\ Phys.\  {\bf 41}, 76 (2000)].

\item L. Freidel \& K. Krasnov. Discrete space--time volume for
three--dimensional BF theory and quantum gravity.
\cqg{16}:351--362,  1999. [hep--th/9804185]

\item L. Freidel,   K. Krasnov \& R. Puzio. BF description of
higher dimensional gravity theories. \atmp{3}:1289--1324,  1999.
[hep--th/9901069]

\item L. Freidel \& K. Krasnov. Spin foam models and the classical
action principle. \atmp{2}:1183--1247,  1999. [hep--th/9807092]

\item R. Gambini,   O. Obregon \& J. Pullin. Yang--Mills analogs
of the Immirzi ambiguity. \pr{D59}:047505,  1999. [gr--qc/9801055]

\item R. Gambini \& J. Pullin. Nonstandard optics from quantum
space--time. By \pr{D59}:124021,  1999. [gr--qc/9809038]

\item R. Gambini \& J. Pullin.  Quantum gravity experimental
physics? \grg{31}:1631--1637,  1999.

\item H.~Garcia-Compean, J.~A.~Nieto, O.~Obregon and C.~Ramirez,
  Dual description of supergravity MacDowell-Mansouri theory,
  \pr{D59}, 124003 (1999)
  [arXiv:hep-th/9812175].

\item V. Husain Apparent horizons,   black hole entropy and loop
quantum gravity. \pr{D59}:084019,  1999. [gr--qc/9806115]

\item V. Husain. Diffeomorphism invariant $SU(N)$ gauge theories.
\cqg{16}:1455--1461,  1999. [hep--th/9901127]

\item V. Husain. Initial data for gravity coupled to scalar,
electromagnetic and Yang--Mills fields. \pr{D59}:044004,  1999.
[gr--qc/9805100]

\item V. Husain \& S. Jaimungal. Topological holography.
\pr{D60}:061501,  1999. [hep--th/9812162]

\item K.V. Krasnov  Quanta of geometry and rotating black holes.
\cqg{16}:L15--L18,  1999. [gr--qc/9902015]

\item K.V. Krasnov  Quantum geometry and thermal radiation from
black holes. \cqg{16}:563--578,  1999. [gr--qc/9710006]

\item J. Lewandowski \& T. Thiemann.   Diffeomorphism invariant
quantum field theories of connections in terms of webs.
\cqg{16}:2299--2322,  1999. [gr--qc/9901015]

\item J. Lewandowski \& J. Wisniewski.  Degenerate sectors of the
Ashtekar gravity. \cqg{16}:3057--3069,  1999. [gr--qc/9902037]

\bibitem{seth} S.~A.~Major,
  A spin network primer,
  {\em Am.\ J.\ Phys.}\  {\bf 67}, 972 (1999).
  [arXiv:gr-qc/9905020].

\bibitem{angles}  S.~A.~Major,
  Operators for quantized directions,
  \cqg{16}, 3859 (1999).
  [arXiv:gr-qc/9905019].

\item  S.~A.~Major,
  Embedded graph invariants in Chern-Simons theory,
  \np{B550}, 531 (1999). [arXiv:hep-th/9810071].

\item M. Montesinos,   H.A. Morales--Tecotl,   L.F. Urrutia \&
J.D. Vergara. Real sector of the nonminimally coupled scalar field
to selfdual gravity. \jmp{40}:1504--1517,  1999. [gr--qc/9903043]

\item M. Montesinos,   H.A. Morales--Tecotl,   L.F. Urrutia, J.D.
Vergara.  Complex canonical gravity and reality constraints.
\grg{31}:719--723,  1999.

\item M. Montesinos,   C. Rovelli \& T. Thiemann. SL(2,  R) mode
with two Hamiltonian constraints. \pr{D60}:044009,  1999.
[gr--qc/9901073]

\item J.M. Mourao,   T. Thiemann \& J.M. Velhinho. Physical
properties of quantum field theory measures. \jmp{40}:2337--2353,
1999. [hep--th/9711139]

\item J.~A.~Nieto and J.~Socorro,
  Self-dual gravity and self-dual Yang-Mills in the context of
  Macdowell-Mansouri formalism,
  \pr{D59}, 041501 (1999)
  [arXiv:hep-th/9807147].

\item
  T.~Ootsuka, S.~Miyagi, Y.~Yasui and S.~Zeze,
  Anti-self-dual Maxwell solutions on hyperKaehler manifold and N = 2
  supersymmetric Ashtekar gravity,
  \cqg{16}, 1305 (1999)
  [arXiv:gr-qc/9809083].

\item C. Rovelli.  The Projector on physical states in loop
quantum gravity. \pr{D59}:104015,  1999. [gr--qc/9806121]

\item L. Smolin. The new universe around the next corner.
\pw{12N12}:79--84,  1999.

\item
  G.~Yoneda and H.~a.~Shinkai,
  Symmetric hyperbolic system in the Ashtekar formulation,
  \prl{82}, 263 (1999)
  [arXiv:gr-qc/9803077].

\newpage
\section*{2000}

\item J. Alfaro,   H.A. Morales--Tecotl,   L.F. Urrutia. Quantum
gravity corrections to neutrino propagation. \prl{84}:2318--2321,
2000. [gr--qc/9909079]

\item M. Arnsdorf \& S. Gupta. Loop quantum gravity on noncompact
spaces. \np{B577}:529--546,  2000. [gr--qc/9909053]

\bibitem{abk} A. Ashtekar,   J.C. Baez \& K. Krasnov. Quantum geometry
of isolated horizons and black hole entropy, \atmp{4}: 1--94,
2000, e--Print Archive: gr--qc/0005126

\bibitem{ack} A. Ashtekar,   A. Corichi \& K. Krasnov. Isolated
horizons: The classical phase space. \atmp{3}:419--478,  2000
e--Print Archive:  gr--qc/9905089

\item A. Ashtekar,   D. Marolf,   J. Mourao \& T. Thiemann.
Constructing hamiltonian quantum theories from path integrals in a
diffeomorphism--invariant context.  AEI--109, CGPG--99--5--1,
2000. 19pp. Published in \cqg{17}:4919--4940, 2000.  e--Print
Archive:  quant--ph/9904094

\item J.C. Baez.  An Introduction to spin foam models of quantum
gravity and BF theory. \lnp{543}:25--94,  2000. [gr--qc/9905087]

\item J.W. Barrett \& L. Crane. A Lorentzian signature model for
quantum general relativity. \cqg{17}:3101--3118,  2000.
[gr--qc/9904025]

\item M. Bojowald. Abelian BF theory and spherically symmetric
electromagnetism. \jmp{41}:4313--4329,  2000. [hep--th/9908170]

\bibitem{lqc1} M. Bojowald. Loop quantum cosmology. I. Kinematics.
\cqg{17}:1489--1508,  2000. [gr--qc/9910103]

\item M. Bojowald.  Loop quantum cosmology. II. Volume operators.
\cqg{17}:1509--1526,  2000. [gr--qc/9910104]

\item M. Bojowald \& H.A. Kastrup.  Quantum symmetry reduction for
diffeomorphism invariant theories of connections.
\cqg{17}:3009--3043,  2000. [hep--th/9907042]

\item G. Burgio,   R. De Pietri,   H.A. Morales--Tecotl,   L.F.
Urrutia,   J.D. Vergara.  The Basis of the physical Hilbert space
of lattice gauge theories. \np{B566}:547--561,  2000.
[hep--lat/9906036]

\item O.~Brodbeck and M.~Zagermann, Dimensionally reduced gravity,
Hermitian symmetric spaces and the  Ashtekar variables
\cqg{17}:2749 (2000)

\item S.~Chakraborty and N.~Chakraborty, Quantum cosmology in
Ashtekar variables, \nc{B115}:537 (2000).

\item R. De Pietri,   L. Freidel,   K. Krasnov \& C. Rovelli.
Barrett--Crane model from a Boulatov--Ooguri field theory over a
homogeneous space. \np{B574}:785--806,  2000. [hep--th/9907154]

\item C. Di Bartolo,   R. Gambini,   J. Griego \& J. Pullin.
Canonical quantum gravity in the Vassiliev invariants arena. 1.
Kinematical structure. \cqg{17}:3211--3238,  2000.
[gr--qc/9911009]

\item C. Di Bartolo,   R. Gambini,   J. Griego \& J. Pullin.
Canonical quantum gravity in the Vassilev invariants arena. 2.
Constraints,   habitats and consistency of the constraint algebra.
\cqg{17}:3239--3264,  2000. [gr--qc/9911010]

\item C. Di Bartolo,   R. Gambini,   J. Griego \& J. Pullin.
Consistent canonical quantization of general relativity in the
space of Vassilev knot invariants. \prl{84}:2314--2317,  2000.
[gr--qc/9909063]

\item C.~Fleischhack,
  Stratification of the Generalized Gauge Orbit Space,
  \cmp{214}, 607 (2000)
  [arXiv:math-ph/0001008].

\item  C.~Fleischhack,
  Gauge Orbit Types for Generalized Connections,
  \cmp{214}, 607 (2000)
  [arXiv:math-ph/0001006].

\item
  C.~Fleischhack,
  Hyphs and the Ashtekar-Lewandowski measure,
  arXiv:math-ph/0001007.

\item L. Freidel \& K. Krasnov. Simple spin networks as Feynman
graphs. \jmp{41}:1681--1690,  2000. [hep--th/9903192]

\item R. Gambini \& J. Pullin.  The Large cosmological constant
approximation to classical and quantum gravity:  Model
examples.\cqg{17}:4515--4540,  2000. [gr--qc/0008032]

\item R. Gambini \& J. Pullin. Making classical and quantum
canonical general relativity computable through a power series
expansion in the inverse cosmological constant.
\prl{85}:5272--5275,  2000. [gr--qc/0008031]

\item M. Gaul \& C. Rovelli. Loop quantum gravity and the meaning
of diffeomorphism invariance. \lnp{541}:277--324,  2000.
[gr--qc/9910079]

\item S. Kauffman \& L. Smolin Combinatorial dynamics in quantum
gravity. \lnp{541}:101--129,  2000. [hep--th/9809161]

\item J. Lewandowski \& A. Okolow.  Two form gravity of the
Lorentzian signature. \cqg{17}:L47--L51,  2000. [gr--qc/9911121]

\item Y. Ling \& L. Smolin Supersymmetric spin networks and
quantum supergravity.  \pr{D61}:044008,  2000. [hep--th/9904016]

\item J. Louko \& C. Rovelli. Refined algebraic quantization in
the oscillator representation of SL(2,  R). \jmp{41}:132--155,
2000. [gr--qc/9907004]

\item  S.~A.~Major,
  Quasilocal energy for spin-net gravity,
  \cqg{17}, 1467 (2000).
  [arXiv:gr-qc/9906052].

\bibitem{Polychronakos:2000mr}
  A.~P.~Polychronakos,
  The area-entropy relation for black holes,
{\it Prepared for Workshop on Current Developments in High-Energy
Physics: HEP 2000, Ioannina, Greece, 20-23 Apr 2000}

\item T. Regge \& R.M. Williams.  Discrete structures in gravity.
\jmp{41}:3964--3984,  2000. [gr--qc/0012035]

\bibitem{Rovelli:1999hz}
C. Rovelli. The Century of the incomplete revolution: Searching
for general relativistic quantum field theory.
\jmp{41}:3776--3800,  2000. [hep--th/9910131]

\item J. Samuel.  Canonical gravity,   diffeomorphisms and
objective histories. \cqg{17}:4645--4654,  2000. [gr--qc/0005094]

\item J. Samuel.  Is Barbero's Hamiltonian formulation a gauge
theory of Lorentzian gravity? \cqg{17}:L141--L148,  2000.
[gr--qc/0005095]

\item L. Smolin A Candidate for a background independent
formulation of M theory. \pr{D62}:086001,  2000. [hep--th/9903166]

\item L. Smolin  A Holographic formulation of quantum general
relativity. \pr{D61}:084007,  2000. [hep--th/9808191]

\item L. Smolin. M theory as a matrix extension of Chern--Simons
theory. \np{B591}:227--242,  2000. [hep--th/0002009]

\item  M.~Tsuda,
  Generalized Lagrangian of N = 1 supergravity and its canonical  constraints
  with the real Ashtekar variable,''
  \pr{D61}, 024025 (2000)
  [arXiv:gr-qc/9906057].

\item
  M.~Tsuda and T.~Shirafuji,
  The canonical formulation of N = 2 supergravity in terms of the  Ashtekar
  variable,''
  \pr{D62}, 064020 (2000)
  [arXiv:gr-qc/0003010].

\item
  M.~Varadarajan,
  Fock representations from U(1) holonomy algebras,
  \pr{D61}, 104001 (2000),
  [arXiv:gr-qc/0001050].

\item M. Varadarajan \& J.A. Zapata. A Proposal for analyzing the
classical limit of kinematic loop gravity. \cqg{17}:4085--4110,
2000. [gr--qc/0001040]

\item J.~M.~Velhinho, A groupoid approach to spaces of generalized
connections, \jgp{41}, 166 (2002) [arXiv:hep-th/0011200].

\item G.~Yoneda and H.~a.~Shinkai,
  Constructing hyperbolic systems in the Ashtekar formulation of general
  relativity,''
  \ijmp{D9}, 13 (2000)
  [arXiv:gr-qc/9901053].

\newpage
\section*{2001}

\item A. Ashtekar \& J. Lewandowski. Relation between polymer and
Fock excitations.
 Jul 2001. 13pp. \cqg{18}:L117--L128,  2001 e--Print Archive:  gr--qc/0107043

\item J.C. Baez \& J.W. Barrett. Integrability for relativistic
spin networks. \cqg{18}:4683--4700,  2001. [gr--qc/0101107]

\item M. Bojowald. The Semiclassical limit of loop quantum
cosmology. \cqg{18}:L109--L116,  2001. [gr--qc/0105113]

\item M. Bojowald. The Inverse scale factor in isotropic quantum
geometry. \pr{D64}:084018,  2001. [gr--qc/0105067]

\item M. Bojowald. Dynamical initial conditions in quantum
cosmology. \prl{87}:121301,  2001. [gr--qc/0104072]

\bibitem{bigbang} M. Bojowald. Absence of singularity in loop quantum
cosmology. \prl{86}:5227--5230,  2001. [gr--qc/0102069]

\item M. Bojowald.  Loop quantum cosmology. 4. Discrete time
evolution. \cqg{18}:1071--1088,  2001. [gr--qc/0008053]

\item M. Bojowald.  Loop quantum cosmology. 3. Wheeler--Dewitt
operators. \cqg{18}:1055--1070,  2001. [gr--qc/0008052]

\item R. Capovilla,   M. Montesinos,   V.A. Prieto \& E. Rojas. BF
gravity and the Immirzi parameter. \cqg{18}:L49--,  2001,
Erratum--ibid.18: 1157,  2001. [gr--qc/0102073]

\item S.~Carlip Quantum gravity:  A Progress report. \rpp{64}:885,
2001. [gr--qc/0108040]

\item S.~Chakraborty, Quantum cosmology with scalar-tensor
theories in Ashtekar variables,  \nc{116B}, 351 (2001).

\item A.~Corichi \& J.M.~Reyes. A Gaussian weave for kinematical
loop quantum gravity. \ijmp{D10}:325--338,  2001. [gr--qc/0006067]

\item L. Crane,   A. Perez \& C. Rovelli.  Perturbative finiteness
in spin--foam quantum gravity. \prl{87}:181301,  2001.

\item
  C.~Fleischhack,
  On the support of physical measures in gauge theories,
  arXiv:math-ph/0109030.

\item
  C.~Fleischhack,
  On the structure of physical measures in gauge theories,
  arXiv:math-ph/0107022.

\item L. Freidel \& D. Louapre. Diffeomorphisms and spin foam
models. \np{B662}:279--298,  2003. [gr--qc/0212001]

\item M. Gaul \& C. Rovelli. A Generalized Hamiltonian constraint
operator in loop quantum gravity and its simplest Euclidean matrix
elements. \cqg{18}:1593--1624,  2001. [gr--qc/0011106]

\item R. Linares,   L.F. Urrutia \& J.D. Vergara. Exact solution
of the Schwinger model with compact U(1). \mpl{A16}:121--134,
2001. [hep--th/0103057]

\item Y. Ling \& L. Smolin. Holographic formulation of quantum
supergravity. \pr{D63}:064010,  2001. [hep--th/0009018]

\item Y. Ling \& L. Smolin. Eleven--dimensional supergravity as a
constrained topological field theory. \np{B601}:191--208, 2001.
[hep--th/0003285]

\item  S.~A.~Major and K.~L.~Setter,
  On the universality of the entropy-area relation,
  \cqg{18}, 5293 (2001).
  [arXiv:gr-qc/0108034].

\item  S.~A.~Major,
  New operators for spin net gravity: Definitions and consequences,
  arXiv:gr-qc/0101032.

\item  S.~A.~Major and K.~L.~Setter,
  Gravitational statistical mechanics: A model,
  \cqg{18}, 5125 (2001).
  [arXiv:gr-qc/0101031].

\item J. Makela \& R.M. Williams. Constraints on area variables in
Regge calculus. \cqg{18}:L43--,  2001. [gr--qc/0011006]

\item A. Mikovic Quantum field theory of spin networks.
\cqg{18}:2827--2850,  2001. [gr--qc/0102110]

\item M. Montesinos \& C. Rovelli. Statistical mechanics of
generally covariant quantum theories:  A Boltzmann -- like
approach. \cqg{18}:555--569,  2001. [gr--qc/0002024]

\item M. Montesinos. Selfdual gravity with topological terms.
\cqg{18}:1847--1852,  2001. [gr--qc/0104068]

\item M. Montesinos \& J.D. Vergara. Gauge invariance of complex
general relativity. \grg{33}:921--929,  2001. [gr--qc/0010113]

\item R. Oeckl \& H. Pfeiffer.  The Dual of pure nonAbelian
lattice gauge theory as a spin foam model. \np{B598}:400--426,
2001. [hep--th/0008095]

\item D. Oriti Space--time geometry from algebra:  Spin foam
models for nonperturbative quantum gravity. \rpp{64}:1489--1544,
2001. [gr--qc/0106091]

\item D. Oriti \& R.M. Williams. Gluing 4 simplices:  A Derivation
of the Barrett--Crane spin foam model for Euclidean quantum
gravity. \pr{D63}:024022,  2001. [gr--qc/0010031]

\item A. Perez \& C. Rovelli.  3+1 spinfoam model of quantum
gravity with space -- like and time -- like components.
\pr{D64}:064002,  2001. [gr--qc/0011037]

\item A. Perez \& C. Rovelli. A Spin foam model without bubble
divergences. \np{B599}:255--282,  2001. [gr--qc/0006107]

\item A. Perez \& C. Rovelli. Spin foam model for Lorentzian
general relativity. \pr{D63}:041501,  2001. [gr--qc/0009021]

\item M.P. Reisenberger \& C. Rovelli.  Space--time as a Feynman
diagram:  The Connection formulation. \cqg{18}:121--140, 2001.
[gr--qc/0002095]

\item H. Sahlmann,   T. Thiemann \& O. Winkler. Coherent states
for canonical quantum general relativity and the infinite tensor
product extension. \np{B606}:401--440,  2001. [gr--qc/0102038]

\item D.C. Salisbury.   Quantum general invariance and loop
gravity. \fp{31}:1105--1118,  2001. [gr--qc/0105097]

\item J. Samuel  Comment on Holst's Lagrangian formulation.
\pr{D63}:068501,  2001.

\item J. Samuel. Comment on Immirzi parameter in quantum general
relativity. \pr{D64}:048501,  2001.

\item M.~Sawaguchi, Canonical formalism of N=1 supergravity with
the real Ashtekar variables, \cqg{18}:1179 (2001).

\item T. Thiemann Gauge field theory coherent states (GCS):  1.
General properties. \cqg{18}:2025--2064,  2001. [hep--th/0005233]

\item T. Thiemann \& O. Winkler. Gauge field theory coherent
states (GCS). 2. Peakedness properties. \cqg{18}:2561--2636, 2001.
[hep--th/0005237]

\item T. Thiemann \& O. Winkler. Gauge field theory coherent
states (GCS):  3. Ehrenfest theorems. \cqg{18}:4629--4682,  2001.
[hep--th/0005234]

\item T. Thiemann \& O. Winkler. Gauge field theory coherent
states (GCS) 4:  Infinite tensor product and thermodynamical
limit. \cqg{18}:4997--5054,  2001. [hep--th/0005235]

\item T. Thiemann Quantum spin dynamics (QSD):  7. Symplectic
structures and continuum lattice formulations of gauge field
theories. \cqg{18}:3293--3338,  2001. [hep--th/0005232]

\item
  M.~Varadarajan,
  Photons from quantized electric flux representations,
  \pr{D64}, 104003 (2001),
  [arXiv:gr-qc/0104051].

\item
  Y.~B.~Wu and Y.~X.~Gui,
  The double coupling of the Ashtekar gravitational field to the Dirac
  spinoral fields,
  {\em Chin.\ Phys.}\  {\bf 10}, 902 (2001).

\item R.~E.~Zimmermann, Classicity from Entangled Ensemble States
of Knotted Spin Networks. A Conceptual Approach, Cf. A. Fokas et al. (eds.), Congress Proceedings, International Press, Boston, 2001, 468 sq.
arXiv:gr-qc/0007024.

\newpage
\section*{2002}

\item S. Alexandrov.  Hilbert space structure of covariant loop
quantum gravity. \pr{D66}:024028,  2002. [gr--qc/0201087]

\item  S. Alexandrov.   On choice of connection in loop quantum
gravity. \pr{D65}:024011,  2002. [gr--qc/0107071]

\item J. Alfaro,   H.A. Morales--Tecotl \& L.F. Urrutia. Loop
quantum gravity and light propagation. \pr{D65}:103509,  2002.
[hep--th/0108061]

\item J. Alfaro,   H.A. Morales--Tecotl \& L.F. Urrutia. Quantum
gravity and spin 1/2 particles effective dynamics.
\pr{D66}:124006,  2002. [hep--th/0208192]

\item J. Alfaro \& G. Palma. Loop quantum gravity corrections and
cosmic rays decays. \pr{D65}:103516,  2002. [hep--th/0111176]

\item A. Ashtekar. Addressing challenges of quantum gravity
through quantum geometry:  black holes and big--bang.  Prepared
for International Conference on Theoretical Physics (TH 2002),
Paris,   France, 22-- 26 Jul 2002. \ahp{4}:S55--S69,  2003

\item A. Ashtekar,   S. Fairhurst \& J.L. Willis. Quantum gravity,
shadow states,   and quantum mechanics. CGPG--02--7--2,   Jul
2002. 35pp.  \cqg{20}:1031--1062,  2003 e--Print Archive:
gr--qc/0207106

\item A. Ashtekar, J. Lewandowski \& H. Sahlmann. Polymer and fock
representations for a scalar field.  ESI--1233, CGPG--02--11--1,
Nov 2002. 13pp. \cqg{20}:L11--1,  2003 e--Print Archive:
gr--qc/0211012

\item J.C. Baez,   J.D. Christensen \& G. Egan. Asymptotics of 10j
symbols. \cqg{19}:6489,  2002. [gr--qc/0208010]

\item J.C. Baez,   J.D. Christensen,   T.R. Halford \& D.C. Tsang.
Spin foam models of Riemannian quantum gravity.
\cqg{19}:4627--4648,  2002. [gr--qc/0202017]

\item J.C. Baez \& J.D. Christensen. Positivity of spin foam
amplitudes. \cqg{19}:2291--2306,  2002. [gr--qc/0110044]

\item M. Bojowald \& F. Hinterleitner. Isotropic loop quantum
cosmology with matter. \pr{D66}:104003,  2002. [gr--qc/0207038]

\bibitem{inflation} M. Bojowald. Inflation from quantum geometry.
\prl{89}:261301,  2002,    [gr--qc/0206054]

\item M. Bojowald Isotropic loop quantum cosmology.
\cqg{19}:2717--2742,  2002. [gr--qc/0202077]

\item M. Bojowald  Quantization ambiguities in isotropic quantum
geometry. \cqg{19}:5113--5230,  2002. [gr--qc/0206053]

\item L.~Bombelli, Statistical geometry of random weave states,
in {\em Proceedings of the Ninth Marcel Grossmann Meeting on
General Relativity}, eds V G Gurzadyan, R T Jantzen and R Ruffini,
World Scientific 2002. arXiv: gr-qc/0101080.

\item S.~Chakraborty, Quantum cosmology in Ashtekar variables with
non-minimally coupled scalar-tensor theory, \pra{58}:1 (2002).

\item J.D. Christensen \& G. Egan. G An Efficient algorithm for
the Riemannian 10j symbols. \cqg{19}:1185--1194,  2002.
[gr--qc/0110045]

\item R. De Pietri,   L. Lusanna,   L. Martucci \& S. Russo.
Dirac's observables for the rest frame instant form of tetrad
gravity in a completely fixed 3 orthogonal gauge.
\grg{34}:877--1033,  2002. [gr--qc/0105084]

\item C. Di Bartolo,   R. Gambini \& J. Pullin. Canonical
quantization of constrained theories on discrete space--time
lattices. \cqg{19}:5275--5296,  2002. [gr--qc/0205123]

\item L. Freidel \& K. Krasnov. The Fuzzy sphere star product and
spin networks. \jmp{43}:1737--1754,  2002. [hep--th/0103070]

\item R. Gambini \& J. Pullin. A Finite spin foam--based theory of
three--dimensional and four--dimensional quantum gravity.
\pr{D66}:024020,  2002. [gr--qc/0111089]

\item L.~J.~Garay and G.~A.~Mena Marugan,
  Immirzi ambiguity in the kinematics of quantum general relativity,
  \pr{D66}, 024021 (2002)
  [arXiv:gr-qc/0205021].

\item  A.M. Gavrilik.  Applying the q algebras U'(q) (so(n)) to
quantum gravity:  Towards q--deformed analog of SO(n) spin
networks. \ujp{47}:213--218,  2002. [gr--qc/0401067]

\item F. Girelli,   R. Oeckl \& A. Perez. Spin foam diagrammatics
and topological invariance. \cqg{19}:1093--1108, 2002.
[gr--qc/0111022]

\item V. Husain. Naked singularities and the Wilson loop.
\mpl{A17}:955--966,  2002. [hep--th/0204180]

\item I.B. Khriplovich.  Entropy and area of black holes in loop
quantum gravity. \mpl{B537}:125--129,  2002. [gr--qc/0109092]

\item P. Kramer \& M. Lorente.   Surface embedding, topology and
dualization for spin networks. \jp{A35}:8563--8574, 2002.
[gr--qc/0401107]

\item R.E. Livine \& D. Oriti. Barrett--Crane spin foam model from
generalized BF type action for gravity. \pr{D65}:044025, 2002.
[gr--qc/0104043] "

\item E.R. Livine. Projected spin networks for Lorentz connection:
Linking spin foams and loop gravity. \cqg{19}:5525--5542,  2002.
[gr--qc/0207084]

\item
  S.~A.~Major and M.~D.~Seifert,
  Modelling space with an atom of quantum geometry,''
  \cqg{19}, 2211 (2002).
  [arXiv:gr-qc/0109056].

\item
  T.~J.~Konopka and S.~A.~Major,
  Observational limits on quantum geometry effects,
  New J.\ Phys.\  {\bf 4}, 57 (2002).
  [arXiv:hep-ph/0201184].

\item A. Mikovic \& R. Picken. Super Chern Simons theory and flat
super connections on a torus. \atmp{5}:243--263,  2002.
[math--ph/0008006]

\item A. Mikovic. Spin foam models of matter coupled to gravity.
\cqg{19}:2335--2354,  2002. [hep--th/0108099]

\item
  G.~A.~Mena Marugan,
  Extent of the Immirzi ambiguity in quantum general
  relativity,
  \cqg{19}, L63 (2002)
  [arXiv:gr-qc/0203027].

\item R. Oeckl The Quantum geometry of supersymmetry and the
generalized group extension problem. \jgp{44}:299--330,  2002.
[hep--th/0106122]

\item D. Oriti Boundary terms in the Barrett--Crane spin foam
model and consistent gluing. \mpl{B532}:363--372,  2002.
[gr--qc/0201077]

\item D. Oriti \& H. Pfeiffer. A Spin foam model for pure gauge
theory coupled to quantum gravity. \pr{D66}:124010,  2002.
[gr--qc/0207041]

\item A. Perez Spin foam quantization of SO(4) Plebanski's action.
\atmp{5}:947--968,  2002,   Erratum--ibid.6: 593--595, 2003.
[gr--qc/0203058]

\item  M. Reisenberger \& C. Rovelli.  Space--time states and
covariant quantum theory. \pr{D65}:125016,  2002. [gr--qc/0111016]

\item C. Rovelli.  Partial observables. \pr{D65}:124013,  2002.
[gr--qc/0110035]

\item D. Sudarsky,   L. Urrutia,   H. Vucetich. New observational
bounds to quantum gravity signals. \prl{89}:231301, 2002.
[gr--qc/0204027]

\item L.F. Urrutia. Quantum gravity corrections to particle
interactions. \mpl{A17}:943--954,  2002. [gr--qc/0205103]

\item
  M.~Varadarajan,
  Gravitons from a loop representation of linearised gravity,
  \pr{D66}, 024017 (2002),
  [arXiv:gr-qc/0204067].

\item J.~M.~Velhinho, Invariance properties of induced Fock
measures for U(1) holonomies, \cmp{227}, 541 (2002)
[arXiv:math-ph/0107002].

\item
J.~M.~Velhinho,
  A groupoid approach to spaces of generalized connections,
  J.\ Geom.\ Phys.\  {\bf 41}, 166 (2002)
  [arXiv:hep-th/0011200].

\item J.~M.~Velhinho, Some properties of generalized connections
in quantum gravity, in
{\em Proceedings of the MGIX MM Meeting} (Volume 2)
The University of Rome "La Sapienza", 2–8 July 2000,
edited by Vahe G Gurzadyan (Yerevan Physics Institute, Armenia), Robert T Jantzen (Villanova University, USA) \& Remo Ruffini (University of Rome "La Sapienza", Italy)
p 1278, 2002. arXiv:hep-th/0101141.

\item J.A. Zapata. Continuum spin foam model for 3--d gravity.
\jmp{43}:5612--5623,  2002. [gr--qc/0205037]

\newpage
\section*{2003}

\item
  A.~Alekseev, A.~P.~Polychronakos and M.~Smedback,
  On area and entropy of a black hole,
  \pl{B574}, 296 (2003)
  [arXiv:hep-th/0004036].

\item S. Alexandrov \& E.R. Livine. SU(2) loop quantum gravity
seen from covariant theory. \pr{D67}:044009,  2003.
[gr--qc/0209105]

\item J. Alfaro,   H.A. Morales--Tecotl,   M. Reyes \& L.F.
Urrutia. On nonAbelian holonomies. \jp{A36}:12097--12107, 2003.
[hep--th/0304223]

\item J. Alfaro \& G. Palma. Loop quantum gravity and
ultrahigh--energy cosmic rays. \pr{D67}:083003,  2003.
[hep--th/0208193]

\item A. Ashtekar,   M. Bojowald \& J. Lewandowski. Mathematical
structure of loop quantum cosmology. \atmp{7}:233--268,  2003.
[gr--qc/0304074]

\item A. Ashtekar \& A. Corichi. Nonminimal couplings,   quantum
geometry and black hole entropy. \cqg{20}:4473--4484,  2003.
[gr--qc/0305082]

\item A. Ashtekar,   J. Lewandowski \& H. Sahlmann.  Polymer and
Fock representations for a scalar field. \cqg{20}:L11--1, 2003.
[gr--qc/0211012]

\item A. Ashtekar,   J. Wisniewski \& O. Dreyer. Isolated horizons
in (2+1) gravity. \atmp{6}:507--555,  2003. [gr--qc/0206024]

\item J. Baez. Quantum gravity:  The quantum of area?
\n{421}:702--703,  2003.

\item J.W. Barrett. Geometrical measurements in three--dimensional
quantum gravity. \ijmp{A18S2}:97--113,  2003. [gr--qc/0203018]

\item J.W. Barrett \& C. M. Steele. Asymptotics of relativistic
spin networks. \cqg{20}:1341--1362,  2003. [gr--qc/0209023]

\item M. Bojowald Homogeneous loop quantum cosmology.
\cqg{20}:2595--2615,  2003. [gr--qc/0303073]

\item M. Bojowald Initial conditions for a universe.
\grg{35}:1877--1883,  2003. [gr--qc/0305069]

\item M. Bojowald \& K. Vandersloot. Loop quantum cosmology,
boundary proposals,   and inflation. \pr{D67}:124023,  2003.
[gr--qc/0303072]

\item D.~Colosi \& C.~Rovelli, A simple background-independent
Hamiltonian quantum model,  \pr{D68}, 104008 (2003)
[arXiv:gr-qc/0306059].

\item A. Corichi,   J. Cortez \& H. Quevedo. Note on canonical
quantization and unitary (in) equivalence in field theory.
\cqg{20}:L83,  2003. [gr--qc/0212023]

\item A. Corichi. On quasinormal modes,   black hole entropy, and
quantum geometry. \pr{D67}, 087502,  2003. [gr--qc/0212126]

\item O. Dreyer. New hints from general relativity.
\ijmp{D12}, 1763,  2003. [gr--qc/0401035]

\bibitem{qnm} O. Dreyer. Quasinormal modes,   the area spectrum,   and
black hole entropy. \prl{90}:081301,  2003. [gr--qc/0211076]

\item
  C.~Fleischhack,
  On the Gribov problem for generalized connections,
  Commun.\ Math.\ Phys.\  {\bf 234}, 423 (2003)
  [arXiv:math-ph/0007001].

\item L. Freidel,   E.R. Livine \& C. Rovelli. Spectra of length
and area in (2+1) Lorentzian loop quantum gravity.
\cqg{20}:1463--1478,  2003. [gr--qc/0212077]

\item L. Freidel \& E.R. Livine. Spin networks for noncompact
groups. \jmp{44}:1322--1356,  2003. [hep--th/0205268]

\item L. Freidel \& D. Louapre. Asymptotics of 6j and 10j symbols.
\cqg{20}:1267--1294,  2003. [hep--th/0209134]

\item L. Freidel \& D. Louapre. Nonperturbative summation over
3--D discrete topologies. \pr{D68}:104004,  2003.
[hep--th/0211026]

\item R. Gambini \& J. Pullin. Canonical quantization of general
relativity in discrete space--times. \prl{90}:021301, 2003.
[gr--qc/0206055]

\item R. Gambini \& J. Pullin. Discrete quantum gravity:  A
Mechanism for selecting the value of fundamental constants.
\ijmp{D12}:1775--1782,  2003. [gr--qc/0306095]

\item R. Gambini \& J. Pullin. Discrete quantum gravity:
Applications to cosmology. \cqg{20}:3341,  2003. [gr--qc/0212033]

\item
  L.~J.~Garay and G.~A.~Mena Marugan,
  Immirzi ambiguity, boosts and conformal frames for black holes,
  Class.\ Quant.\ Grav.\  {\bf 20}, L115 (2003).
  [gr-qc/0304055].

\item
  F.~Hinterleitner and S.~Major,
  Isotropic loop quantum cosmology with matter. II: The Lorentzian
  constraint,
  \pr{D68}, 124023, 2003.
  [arXiv:gr-qc/0309035].

\item K. Krasnov.  Black hole thermodynamics and Riemann surfaces.
\cqg{20}:2235--2250,  2003. [gr--qc/0302073]

\item G. Lambiase.  Cerenkov's effect and neutrino oscillations in
loop quantum gravity. \mpl{A18}:23--30,  2003. [gr--qc/0301058]

\item G. Lambiase.  MSW effect in loop quantum gravity and
constraints on parameters from neutrino antineutrino transitions.
\mpl{A18}:1397--1401,  2003.

\item G. Lambiase.  Spin flavor conversion of neutrinos in loop
quantum gravity. \cqg{20}:4213--4220,  2003. [gr--qc/0302053]

\item G. Lambiase \& P. Singh.  Matter antimatter asymmetry
generated by loop quantum gravity. \mpl{B565}:27--32, 2003.
[gr--qc/0304051]

\item
  Y.~Ling and H.~b.~Zhang,
  Quasinormal modes prefer supersymmetry?,
  \pr{D68}, 101501 (2003)
  [arXiv:gr-qc/0309018].

\item E.R. Livine \& D. Oriti. Implementing causality in the spin
foam quantum geometry. \np{B663}:231--279,  2003. [gr--qc/0210064]

\item E.R. Livine,   A. Perez \& C. Rovelli. 2D
manifold--independent spinfoam theory. \cqg{20}:4425--4445, 2003.

\item A. Mikovic. Quantum field theory of open spin networks and
new spin foam models. \ijmp{A18S2}:83--96,  2003. [gr--qc/0202026]

\item A. Mikovic. Quantum gravity vacuum and invariants of
embedded spin networks. \cqg{20}:3483--3492,  2003.
[gr--qc/0301047]

\item A. Mikovic. Spin foam models of Yang--Mills theory coupled
to gravity. \cqg{20}:239--246,  2003. [gr--qc/0210051]

\item M. Montesinos. Noether currents for BF gravity.
\cqg{20}:3569--3575,  2003.

\item K. Noui.  \& P. Roche.  Cosmological deformation of
Lorentzian spin foam models.  \cqg{20}:3175--3214,  2003.
[gr--qc/0211109]

\item R. Oeckl. A 'General boundary' formulation for quantum
mechanics and quantum gravity. \mpl{B575}:318--324,  2003.
[hep--th/0306025]

\item R. Oeckl. Generalized lattice gauge theory,   spin foams and
state sum invariants. \jgp{46}:308--354,  2003. [hep--th/0110259]

\item R. Oeckl. Renormalization of discrete models without
background. \np{B657}:107--138,  2003. [gr--qc/0212047]

\item R. Oeckl. Schroedinger's cat and the clock:  Lessons for
quantum gravity. \cqg{20}:5371--5380,  2003. [gr--qc/0306007]

\item A. Okolow \& J. Lewandowski.  Diffeomorphism covariant
representations of the holonomy flux *algebra.
\cqg{20}:3543--3568,  2003. [gr--qc/0302059]

\item A. Perez.  Spin foam models for quantum gravity.
\cqg{20}:R43,  2003. [gr--qc/0301113]

\bibitem{pullin2} J. Pullin,
  Canonical quantization of general relativity: The last 18 years in a
  nutshell {\em AIP Conf.\ Proc.}  {\bf 668} 141 (2003)
  ({\it Preprint} gr-qc/0209008)

\item C. Rovelli. Loop quantum gravity, \pw{16N11}:37--41, 2003.

\bibitem{Rovelli:2003wd}
  C.~Rovelli, A dialog on quantum gravity,
  {\em Int.\ J.\ Mod.\ Phys.\ D} {\bf 12}, 1509 (2003).
  ({\it Preprint} hep-th/0310077)

\item C. Rovelli \& S. Speziale.  Reconcile Planck scale
discreteness and the Lorentz--Fitzgerald contraction.
\pr{D67}:064019,  2003. [gr--qc/0205108]

\item L.~Shao, H.~Noda, D.~Shao and C.~G.~Shao, The curvature
excitation of quantum Wilson loop in (R + R**2) gravity,
\grg{35}:527 (2003).

\item
  L.~Smolin and A.~Starodubtsev,
  General relativity with a topological phase: An action principle,
  arXiv:hep-th/0311163.

\item J. Swain.   The Pauli exclusion principle and SU(2) versus
SO(3) in loop quantum gravity. \ijmp{D12}:1729--1736, 2003.
[gr--qc/0305073]

\bibitem{Thiemann:2002nj} T. Thiemann. Lectures on loop quantum gravity.
\lnp{631}:41--135,  2003. [gr--qc/0210094]

\item L.~F.~Urrutia, Loop quantum gravity induced modifications to
particle dynamics, \aipcp{670}:289 (2003) [arXiv:hep-ph/0303189].

\newpage

\section*{Preprints older than 12 months}

\item P. Aichelburg and A. Ashtekar. Mathematical Problems of
Quantum Gravity. Abstracts of seminars given at the {\em Quantum
Gravity Workshop at the Erwin Schrodinger Institute, Viena}.
arXiv: gr-qc/9701042.

\item S.~Alexander, L.~Crane and M.~D.~Sheppeard,
  The geometrization of matter proposal in the Barrett-Crane model and
  resolution of cosmological problems,
  arXiv:gr-qc/0306079.

\item G.~Amelino--Camelia, On the fate of Lorentz symmetry in loop
quantum gravity and  noncommutative spacetimes,
arXiv:gr-qc/0205125.

\item M.~Arnsdorf, Approximating connections in loop quantum
gravity, arXiv:gr-qc/9910084.

\item M.~Arnsdorf, Loop quantum gravity and asymptotically flat
spaces, arXiv:gr-qc/0008038.

\item J. M. Aroca, H. Fort and R. Gambini Path Integral for
Lattice Staggered Fermions in the Loop Representation.
hep-lat/9607050.

\item
H.~Baumgartel, On a theorem of Ashtekar and Lewandowski in the
mathematical  framework of
  canonical quantization in quantum gravity,''
{\it Prepared for Conference on New Insights in Quantum Mechanics,
Goslar, Germany, 31 Aug - 3 Sep 1998}

\item C.~Beetle and A.~Corichi, Bibliography of publications
related to classical and quantum gravity  in terms of connection
and loop variables, arXiv:gr-qc/9703044.

\item I. Bengtsson. Ashtekar's variables and the cosmological
constant. Goteborg preprint, 1991.

\item I. Bengtsson. Curvature tensors in an exact solution of
Capovilla's equations. Goteborg-91-5 (February 1991).

\item I. Bengtsson. Form connections. gr-qc/9305004

\item I. Bengtsson. Form Geometry and the `t Hooft-Plebanski
action. gr-qc/950210

\item M.~Bojowald, Angular momentum in loop quantum gravity,
arXiv:gr-qc/0008054.

\item M.~Bojowald and A. Perez. SPin foam quantization and
anomalies, arXiv:gr-qc/0303026.

\item Ola Bostr\"om. Some new results about the cosmological
constants. G\"oteborg preprint ITP91-34

\item O. Bostr\"om, M. Miller and L. Smolin. A new discretization
of classical and quantum general relativity. G\"oteborg ITP 94-5,
SU-GP-93-4-1, CGPG-94/3-3, gr-qc/9304005.

\item R. Brooks. Diff($\Sigma$) and metrics from Hamiltonian-TQFT.
MIT preprint CTPH2175

\item Greorgy Burnett, Joseph D. Romano, and Ranjeet S. Tate.
Polynomial coupling of matter to gravity using {A}shtekar
variables. Syracuse preprint.

\item Steven Carlip. 2+1 dimensional quantum gravity and the Braid
group. Talk given at the Workshop on Physics, Braids and Links,
Banff Summer School in Theoretical Physics, August 1989.

\item L. Crane. Categorical physics. Preprint ???.

\item L.~Crane and D.~Yetter, A more sensitive Lorentzian state sum,
  arXiv:gr-qc/0301017.

\item A.~Dasgupta,  Counting the apparent horizon, arXiv:hep-th/0310069.

\item A.~Doring and H.~F.~de Groote, The kinematical frame of loop
quantum gravity. I, arXiv:gr-qc/0112072.

\item R. Floreanini, R. Percacci and E. Spallucci. Why is the
metric non-degenerate? SISSA 132/90/EP preprint (October 1990).
Proceedings of the 6-th Marcel Grossmann Meeting, Kyoto 1991, p. 323-325

\item R. Gambini, and L. Leal. Loop space coordinates, linear
representations of the diffeomorphism group and knot invariants.

\item J.~M.~Garcia--Islas, Spin foam models of n-dimensional
quantum gravity, and non-archimedean  and non-commutative
formulations, arXiv:gr-qc/0401058.

\item J. Griego. On the Extended Loop Calculus. gr-qc/9512011.

\item B.~Grossmann. General relativity and a non-topological phase
of topological {Y}ang-{M}ills theory. Inst. for Advanced Studies,
Princeton, 1990 preprint.

\item Y. Hashimoto, Y. Yasui, S. Miyagi, T. Otsuka. Applications
of the Ashtekar gravity to four dimensional hyper-K\"ahler
geometry and Yang-Mills instantons. hep-th/9610069.

\item G.~Helesfai and G.~Bene, A numerical study of spectral
properties of the area operator in loop quantum gravity,
arXiv:gr-qc/0306124.

\item W. Kalau. Ashtekar formalism with real variables. U. Of
Wuppertal NIKHEF-H/91-03 Amsterdam preprint (December 1990).

\item K. Kamimura and T. Fukuyama. Massive analogue of
Ashtekar-CDJ action. gr-qc/9208010

\item V. Khatsymovsky. On polynomial variables in general
relativity. BINP 93-41, gr-qc/9310005.

\item S. Koshti and N. Dadhich. Gravitational instantons with
matter sources using Ashtekar variables. Inter Univ. Centre for
Astron. and Astrophysics, Pune, India. June 1990 preprint.

\item Sucheta Koshti and Naresh Dadhich. The General Self-dual
solution of the Einstein Equations.  IUCAA 94/29, gr-qc/9409046.

\item R.~Loll. An example of loop quantization. CGPG-94/7-2.

\item R.~Loll. Chromodynamics and gravity as theories on loop
space. CGPG-93/9-1, hep-th/9309056.

\item R.~Loll. Latticing quantum gravity. Contributed to 2nd
meeting on constrained dynamics and quantum gravity, Santa
Margherita, Italy, Sep 1996. gr-qc/9701007

\item R.~Loll. Quantizing canonical gravity in the real domain.
gr-qc/9701031.

\item A.M.R. Magnon. Self duality and CP violation in gravity.
Univ. Blaise Pascal (France) preprint (1990).

\item S. Major and L. Smolin. Mixmaster Quantum Cosmology in Terms
of Physical Dynamics. gr-qc/9607020.

\item J. Maluf. Fermi coordinates and reference frames in the ECSK
theory. SU-GP-92/1-2

\item
  S.~K.~Maran,
  Spin foam models of gravity and BF theory as
evolution of spin networks,
  arXiv:gr-qc/0310123.

\item
  S.~K.~Maran,
  Ashtekar formulation with temporal foliations,
  arXiv:gr-qc/0312111.

\item L.~J. Mason and J{\"o}rg Frauendiener. The {S}parling
3-form, {A}shtekar variables and quasi-local mass, in {\em Twistors in mathematics and physics}, eds R.Baston \& T.Bailey, Cambridge U. Press, p.189-217 (1989)

\item P.~Mora, A note in cosmology and loop quantum gravity,
arXiv:gr-qc/0008048.

\item O. Moritsch, M. Schweda, T. Sommer, L. Tataru and H.
Zerrouk. BRST cohomology of Yang--Mills gauge fields in the
presence of gravity in Ashtekar variables. TUW 94-17,
hep-th/9409081.

\item N. O'Murchadha and M. Vandyck. Gravitational degrees of
freedom in Ashtekar's formulation of General Relativity. Univ. of
Cork preprint - 1990

\item P. L. Paul. Topological Symmetries of twisted N=2 chiral
supergravity in Ashtekar formalism. hep-th/9504144.

\item M.~Rainer, Is loop quantum gravity a QFT?,
arXiv:gr-qc/9912011.

\item J. Rasmussen and M. Weis. Induced Topology on the hoop
group. NBI-HE-94-46, hep-th/9410194.

\item M.~Reisenberger, C.~Rovelli: ``Spinfoams as Feynman
        diagrams", gr-qc/ 0002083.

\item C. Rovelli and L. Smolin. Finiteness of diffeomorphism
invariant operators in nonperturbative quantum gravity. Syracuse
University preprint SU-GP-91/8-1, August 1991.

\item Carlo Rovelli and Lee Smolin. Loop representation for
lattice gauge theory. 1990 Pittsburgh and Syracuse preprint.

\bibitem{Rovelli:1998gg}
 C.~Rovelli and P.~Upadhya, Loop quantum gravity and quanta of
space: A primer, arXiv:gr-qc/9806079.

\item M.P. Ryan, Jr. Cosmological ``ground state'' wave functions
in gravity and electromagnetism.  in {\em Proceedings of
VIIth Marcel Grossman Meeting on General Relativity, 1994}
gr-qc/9312024

\item H.~Sahlmann, Some comments on the representation theory of
the algebra underlying  loop quantum gravity, arXiv:gr-qc/0207111.

\item H.~Sahlmann, When do measures on the space of connections
support the triad  operators of loop quantum gravity?,
arXiv:gr-qc/0207112.

\item J. Schirmer. Triad formulations of canonical gravity without
a fixed reference frame. gr-qc/9503037.

\item L. Smolin. Fermions and topology. GCPG-93/9-4,
gr-qc/9404010.

\item L. Smolin. The classical limit and the form of the
Hamiltonian constraint in nonperturbative quantum general
relativity. gr-qc/9609034.

\item Lee Smolin. The Problem of Quantum Gravity: a status report
(Address to the AAAS meeting, Washington D.C., February 1991).
Syracuse preprint SU-GP-91/2-1.

\item L. Smolin. Three dimensional strings a collective
coordinates of four-dimensional nonperturbative quantum gravity.
gr-qc/9609031.

\bibitem{Smolin:2003rk}
  L. Smolin, How far are we from the quantum theory of gravity?
  {\tt Preprint} hep-th/0303185.

\item L. Smolin and M. Varadarajan. Degenerate solutions and the
instability of the perturbative vacuum in nonperturbative
formulations of quantum gravity. Syracuse University preprint
SU-GP-91/8-3, August 1991.

\item A.~Starodubtsev, String theory in a vertex operator
representation: A simple model for testing loop quantum gravity,
arXiv:gr-qc/0201089.

\bibitem{Thiemann:2001yy} T. Thiemann, Introduction to modern canonical
quantum general relativity, {\it Preprint} gr-qc/0110034

\item T.~Thiemann, The LQG string: Loop quantum gravity
quantization of string theory. I:  Flat target space,
arXiv:hep-th/0401172.

\item T.~Thiemann, The Phoenix project: Master constraint
programme for loop quantum  gravity, arXiv:gr-qc/0305080.

\item M. Tsuda. Consistency of matter field equations in Ashtekar
formulation. gr-qc/9602022.

\item M. Tsuda. General considerations of matter coupling with
selfdual connection. gr-qc/9505019.

\item M. Tsuda, T. Shirafuji and H.-J. K Xie. Ashtekar Variables
and Matter Coupling. STUPP-95-138, gr-qc/9501021.

\item
  M.~Tsuda, T.~Shirafuji and H.~J.~Xie,
  Ashtekar variables and matter coupling,
  arXiv:gr-qc/9501021.

\item M. Tsuda and T. Shirafuji. Consistency of matter field
equations in Ashtekar formulation. gr-qc/9602022.

\item T.~Tsushima, The expectation value of the metric operator
with respect to Gaussian weave state in loop quantum gravity,
arXiv:gr-qc/0212117.

\item L.~F.~Urrutia, Flat space modified particle dynamics induced
by loop quantum gravity, arXiv:hep-ph/0402271.

\item L.~F.~Urrutia, Loop quantum gravity induced corrections to
fermion dynamics in flat space, arXiv:hep-ph/0402004.

\item L.F. Urrutia. Towards a Loop Representation of Connection
Theories Defined Over a Super-Lie Algebra. hep-th/9609010.

\item R.~E.~Zimmermann, Loops and knots as topoi of substance:
Spinoza revisited, arXiv:gr-qc/0004077.

\item R.~E.~Zimmermann, Recent conceptual consequences of loop
quantum gravity. Part I:  Foundational aspects,
arXiv:physics/0107061.

\item R.~E.~Zimmermann, Recent conceptual consequences of loop
quantum gravity. II: Holistic aspects, arXiv:physics/0107081.

\item R.~E.~Zimmermann, Recent conceptual consequences of loop
quantum gravity.  III: A postscript on time,
arXiv:physics/0108026.

\newpage

\section*{2004}

\item S.~Alexander, J.~Malecki and L.~Smolin, Quantum gravity and inflation,
 \pr{D70}, 044025 (2004),  [arXiv:hep-th/0309045].

\item S.Y. Aleksandrov.  Lorentz--covariant loop quantum gravity.
\tmp{139}:751--765,  2004,   \tmf{139}:363--380,  2004.

\item J.~Alfaro, M.~Reyes, H.~A.~Morales-Tecotl and L.~F.~Urrutia,
On alternative approaches to Lorentz violation in loop quantum
gravity inspired models, \pr{D70}:084002 (2004)
[arXiv:gr-qc/0404113].

\item G.~Amelino-Camelia, M.~Arzano and A.~Procaccini, Severe
constraints on loop-quantum-gravity energy-momentum dispersion
relation from black-hole area-entropy law, \pr{D70}:107501 (2004)

\item
  G.~Amelino-Camelia, L.~Smolin and A.~Starodubtsev,
  Quantum symmetry, the cosmological constant and Planck scale
  phenomenology,
  \cqg{21}, 3095 (2004),
  [arXiv:hep-th/0306134].

\bibitem{AL:review}
 A. Ashtekar \& J. Lewandowski.  Background independent quantum
gravity:  A Status report. \cqg{21}:R53,  2004. [gr--qc/0404018]

\item M.~Bojowald. Spherically Symmetric Quantum Geometry: States and Basic
Operators, \cqg{21}, 3733--3753 (2004).

\item M. Bojowald. Quantum Gravity and the Big Bang, in
 {\em Where Cosmology and Fundamental Physics Meet}, Le Brun,
 V.\ and Basa, S.\ and Mazure, A. Eds.,  IUFM, Marseille, Frontier Group,
 2004, Pp. 54--58. ArXiv:astro-ph/0309478

\item M.~Bojowald. Loop Quantum Cosmology: Recent Progress, in
Proceedings of the International Conference on Gravitation and
Cosmology (ICGC 2004), Cochin, India. {\em Pramana} {\bf 63},
765--776 (2004). [gr-qc/0402053].

\item M. Bojowald,   G. Date \& G.M. Hossain. The Bianchi IX model
in loop quantum cosmology. \cqg{21}:3541--3570,  2004.
[gr--qc/0404039]

\item M. Bojowald \& G. Date. Consistency conditions for
fundamentally discrete theories. \cqg{21}:121--143,  2004.
[gr--qc/0307083]

\item M. Bojowald \& G. Date. Quantum suppression of the generic
chaotic behavior close to cosmological singularities.
\prl{92}:071302,  2004. [gr--qc/0311003]

\item M. Bojowald,   G. Date \& K. Vandersloot. Homogeneous loop
quantum cosmology:  The Role of the spin connection.
\cqg{21}:1253--1278,  2004. [gr--qc/0311004]

\item M. Bojowald,   R. Maartens \& P. Singh.  Loop quantum
gravity and the cyclic universe. \pr{D70}:083517,  2004.
[hep--th/0407115]

\bibitem{lqc-rev} M. Bojowald \& H.A. Morales--Tecotl. Cosmological
applications of loop quantum gravity. \lnp{646}:421--462,  2004.
[gr--qc/0306008]

\item M. Bojowald,   J.E. Lidsey,   D.J. Mulryne,   P. Singh \& R.
Tavakol.  Inflationary cosmology and quantization ambiguities in
semiclassical loop quantum gravity. \pr{D70}:043530,  2004.
[gr--qc/0403106]

\item
  M.~Bojowald, P.~Singh and A.~Skirzewski,
  Time dependence in quantum gravity,
  \pr{D70}, 124022 (2004)
  [arXiv:gr-qc/0408094].

\item M.~Bojowald, and A.~Skirzewski. The Volume Operator in
Spherically Symmetric Quantum Geometry, \cqg{21}, 4881--4900
(2004). [arXiv:gr-qc/0407018].

\item
  D.~Cartin, G.~Khanna and M.~Bojowald,
  Generating function techniques for loop quantum cosmology,
  \cqg{21}, 4495 (2004)
  [arXiv:gr-qc/0405126].

\item F. Conrady,   L. Doplicher,   R. Oeckl,   C. Rovelli \& M.
Testa. Minkowski vacuum in background independent quantum gravity.
\pr{D69}:064019,  2004. [gr--qc/0307118]

\item A.~Corichi, Comments on area spectra in loop quantum
gravity, \rmf{50}:549 (2004) [arXiv:gr-qc/0402064].

\item A. Corichi \& J. Cortez. Note on selfduality and the Kodama
state. \pr{D69}:047702,  2004. [hep--th/0311089]

\bibitem{lew-bh} M.~Domagala and J.~Lewandowski, Black hole entropy from
quantum geometry, \cqg{21}, 5233 (2004) [arXiv:gr-qc/0407051].

\item W.~Fairbairn and C.~Rovelli, Separable Hilbert space in loop
quantum gravity, \jmp{45}:2802 (2004) [arXiv:gr-qc/0403047].

\item L. Freidel,   J. Kowalski--Glikman \& L. Smolin. 2+1 gravity
and doubly special relativity.  \pr{D69}:044001,  2004.
[hep--th/0307085]

\item L.~Freidel and D.~Louapre, Ponzano-Regge model revisited. I:
Gauge fixing, observables and  interacting spinning particles,
\cqg{21}, 5685 (2004) [arXiv:hep-th/0401076].

\item L.~Freidel and L.~Smolin, The linearization of the Kodama
state, \cqg{21}, 3831 (2004) [arXiv:hep-th/0310224].

\item R. Gambini,   R.A. Porto \& J. Pullin. Decoherence from
discrete quantum gravity. \cqg{21}:L51--L57,  2004.
[gr--qc/0305098]

\item R.~Gambini, R.~Porto and J.~Pullin,
  A relational solution to the problem of time in quantum mechanics and
  quantum gravity induces a fundamental mechanism for quantum  decoherence,
  {\em New J.\ Phys.}\  {\bf 6}, 45 (2004)
  [arXiv:gr-qc/0402118].

\item R.~Gambini and J.~Pullin,
  Canonical quantum gravity and consistent discretizations,
  {\em Pramana} {\bf 63}, 755 (2004)
  [arXiv:gr-qc/0402062].

\item J.M. Garcia--Islas.  (2+1)--dimensional quantum gravity,
spin networks and asymptotics. \cqg{21}:445--464, 2004.
[gr--qc/0307054]

\item J.~M.~Garcia-Islas, Observables in 3--dimensional quantum
gravity and topological invariants,  \cqg{ 21}:3933 (2004)
[arXiv:gr-qc/0401093].

\item F. Girelli \& E.R. Livine. Quantizing speeds with the
cosmological constant. \pr{D69}:104024,  2004. [gr--qc/0311032]

\item G.~Gour and V.~Suneeta, Comparison of area spectra in loop
quantum gravity, \cqg{21}:3405 (2004) [arXiv:gr-qc/0401110].

\item
  D.~Green and W.~G.~Unruh,
  Difficulties with closed isotropic loop quantum cosmology,
  \pr{D70}, 103502 (2004)
  [arXiv:gr-qc/0408074].

\item V. Husain \& O. Winkler. On singularity resolution in
quantum gravity. \pr{D69}:084016,  2004. [gr--qc/0312094]

\item V. Husain \& O. Winkler. Discrete Hamiltonian evolution and
quantum gravity. \cqg{21}:941--950,  2004. [gr--qc/0308011]

\item
  I.~B.~Khriplovich,
  Quantized black holes, correspondence principle, and holographic bound,
  arXiv:gr-qc/0409031.

\item
  I.~B.~Khriplovich,
  Spectrum of quantized black hole, correspondence principle, and
  holographic bound,''
  {\em Sov.\ Phys.\ } {\bf JETP}, 99:460 (2004)
  [Zh.\ Eksp.\ Teor.\ Fiz.\  {\bf 126}, 527 (2004)]
  [arXiv:gr-qc/0404083].

\item C.~N.~Kozameh and M.~F.~Parisi, Lorentz invariance and the
semiclassical approximation of loop quantum gravity, \cqg{21}:2617
(2004) [arXiv:gr-qc/0310014].

\item T. Levy.  Wilson loops in the light of spin networks.
\jgp{52}:382--397,  2004. [math--ph/0306059]

\item
  J.~E.~Lidsey,
  Early universe dynamics in semi-classical loop quantum cosmology,
  JCAP {\bf 0412}, 007 (2004)
  [arXiv:gr-qc/0411124].

\item E.R. Livine \& D. Oriti. About Lorentz invariance in a
discrete quantum setting. \jhep{0406}:050,  2004. [gr--qc/0405085]

\item E.R. Livine \& R. Oeckl. Three--dimensional quantum
supergravity and supersymmetric spin foam models.
\atmp{7}:951--1001,  2004. [hep--th/0307251]

\item S.~K.~Maran.
  Relating spin foams and canonical quantum gravity:
  (n-1)+1 formulation  of nD spin foams,''
  \pr{D70}, 124004 (2004)
  [arXiv:gr-qc/0412011].

\bibitem{meissner} K.~A.~Meissner, Black hole entropy in loop quantum
gravity, \cqg{21}:5245 (2004) [arXiv:gr-qc/0407052].

\item A. Mikovic. Flat space--time vacuum in loop quantum gravity.
\cqg{21}:3909--3922,  2004. [gr--qc/0404021]

\item  L. Modesto.  Disappearance of black hole singularity in
quantum gravity. \pr{D70}:124009,  2004. [gr--qc/0407097]

\item R. Oeckl, The general boundary approach to quantum gravity,
in: {\em Proceedings of the First International Conference on Physics},
Amirkabir
University, Tehran, 2004, pp. 257-265; gr-qc/0312081.

\item
  A.~P.~Polychronakos,
  Area spectrum and quasinormal modes of black holes,
  \pr{D69}, 044010 (2004)
  [arXiv:hep-th/0304135].

\item D. Shao,   L. Shao,   M. Saito,   C.G. Shao \& H. Noda.
Cosmological term and extended loop representation of quantum
gravity. \ijmp{A19}:4659--4670,  2004.

\item
  P.~Singh and A.~Toporensky,
``Big crunch avoidance in k = 1 loop quantum cosmology,
  \pr{D69}, 104008 (2004)
  [arXiv:gr-qc/0312110].

\item L. Smolin. Atoms of space and time. \sa{290N1}:66--75, 2004.

\item
  L.~Smolin,
  Quantum theories of gravity: Results and prospects,
In *Barrow, J.D. (ed.) et al.: Science and ultimate reality*
492-527, 2004.

\item  S. Tsujikawa,   P. Singh \& R. Maartens.  Loop quantum
gravity effects on inflation and the CMB. \cqg{21}:5767--5775,
2004. [astro--ph/0311015]

\item J.~M.~Velhinho. Comments on the kinematical structure of
loop quantum cosmology  \cqg{21}:L109 (2004)
[arXiv:gr-qc/0406008].

\item J.~M.~Velhinho. On the structure of the space of generalized
connections,  \ijgmmp{1}:311 (2004) [arXiv:math-ph/0402060].

\item
  G.~V.~Vereshchagin,
  Qualitative approach to semi-classical loop quantum cosmology,
  {\em JCAP} {\bf 0407}, 013 (2004)
  [arXiv:gr-qc/0406108].

\item J.A. Zapata.  Loop quantization from a lattice gauge theory
perspective. \cqg{21}:L115--L122,  2004. [gr--qc/0401109]

\newpage
\section*{2005}

\item A.~Alekseev, A.~P.~Polychronakos and M.~Smedback, Remarks on
the black hole entropy and Hawking spectrum in loop quantum
gravity, \pr{D71}:067501 (2005) [arXiv:hep-th/0405036].

\item J.~Alfaro,
  Quantum gravity induced Lorentz invariance violation in the standard model:
  Hadrons,''
  \pr{72}, 024027 (2005)
  [arXiv:hep-th/0505228].

\item J.~Alfaro,
  Quantum gravity and Lorentz invariance deformation in the standard
  model,
  \prl{94}, 221302 (2005)
  [arXiv:hep-th/0412295].

\item A.~Ashtekar and M.~Bojowald, Black hole evaporation: A
paradigm, \cqg{22}, 3349 (2005). arXiv:gr-qc/0504029.

\item A.~Ashtekar, L.~Bombelli and A.~Corichi,
 Semiclassical states for constrained systems,
  \pr{D72}, 025008 (2005)
  [arXiv:gr-qc/0504052].

\item A.~Ashtekar, J.~Engle and C.~Van Den Broeck, Quantum
horizons and black hole entropy: Inclusion of distortion and
rotation, \cqg{22}, L27 (2005) [arXiv:gr-qc/0412003].

\item J.~W.~Barrett. Feynman loops and three-dimensional quantum gravity,
  \mpl{A20}, 1271 (2005)
  [arXiv:gr-qc/0412107].

\item M.~Bojowald, Original Questions, \n{436}, 920--921 (2005).

\item
  M.~Bojowald,
  Nonsingular black holes and degrees of freedom in quantum
  gravity,\prl{95}, 061301 (2005).
  arXiv:gr-qc/0506128.

\item M.~Bojowald, R.~Goswami, R.~Maartens and P.~Singh
  A black hole mass threshold from non-singular quantum gravitational
  collapse,
 \prl{95}, 091302 (2005)
  [arXiv:gr-qc/0503041].

\item
  M.~Bojowald and A.~Rej,
  Asymptotic properties of difference equations for isotropic loop quantum
  cosmology, \cqg{22}, 3399 (2005).
  arXiv:gr-qc/0504100.

\item M.~Bojowald, H.~A.~Morales-Tecotl and H.~Sahlmann, On loop
quantum gravity phenomenology and the issue of Lorentz invariance,
\pr{D71}:084012 (2005) [arXiv:gr-qc/0411101].

\item M.~Bojowald and R.~Swiderski, Spherically Symmetric Quantum
Horizons, \pr{D71}, 081501(R) (2005). [arXiv:gr-qc/0410147]

\item
  L.~Bombelli, A.~Corichi and O.~Winkler,
  Semiclassical quantum gravity: Statistics of combinatorial Riemannian
  geometries,'' \Ap{14}, 499-519 (2005).
  arXiv:gr-qc/0409006.

\item D.~Colosi, L.~Doplicher, W.~Fairbairn, L.~Modesto, K.~Noui
and C.~Rovelli, Background independence in a nutshell: The
dynamics of a tetrahedron, \cqg{22}, 2971 (2005)
[arXiv:gr-qc/0408079].

\item F.~Conrady, Free vacuum for loop quantum gravity, \cqg{22},
3261 (2005). arXiv:gr-qc/0409036.

\item L.~Crane,
  Relational topology as the foundation for quantum gravity,
  \mpl{A20}, 1261 (2005).

\item
  G.~Date,
  Absence of the Kasner singularity in the effective dynamics from loop
  quantum cosmology,
  \pr{D71}, 127502 (2005)
  [arXiv:gr-qc/0505002].

\item C.~Di Bartolo, R.~Gambini and J.~Pullin,
  Consistent and mimetic discretizations in general relativity,
  \jmp{46}, 032501 (2005)
  [arXiv:gr-qc/0404052].

\item R.~Gambini and J.~Pullin,
  Consistent discretization and loop quantum geometry,
  {\em Phys.\ Rev.\ Lett.}\  {\bf 94}, 101302 (2005)
  [arXiv:gr-qc/0409057].

\item
  F.~Girelli and E.~R.~Livine,
  Reconstructing quantum geometry from quantum information: Spin
  networks  as
  harmonic oscillators, \cqg{22}, 3295 (2005).
  arXiv:gr-qc/0501075.

\item F.~Girelli and E.~R.~Livine,
  Physics of Deformed Special Relativity,''
  {\em Braz.\ J.\ Phys.}  {\bf 35}, 432 (2005)
  [arXiv:gr-qc/0412079].

\item
  F.~Girelli, E.~R.~Livine and D.~Oriti,
  Deformed special relativity as an effective flat limit of quantum
  gravity,''
  \np{B708}, 411 (2005)
  [arXiv:gr-qc/0406100].

\item
  G.~M.~Hossain,
  On energy conditions and stability in effective loop quantum cosmology,
  \cqg{22}, 2653 (2005)
  [arXiv:gr-qc/0503065].

\item I.~B.~Khriplovich,
  Radiation of quantized black hole,
  {\em J.\ Exp.\ Theor.\ Phys.}  {\bf 100}, 1075 (2005)
  [{\em Zh.\ Eksp.\ Teor.\ Fiz.}  {\bf 100}, 1223 (2005)]
  [arXiv:gr-qc/0412121].

\item I.~B.~Khriplovich,
  Quasinormal modes, quantized black holes, and correspondence principle,
  \ijmp{D14}, 181 (2005)
  [arXiv:gr-qc/0407111].

\item
  E.~R.~Livine,
  Some Remarks on the Semi-Classical Limit of Quantum Gravity,
  {\em Braz.\ J.\ Phys.}\  {\bf 35}, 442 (2005)
  [arXiv:gr-qc/0501076].

\item
 D.~J.~Mulryne, N.~J.~Nunes, R.~Tavakol and J.~E.~Lidsey,
  Inflationary cosmology and oscillating universes in loop quantum
  cosmology,
  \ijmp{A20}, 2347 (2005)
  [arXiv:gr-qc/0411125].

\item D.~J.~Mulryne, R.~Tavakol, J.~E.~Lidsey and G.~F.~R.~Ellis.
  An emergent universe from a loop,
  \pr{D71}, 123512 (2005)
  [arXiv:astro-ph/0502589].

\item
  J.~A.~Nieto,
  Towards an Ashtekar formalism in eight dimensions,
  \cqg{22}, 947 (2005)
  [arXiv:hep-th/0410260].

\bibitem{Nicolai:2005mc}
 H.~Nicolai, K.~Peeters and M.~Zamaklar, Loop quantum gravity: An
outside view, \cqg{22}, R193-R247 (2005).  [arXiv:hep-th/0501114].

\item
J.~A.~Nieto, Equivalence between various versions of the self-dual
action of the Ashtekar formalism, \mpl{A20}, 2157-2163, 2005
[arXiv: hep-th/0411124]

\item K.~Noui and A.~Perez, Three dimensional loop quantum
gravity: Physical scalar product and  spin foam models,
\cqg{22}:1739 (2005) [arXiv:gr-qc/0402110].

\item K.~Noui, A.~Perez and K.~Vandersloot, On the physical
Hilbert space of loop quantum cosmology, \pr{D71}, 044025 (2005)
[arXiv:gr-qc/0411039].

\item R.~Oeckl,
  States on timelike hypersurfaces in quantum field theory,
  \pl{B622}, 172 (2005)
  [arXiv:hep-th/0505267].

\item R. Oeckl, Renormalization for spin foam models of quantum gravity,
in: {\em Proceedings of the Tenth Marcel Grossmann Meeting on General
Relativity}
(Rio de Janeiro, 2003), World Scientific, Singapore, 2005; gr-qc/0401087.

\item A.~Okolow and J.~Lewandowski, Automorphism covariant
representations of the holonomy-flux *-algebra, \cqg{22}, 657
(2005) [arXiv:gr-qc/0405119].

\item
  D.~Oriti,
  The Feynman propagator for quantum gravity: Spin foams, proper time,
  orientation, causality and timeless-ordering,''
  {\em Braz.\ J.\ Phys.}\  {\bf 35}, 481 (2005)
  [arXiv:gr-qc/0412035].

 \item
  D.~Oriti,
  The Feynman propagator for spin foam quantum gravity,
  \prl{94}, 111301 (2005)
  [arXiv:gr-qc/0410134].

\item D.~Oriti, C.~Rovelli and S.~Speziale, Spinfoam 2d quantum
gravity and discrete bundles, \cqg{22}, 85 (2005)
[arXiv:gr-qc/0406063].

\item M.~Saito, H.~Noda and T.~Tashiro, Structure of extended loop
wave function in quantum gravity and  operator formalism,
\ijmp{A20}:907 (2005) [arXiv:gr-qc/0405106].

\item J.~Swain, The Pauli exclusion principle, spin, and
statistics in loop quantum  gravity: SU(2) versus SO(3),
Proceedings of the Tenth Marcel Grossmann Meeting on General Relativity,
edited by M. Novello, S. Perez-Bergliaffa and R. Ruffini,
World Scientific, Singapore, 2005.
arXiv:gr-qc/0402091.

\item
  K.~Vandersloot,
  On the Hamiltonian constraint of loop quantum cosmology,
  \pr{D71}, 103506 (2005)
  [arXiv:gr-qc/0502082].

\item M.~Varadarajan, The graviton vacuum as a distributional
state in kinematic loop quantum gravity, \cqg{22}:1207 (2005)
[arXiv:gr-qc/0410120].

\item
  J.~M.~Velhinho,
  Functorial aspects of the space of generalized connections,''
  \mpl{A20}, 1299 (2005)
  [arXiv:math-ph/0411073].

\item J.~M.~Velhinho, Denseness of Ashtekar-Lewandowski states and
a generalized cut-off in loop quantum gravity, \cqg{22}, 3061
(2005). arXiv:gr-qc/0502038.

\newpage

\section*{Recent preprints}

\item S.~Alexander,
  A quantum gravitational relaxation of the cosmological constant,
  arXiv:hep-th/0503146.

\item S.~Alexander, K.~Schleich and D.~M.~Witt,
  Fermionic sectors for the Kodama state.
  arXiv:gr-qc/0503062.

\item S.~Alexandrov, On the counting of black hole states in loop
quantum gravity, arXiv:gr-qc/0408033.

\item J.~Alfaro and G.~A.~Palma, Loop quantum gravity effects on
the high energy cosmic ray spectrum, arXiv:hep-th/0501116.

\item
  M.~H.~Ansari and L.~Smolin,
  Self-organized criticality in quantum gravity,
  arXiv:hep-th/0412307.

\bibitem{AA:NJP} A.~Ashtekar, Gravity and the quantum
  arXiv:gr-qc/0410054

\item A.~Ashtekar, M.~Bojowald, Quantum geometry and the
Schwarzschild singularity,  arXiv:gr-qc/0509075

\item J.~W.~Barrett.
  Feynman diagams coupled to three-dimensional quantum gravity,
  arXiv:gr-qc/0502048.

\item J.~W.~Barrett, J.~M.~Garcia--Islas and J.~F.~Martins,
Observables in the Turaev-Viro and Crane-Yetter models,
arXiv:math.qa/0411281.

\item
  M.~Bojowald,
  The early universe in loop quantum cosmology,
  arXiv:gr-qc/0503020.

\item
  M.~Bojowald,
  Elements of loop quantum cosmology,
  arXiv:gr-qc/0505057.

\bibitem{bojo:response} M.~Bojowald, Degenerate Configurations,
Singularities and the Non-Abelian Nature of Loop Quantum
Gravity,''
  arXiv:gr-qc/0508118.

\item J.~Brunnemann and T.~Thiemann, Simplification of the
spectral analysis of the volume operator in loop quantum gravity,
arXiv:gr-qc/0405060.

\bibitem{tt-lqc} J.~Brunnemann and T.~Thiemann, On (cosmological)
singularity avoidance in loop quantum gravity,
arXiv:gr-qc/0505032.

\item J.~Brunnemann and T.~Thiemann, Unboundedness of triad - like
operators in loop quantum gravity, arXiv:gr-qc/0505033.

\item
  D.~Cartin and G.~Khanna,
  Separable wave functions in Bianchi I loop quantum cosmology,
  arXiv:gr-qc/0506024.

\item W.~Cherrington,  Finiteness and Dual Variables for Lorentzian
Spin Foam Models. arXiv:gr-qc/0508088.

\item W.~Cherrington \& J.~D.~Christensen, Positivity in Lorentzian
Barrett-Crane Models of Quantum Gravity. arXiv:gr-qc/0509080.

\item C.~H.~Chou, R.~S.~Tung and H.~L.~Yu,
  Origin of the Immirzi Parameter,
  arXiv:gr-qc/0509028.

\item  J.~D.~Christensen and L.~Crane,
  Causal sites as quantum geometry,
  arXiv:gr-qc/0410104.

\item I.~B.~Khriplovich and A.~A.~Pomeransky,
  Remark on Immirzi Parameter, Torsion, and Discrete Symmetries,
  arXiv:hep-th/0508136.

\item F.~Conrady,
  Geometric spin foams, Yang-Mills theory and background-independent
  models,
  arXiv:gr-qc/0504059.

\item
  A.~Corichi,
  Loop quantum geometry: A primer,
  arXiv:gr-qc/0507038.

\item
  A.~Corichi and D.~Sudarsky,
  Towards a new quantum gravity phenomenology,
  arXiv:gr-qc/0503078.

\item
  G.~Date,
  Pre-classical solutions of the vacuum Bianchi I loop quantum cosmology,
  arXiv:gr-qc/0505030.

\item B.~Dittrich,
  Partial and Complete Observables for Canonical General Relativity,
  arXiv:gr-qc/0507106.

\item B.~Dittrich and T.~Thiemann, Testing the master constraint
programme for loop quantum gravity. I:  General framework,
arXiv:gr-qc/0411138.

\item B.~Dittrich and T.~Thiemann, Testing the master constraint
programme for loop quantum gravity. II:  Finite dimensional
systems, arXiv:gr-qc/0411139.

\item B.~Dittrich and T.~Thiemann, Testing the master constraint
programme for loop quantum gravity. III: SL(2,R) models,
arXiv:gr-qc/0411140.

\item B.~Dittrich and T.~Thiemann, Testing the master constraint
programme for loop quantum gravity. IV:  Free field theories,
arXiv:gr-qc/0411141.

\item B.~Dittrich and T.~Thiemann, Testing the master constraint
programme for loop quantum gravity. V: Interacting field theories,
arXiv:gr-qc/0411142.

\item O.~Dreyer, F.~Markopoulou and L.~Smolin, Symmetry and
entropy of black hole horizons, arXiv:hep-th/0409056.

\item J.~Engle,
  Quantum geometry and black hole entropy: inclusion of distortion and
  rotation,
  arXiv:gr-qc/0509033.

\item M.~Frasca, Existence of a semiclassical approximation in
loop quantum gravity, arXiv:hep-th/0411245.

\item L.~Freidel, Group field theory: An overview,
  arXiv:hep-th/0505016.

\item L.~Freidel and D.~Louapre, Ponzano-Regge model revisited.
II: Equivalence with Chern-Simons, arXiv:gr-qc/0410141.

\item L.~Freidel, R.~B.~Mann and E.~M.~Popescu, Canonical analysis
of the BCEA topological matter model coupled to gravitation in
(2+1) dimensions, arXiv:gr-qc/0411117.

\item L.~Freidel and E.~R.~Livine, Ponzano-Regge model revisited.
III: Feynman diagrams and effective field theory,
arXiv:hep-th/0502106.

\bibitem{eft2} L.~Freidel, D.~Minic and T.~Takeuchi,
  Quantum gravity, torsion, parity violation and all that,
  arXiv:hep-th/0507253.

\item L.~Freidel, D.~Oriti and J.~Ryan, A group field theory for
3d quantum gravity coupled to a scalar field, arXiv:gr-qc/0506067.

\item L.~Freidel and A.~Starodubtsev, Quantum gravity in terms of
topological observables, arXiv:hep-th/0501191.

\item R.~Gambini, and J.~Pullin, Discrete space-time. e-Print
Archive: gr-qc/0505023.

\item R.~Gambini, S.~J.~Olson and J.~Pullin, Unified model of loop
quantum gravity and matter, arXiv:gr-qc/0409045.

\item
  K.~Giesel and T.~Thiemann,
``Consistency check on volume and triad operator quantisation in
loop  quantum gravity. II,''
  arXiv:gr-qc/0507037.

\item
K.~Giesel and T.~Thiemann, Consistency check on volume and triad
operator quantisation in loop quantum  gravity. I,''
  arXiv:gr-qc/0507036.

\item F.~Girelli and E.~R.~Livine,
  Physics of deformed special relativity: Relativity principle revisited,
  arXiv:gr-qc/0412004.

\item
  R.~Goswami, P.~S.~Joshi and P.~Singh,
  Quantum evaporation of a naked singularity,
  arXiv:gr-qc/0506129.

\item  M.~Han, W.~Huang, Y.~Ma, Fundamental Structure of Loop Quantum
Gravity. arXiv:gr-qc/0509064.

\item R.~C.~Helling and G.~Policastro,
  String quantization: Fock vs. LQG representations,
  arXiv:hep-th/0409182.

\item
  G.~M.~Hossain,
  Large volume quantum correction in loop quantum cosmology: Graviton
  illusion?,
  arXiv:gr-qc/0504125.

\item V.~Husain and O.~Winkler,
  Quantum resolution of black hole singularities,
  arXiv:gr-qc/0410125.

\item V.~Husain and O.~Winkler,
  Quantum black holes,
  arXiv:gr-qc/0412039.

\item V.~Husain and O.~Winkler, How red is a quantum black hole?
arXiv:gr-qc/0503031.

\item W.~Kaminski, J.~Lewandowski, M.~Bobienski, Background independent
quantizations: the scalar field I. arXiv:gr-qc/0508091.

\item K.~Krasnov,
  Quantum gravity with matter via group field theory,
  arXiv:hep-th/0505174.

\item F.~Girelli and E.~R.~Livine,
  Physics of deformed special relativity: Relativity principle revisited,
  arXiv:gr-qc/0412004.

\item
  I.~B.~Khriplovich,
  Quantized Black Holes, Their Spectrum and Radiation,
  arXiv:gr-qc/0506082.

\item J.~Lewandowski, A.~Okolow, H.~Sahlmann and T.~Thiemann,
Uniqueness of diffeomorphism invariant states on holonomy-flux
algebras, arXiv:gr-qc/0504147.

\item T.~Liko and L.~H.~Kauffman,
  Knot theory and a physical state of quantum gravity,
  arXiv:hep-th/0505069.

\item  E. R. Livine and D. R. Terno, Quantum Black Holes: Entropy and
Entanglement on the Horizon. arXiv:gr-qc/0508085.

\item S.~K.~Maran,
Spin foams of real general relativity from that of complex general
relativity using a reality constraint.
  arXiv:gr-qc/0504092.

\item S.~K.~Maran, Real general relativity from complex
general relativity with a reality constraint.
  arXiv:gr-qc/0504091.

\item K.~A.~Meissner,
  Eigenvalues of the volume operator in loop quantum gravity.
  arXiv:gr-qc/0509049.

\item J.~Martinez, C.~Meneses and J.~A.~Zapata,
  Geometry of C-flat connections, coarse graining and the continuum limit,
  arXiv:hep-th/0507039.

\item F.~Mattei, C.~Rovelli, S.~Speziale and M.~Testa,
From 3-geometry transition amplitudes to graviton states,
  arXiv:gr-qc/0508007.

\item L.~Modesto, The Kantowski-Sachs space-time in loop quantum
gravity, arXiv:gr-qc/0411032.

\item L.~Modesto and C.~Rovelli, Particle scattering in loop
quantum gravity, arXiv:gr-qc/0502036.

\item
  L.~Modesto,
  Quantum gravitational collapse,
  arXiv:gr-qc/0504043.

\item
  L.~Modesto, Loop quantum black hole, arXiv:gr-qc/0509078.

\item
  J.~A.~Nieto,
  Ashtekar formalism and two time physics,
  arXiv:hep-th/0506253.

\item K.~Noui and A.~Perez, Dynamics of loop quantum gravity and
spin foam models in three  dimensions, arXiv:gr-qc/0402112.

\item K.~Noui and A.~Perez, Observability and geometry in three
dimensional quantum gravity, arXiv:gr-qc/0402113.

\item K.~Noui and A.~Perez, Three dimensional loop quantum
gravity: Coupling to point particles, arXiv:gr-qc/0402111.

\item N.J.~Nunes. Inflation: A graceful entrance from Loop Quantum
Cosmology, arXiv:astro-ph/0507683.

\bibitem{Perez:2004hj}
 A.~Perez, Introduction to loop quantum gravity and spin foams,
arXiv:gr-qc/0409061.

\bibitem{eft1} A.~Perez and C.~Rovelli, Physical effects of the Immirzi
parameter, arXiv:gr-qc/0505081.

\item
  A.~Randono,
  A generalization of the Kodama state for arbitrary values of the Immirzi
  parameter,
  arXiv:gr-qc/0504010.

\item C.~Rovelli,
  Graviton propagator from background-independent quantum gravity,
  arXiv:gr-qc/0508124.

\item C.~Rovelli and S.~Speziale,
On the perturbative expansion of a quantum field theory around a
topological sector,
  arXiv:gr-qc/0508106.

\item A.~Shojai and F.~Shojai, Causal loop quantum gravity and
cosmological solutions, arXiv:gr-qc/0409020.

\item F.~Shojai and A.~Shojai, Constraint algebra in causal loop
quantum gravity, arXiv:gr-qc/0409035.

\item
  P.~Singh,
  Effective state metamorphosis in semi-classical loop quantum cosmology,
  arXiv:gr-qc/0502086.

\item
  P.~Singh and K.~Vandersloot,
  Semi-classical states, effective dynamics and classical emergence in loop
  quantum cosmology,''
  arXiv:gr-qc/0507029.

\bibitem{Smolin:2004sx}
 L.~Smolin, An invitation to loop quantum gravity,
arXiv:hep-th/0408048.

 \item
  L.~Smolin,
  Falsifiable predictions from semiclassical quantum gravity,
  arXiv:hep-th/0501091.

\item J.~Swain, Entropy and area in loop quantum gravity,
arXiv:gr-qc/0505111.

\item T.~Tamaki, H.~Nomura, Ambiguity of black hole entropy
in loop quantum gravity. arXiv:hep-th/0508142.

\item
  T.~Thiemann,
  Reduced phase space quantization and Dirac observables,
  arXiv:gr-qc/0411031.

\item H.~Yepez, J.~M.~Romero and A.~Zamora, Corrections to the
Planck's radiation law from loop quantum gravity,
arXiv:hep-th/0407072.

\newpage

\end{enumerate}

\end{document}